\PassOptionsToPackage{prologue,usenames,dvipsnames,table}{xcolor}
\PassOptionsToPackage{bookmarks,unicode,colorlinks=true}{hyperref}

\newif\ifanonymous

\anonymousfalse

\ifanonymous
	\documentclass[acmsmall,review,anonymous]{acmart}
\else
	\documentclass[acmsmall]{acmart}
\fi

\makeatletter
\renewcommand\@formatdoi[1]{\ignorespaces}
\makeatother
\renewcommand\footnotetextcopyrightpermission[1]{}

\usepackage{amssymb,amsmath,mathtools}
\usepackage{amsthm}
\usepackage{thm-restate}
\usepackage{semantic} 
\usepackage{nicefrac}

\usepackage{verbatim}
\usepackage[only,fatsemi]{stmaryrd}
\usepackage{newtxmath} 
\usepackage{stackengine} 

\setstackgap{L}{1.9\normalbaselineskip} 
\stackMath
\usepackage{actuarialangle}
\usepackage{pifont}
\usepackage{wasysym} 
\usepackage[inline]{enumitem}
\usepackage[ruled,vlined,linesnumbered]{algorithm2e}
\usepackage[font=scriptsize,labelfont=bf,skip=4pt]{caption}
\usepackage{subcaption}
\usepackage{comment}
\usepackage{csquotes}
\usepackage{centernot}
\usepackage[all]{xy}
\usepackage{overpic}
\usepackage{epic,eepic}
\usepackage{makecell}
\usepackage{booktabs}
\usepackage{tensor}
\usepackage[usenames,dvipsnames,table]{xcolor}

\usepackage{pgf}
\usepackage{tikz}
\usetikzlibrary{arrows,shapes,fit,tikzmark,snakes,automata,backgrounds,petri,calc,hobby,positioning,fadings,through,arrows.meta,decorations.pathmorphing}

\usepackage{etoolbox}
\usepackage{adjustbox}
\usepackage{scalerel} 
\usepackage{xsavebox} 

\usepackage[normalem]{ulem}  

\usepackage{tabularx}
\usepackage{array}
\usepackage{multirow}
\usepackage{bm}

\usepackage[title,page]{appendix}

\usepackage{graphicx}
\graphicspath{{./figures/}}
\DeclareGraphicsExtensions{.pdf,.eps}
\usepackage{wrapfig}

\usepackage{picinpar} 

\usepackage[most]{tcolorbox}

\usepackage{relsize}

\usepackage{listings}
\usepackage[capitalize,nameinlink]{cleveref}
\crefname{lemma}{Lem.}{Lem.}
\crefname{example}{Exmp.}{Exmp.}
\crefname{section}{Sect.}{Sect.}
\Crefname{appendix}{Appx.}{Appx.}
\crefname{definition}{Def.}{Def.}
\crefname{theorem}{Thm.}{Thm.}
\crefname{corollary}{Cor.}{Cor.}
\crefname{algorithm}{Alg.}{Alg.}
\crefname{equation}{}{}
\crefname{enumi}{}{}

\crefname{lstinputlisting}{program}{programs}
\Crefname{lstinputlisting}{Program}{Programs}

\crefname{prob}{problem}{problems}
\Crefname{prob}{Problem}{Problems}
\creflabelformat{prob}{#2\textup{(#1)}#3}

\crefname{cha}{challenge}{challenges}
\Crefname{cha}{Challenge}{Challenges}

\usepackage{footnote}
\makesavenoteenv{figure}
\makesavenoteenv{table}

\theoremstyle{plain}
\newtheorem*{thm*}{Theorem}
\newtheorem*{exmp*}{Example}
\newtheorem*{counterexmp*}{Counterexample}

\newtheoremstyle{ourstyle}{}{}{\itshape}{}{\bfseries}{.}{ }{\thmname{#1}\thmnumber{ #2}\thmnote{ (#3)}}
\theoremstyle{ourstyle}

\newtheorem{theorem}{Theorem}
\newtheorem{definition}[theorem]{Definition}
\newtheorem{lemma}[theorem]{Lemma}

\newtheoremstyle{exmpstyle}{}{}{}{}{\bfseries}{.}{ }{\thmname{#1}\thmnumber{ #2}\thmnote{ (#3)}}
\theoremstyle{exmpstyle}

\newtheorem{example}[theorem]{Example}

\newtheoremstyle{rmkstyle}{}{}{}{}{\bfseries}{.}{ }{\thmname{#1}\thmnote{ (#3)}}
\theoremstyle{rmkstyle}

\newtheorem{remark}{Remark}
\input{macros}

\acmJournal{PACMPL}
\acmVolume{1}
\acmNumber{OOPSLA} 
\acmArticle{1}
\acmYear{2023}
\acmMonth{1}
\acmDOI{10.1145/nnnnnnn.nnnnnnn}
\startPage{1}

\setcopyright{none}

\makeatletter
\def\@fnsymbol#1{\ensuremath{\ifcase#1\or \dagger\or \ddagger\or
		\mathsection\or \mathparagraph\or \|\or **\or \dagger\dagger
		\or \ddagger\ddagger \else\@ctrerr\fi}}
\makeatother

\renewcommand{\rev}[1]{#1}

\begin{document}


\title[Lower Bounds for Possibly Divergent Probabilistic Programs]{Lower Bounds for Possibly Divergent Probabilistic Programs}         



\author{Shenghua Feng}
\orcid{0000-0002-5352-4954}             
\affiliation{
  \institution{SKLCS, Institute of Software, University of Chinese Academy of Sciences}            
  \city{Beijing}
  \country{China}                    
}
\email{fengsh@ios.ac.cn}          

\author{Mingshuai Chen}
\authornote{The corresponding author}          
\orcid{0000-0001-9663-7441}             
\affiliation{
	\institution{Zhejiang University}            
	\city{Hangzhou}
	\country{China}                    
}
\email{m.chen@zju.edu.cn}          

\author{Han Su}
\orcid{0000-0003-4260-8340}             
\affiliation{
	\institution{SKLCS, Institute of Software, University of Chinese Academy of Sciences}            
	\city{Beijing}
	\country{China}                    
}
\email{suhan@ios.ac.cn}          

\author{Benjamin Lucien Kaminski}
\email{b.kaminski@ucl.ac.uk}
\orcid{0000-0001-5185-2324}
\affiliation{%
	\institution{Saarland University, Saarland Informatics Campus}
	\city{Saarbr\"{u}cken}
	\country{Germany}
}
\affiliation{%
	\institution{University College London}
	\city{London}
	\country{United Kingdom}
}

\author{Joost-Pieter Katoen}
\orcid{0000-0002-6143-1926}             
\affiliation{
	\institution{RWTH Aachen University}            
	\city{Aachen}
	\country{Germany}                    
}
\email{katoen@cs.rwth-aachen.de}          

\author{Naijun Zhan}
\orcid{0000-0003-3298-3817}             
\affiliation{
	\institution{SKLCS, Institute of Software, University of Chinese Academy of Sciences}            
	\city{Beijing}
	\country{China}                    
}
\email{znj@ios.ac.cn}          


%
\begin{abstract}
We present a new proof rule for verifying lower bounds on quantities of probabilistic programs. Our proof rule is not confined to almost-surely terminating programs -- as is the case for existing rules -- and can be used to establish non-trivial lower bounds on, e.g., termination probabilities and expected values, for possibly \emph{divergent} probabilistic loops, e.g., the well-known three-dimensional random walk on a lattice.
\end{abstract}



\maketitle
\renewcommand{\shortauthors}{S.~Feng, M.~Chen, H.~Su, B.~L.~Kaminski, J.-P.~Katoen, and N.~Zhan}

\setlength{\floatsep}{1\baselineskip}
\setlength{\textfloatsep}{1\baselineskip}
\setlength{\intextsep}{1\baselineskip}
%
%
%
\section{Introduction}
Probabilistic programs~\cite{DBLP:journals/jcss/Kozen81,ACM:conf/fose/Gordon14,DBLP:journals/corr/abs-1809-10756,saheb1978probabilistic} extend deterministic programs with stochastic behaviors, e.g., random sampling, probabilistic choice, and conditioning (via posterior observations).
Probabilistic programs have witnessed numerous applications in various domains: They steer autonomous robots and self-driving cars~\cite{agentmodels,DBLP:conf/mfi/ShamsiFGN20}, are key to describe security~\cite{DBLP:journals/toplas/BartheKOB13} and quantum~\cite{DBLP:journals/toplas/Ying11} mechanisms, intrinsically code up randomized algorithms for solving NP-hard or even deterministically unsolvable problems (in, e.g., distributed computing~\cite{DBLP:journals/csur/Schneider93,DBLP:journals/jal/AspnesH90}), and are at the heart of modern machine learning
and approximate computing~\cite{DBLP:journals/cacm/CarbinMR16}. See~\cite{barthe_katoen_silva_2020} for recent advancements in probabilistic programming.

Probabilistic programs, though typically relatively small in size, are hard to grasp: The crux of probabilistic programming 
is to treat normal-looking programs as if they were \emph{probability distributions}~\cite{saheb1978probabilistic,Hicks-blog2014}.
Such a lift from deterministic program states to possibly infinite-support distributions (over states) renders the verification problem of probabilistic programs notoriously hard~\cite{DBLP:journals/acta/KaminskiKM19}. In particular, given a random variable $f$ (mapping program states to numbers), a key verification task is to reason about the \emph{expected value} of $f$ after termination of a program $\cc$ on input $\pstate$. If $f$ is the indicator function of an event $A$, then this expected value is the \emph{probability} that $A$ occurs upon termination of $\cc$. In case of a potentially \emph{unbounded loopy program} $\cc$, the expected value of $f$ is often characterized as the \emph{least fixed point} of some monotonic operator capturing the semantics of $\cc$ w.r.t.\ $f$. Computing the exact expected value of $f$ hence amounts to inferring the least fixed point which is in general highly intractable.

As a consequence, existing verification techniques for reasoning about probabilistic loops are mostly concerned with proving \emph{upper} and/or \emph{lower bounds} on expected values, i.e., on least fixed points. Verifying lower bounds is notably essential for establishing \emph{total correctness} of probabilistic programs \cite{DBLP:conf/birthday/KatoenGJKO15,DBLP:journals/tcs/McIverM01a} and for assessing the quality and tightness of upper bounds. For verifying a candidate upper bound $u$, the well-known principle of \emph{Park induction} \cite{park1969fixpoint,DBLP:journals/jcss/Kozen85}, or more generally, \emph{$\kappa$-induction} \cite{DBLP:conf/cav/BatzCKKMS21}, suffices by \enquote{pushing $u$ through the loop semantics} \emph{once}. Whereas for lower bounds on least fixed point, a \enquote{dual} version of Park induction is \emph{unsound} (see \cref{subsec:lower-induction-rules}).

Existing (sound) lower induction rules for probabilistic programs are confined to either
\begin{enumerate*}[label=(\roman*)]
	\item \emph{bounded} random variables with a priori knowledge on the \emph{termination probability} of the program \cite{DBLP:series/mcs/McIverM05}; or
	\item (universally) \emph{almost-surely terminating} (AST) programs (i.e., programs that terminate with probability 1 on all inputs) and \emph{uniformly integrable} random variables -- a notion from stochastic processes, which requires reasoning about looping times and/or bounds on random variables \cite{DBLP:journals/pacmpl/HarkKGK20}.
\end{enumerate*}
In contrast to Park induction for upper bounds, applying these lower induction rules requires heavy proof efforts in, e.g., looking for supermartingales~\cite{chatterjeeFOPP20} witnessing AST, checking uniform integrability, and inferring termination probabilities. \emph{In particular, none of these rules is capable of inferring lower bounds on termination probabilities strictly less than 1}, i.e., for non-AST (aka, \emph{divergent}) programs. Consider, e.g., the following probabilistic loop $\cc_{\textnormal{3dsrw}}$ modelling the well-known \emph{three-dimensional (3-D) random walk} on the lattice over $\Ints^3$.\footnote{The iterated symbol $\oplus$ is shorthand for discrete uniform choice (in this case, with probability $\nicefrac{1}{6}$ each).}
\begin{align*}
	\cc_{\textnormal{3dsrw}}\colon
	\quad
	&\WHILESYMBOL\left(\,x\neq 0\vee y\neq 0\vee z\neq 0\,\right)\left\{\,\right.\\
	&\quad \left.\ASSIGN{x}{x-1}~\oplus~\ASSIGN{x}{x+1}~\oplus~\ASSIGN{y}{y-1}~\oplus~\ASSIGN{y}{y+1}~\oplus~\ASSIGN{z}{z-1}~\oplus~\ASSIGN{z}{z+1}\,\right\}~.
\end{align*}%
The random nature underneath $\cc_{\textnormal{3dsrw}}$ is fundamentally \emph{different} from its 1- and 2-D counterparts: \citet{Polya1921} proved that the probability $\mathcal{P}$ that such a random walk returns to the origin at $(0, 0, 0)$ is strictly less than 1, indicating that $\cc_{\textnormal{3dsrw}}$ does \emph{not} terminate almost-surely. More precisely, the termination probability of $\cc_{\textnormal{3dsrw}}$ starting from any neighbor location of the origin is
\begin{align}\label{eq:3d-return-prob-intro}
	\mathcal{P} \eeq 1 - \left( \frac{3}{(2\pi)^3} \int^\pi_{-\pi} \int^\pi_{-\pi} \int^\pi_{-\pi} \frac{\dif\, x \dif\, y \dif\, z}{3 - \cos{x} - \cos{y} - \cos{z}}\right)^{-1}\! \eeq 0.3405373296\ldots\tag{$\dagger$}
\end{align}%
Existing verification techniques cannot tackle $\cc_{\textnormal{3dsrw}}$ due to its complex nature of divergence.
%
%

In this paper, we present a new proof rule, termed the \emph{guard-strengthening rule} for verifying lower bounds on the expected value of a potentially \emph{unbounded} random variable $f$ for a possibly \emph{divergent} probabilistic loop $\cloop= \WHILEDO{\guard}{\cc}$. \emph{Our proof rule employs reduction:} Suppose we aim to certify $l$ as a lower bound on the expected value of $f$ after termination of $\cloop$. We first forge a new loop $\pcloop= \WHILEDO{\guard'}{\cc}$ out of $\cloop$ by \emph{strengthening} its loop guard $\guard$ to $\guard'$, yielding a reduced problem where
\begin{enumerate*}[label=(\roman*)]
	\item\label{item:reduction-1}\!\! the modified loop $\pcloop$ features a \emph{stronger} termination property (e.g., provably AST), and
	\item\label{item:reduction-2}\!\! both the uniform integrability of $l$ and the boundedness conditions are \emph{easier} to verify.
\end{enumerate*}
Our proof rule then asserts -- by exploiting the \enquote{difference} between $\pcloop$ and $\cloop$ w.r.t.\ $f$ in terms of the \emph{weakest preexpectation calculus} \cite{DBLP:series/mcs/McIverM05,DBLP:journals/jcss/Kozen85} -- that \emph{a lower bound $l$ for $\pcloop$ (w.r.t.\ a restricted form of $f$) also suffices as a lower bound for $\cloop$ (w.r.t.\ $f$)}. The former, due to \cref{item:reduction-1,item:reduction-2} by guard strengthening, can often be established by applying the aforementioned lower induction rules or -- if $\pcloop$ has a \emph{finite} state space -- probabilistic model checking \cite{baier2008principles,DBLP:conf/lics/Katoen16,DBLP:conf/lics/Kwiatkowska03}. In this case, our proof rule can be (partially) \emph{automated} to derive \emph{increasingly tighter} lower bounds -- as $\guard'$ \enquote{approaches} $\guard$ -- on, e.g., the termination probability $\mathcal{P}$ of the 3-D random walk in \cref{eq:3d-return-prob-intro}, see details in \cref{ex:3drw}.

\smallskip
The main results of this paper are the following:
\begin{enumerate}[label=\textnormal{(\alph*)}]
	\item We present a new proof rule via \emph{guard strengthening} for verifying lower bounds on expected values of probabilistic programs. To the best of our knowledge, this is the first lower bound rule that admits \emph{divergent} probabilistic loops with \emph{unbounded} expected values.
	\item We show that the modified loops with strengthened guards feature \emph{easily provable} almost-sure termination and uniform integrability. This eases and enlarges the use of existing proof rules for lower bounds; Moreover, we propose a novel sufficient criterion for proving uniform integrability which recognizes cases that are \emph{out-of-reach} by existing sufficient conditions based on the optional stopping theorem \cite{DBLP:journals/pacmpl/HarkKGK20}.
	\item We show that the approximation error incurred by our guard-strengthening technique can be \emph{arbitrarily small} thereby yielding \emph{tight} lower bounds.
	\item We identify scenarios where our proof rule facilitates inferring quantitative properties of \emph{infinite}-state probabilistic programs by model checking \emph{finite}-state probabilistic models.
\end{enumerate}
We demonstrate the effectiveness of our proof rule on a collection of examples, including the 3-D random walk and a real-world randomized networking protocol.

\paragraph*{Paper Structure.} 
\cref{sec:overview} gives an overview of our approach via a simple example.
\cref{sec:wp-reasoning} recaps the weakest preexpectation ($\wpsymbol$) calculus as our semantic foundation. We formalize our problem in \cref{sec:problem} in position to existing proof rules. 
In \cref{sec:diff_wp_loop}, we exploit $\wpsymbol$-difference between loops which gives rise to our lower bound rule in \cref{sec:proof_rule_lower_bound}. 
We demonstrate the effectiveness of our proof rule through case studies in \cref{sec:case-studies}. \rev{The limitations of our guard-strengthening principle are addressed in \cref{sec:limitations}.} We discuss related work in \cref{sec:related-work} and draw conclusions in \cref{sec:conclusion}. Basic concepts in measure theory, e.g., (sub)probability measures, random variables, (stopped) stochastic processes, and stopping times, are introduced in \cref{app:preliminary}. Additional background materials, elaborated proofs, and details on the examples can be found in \cref{app:compositional-wp,
app:proofs,app:subinvariance}, respectively.
%
\section{Overview of Our Approach}\label{sec:overview}

In a nutshell, our idea is to transform a given potentially non-AST loop $\cloop$ into a provably AST loop $\pcloop$ and then certify lower bounds for $\pcloop$. Our transformation is performed in a way s.t.\ the expected outcome of $\pcloop$ is guaranteed to be a lower bound on the expected outcome of $\cloop$. Thus, as encoded in our proof rule, a lower bound for $\pcloop$ suffices as a lower bound for $\cloop$.

Let us demonstrate our approach by analyzing one of the most basic expected outcomes of a probabilistic loop: termination. Consider the 1-D random walk $\cc_{\textnormal{1dbrw}}$ on $\Ints$ (shown below on the left) with \emph{biased} probability $\nicefrac{1}{3}$ moving to the left and probability $\nicefrac{2}{3}$ moving to the right. Due to its biased nature, this loop does not terminate almost-surely and \emph{none of the existing proof rules\footnote{\rev{Referring to syntactic proof rules in the expectation-based program logic \cite{DBLP:series/mcs/McIverM05,DBLP:journals/jcss/Kozen85}.}} suffices} to establish non-trivial lower bounds on its termination probability; see details in \cref{ex:biased_RW}.

\vspace*{-1.7mm}
\begin{center}
\begin{minipage}[c]{0.45\textwidth}
	\begin{align*}
		\cc_{\textnormal{1dbrw}}\colon
		\quad
		&\WHILESYMBOL\left(\,0<n\,\right)\{\\
		&\quad	\ASSIGN{n}{n-1}~[\nicefrac{1}{3}]~{\ASSIGN{n}{n+1}}\\
		&\}
	\end{align*}%
\end{minipage}%
\qquad%
\begin{minipage}[c]{0.45\textwidth}
	\begin{align*}
		\cc^{\maroon{M}}_{\textnormal{1dbrw}}\colon
		\quad
		&\WHILESYMBOL\left(\,0<n~\maroon{<M}\,\right)\{\\
		&\quad	\ASSIGN{n}{n-1}~[\nicefrac{1}{3}]~{\ASSIGN{n}{n+1}}\\
		&\}
	\end{align*}%
\end{minipage}%
\end{center}%
\vspace*{-.5mm}

\noindent
Above right, we see the modified version of this loop, $\cc^M_{\textnormal{1dbrw}}$, which is obtained from $\cc_{\textnormal{1dbrw}}$ by introducing an artificial upper bound $M \in \NN$ on $n$ in the loop guard. This modified loop \emph{does} terminate almost-surely and can moreover visit only finitely many different states.

\begin{figure}[t]
	\centering
	\begin{minipage}[b]{.49\linewidth}
		\centering
		\begin{adjustbox}{max width=1\linewidth}
			\scalebox{1.7}{
				\begin{tikzpicture}[>=stealth',dot/.style={draw,circle,minimum size=1mm,inner sep=0pt,outer sep=0pt,fill=black},
					position label/.style={
						below = 3pt,
						text height = 1.5ex,
						text depth = 1ex
					},
					foo/.style={%
						->,
						shorten >=1pt,
						shorten <=1pt,
						decorate,
						decoration={%
							snake,
							segment length=1.64mm,
							amplitude=0.2mm,
							pre length=2pt,
							post length=2pt,
						}
					}
					]
					
					\draw (1,0) -- (7,0);
					\foreach \x in {1,2,3,4,5,7}
					\draw (\x cm,3pt) -- (\x cm,0pt);
					
					\draw[draw=Maroon, thick] (6 cm,27pt) -- (6 cm,0pt);
					\foreach \y in {0,3,6,9,12,15,18,21,24}
					\draw[draw=Maroon, thick] (6 cm,\y pt) --++(0.16cm,1.6pt);
					
					\node [position label] (dotsleft) at (1,0) {\small $\cdots$};
					\node [position label] (dotsright) at (7,0) {\small $\cdots$};
					\node [position label] (0) at (2,0) {\small $0$};
					\node [position label] (barrier) at (6,0) {\maroon{\small $M$}};
					\node [position label] (n) at (4,0) {\small $n$};
					
					\node (flag) at (2.04,.6) {\includegraphics[width=.06\linewidth]{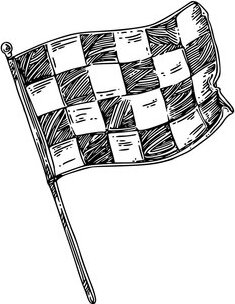}};
					
					\node[dot,draw=blue,fill=blue] (dot0) at (2,.2) {};
					\node[dot] (dot1) at (3,.2) {};
					\node[dot,draw=ForestGreen,fill=ForestGreen] (dot3) at (4,.2) {};
					\node[dot] (dot4) at (5,.2) {};
					\node[dot,draw=Gray,fill=Gray] (dot5) at (6,.2) {};
					\node[dot,draw=Gray,fill=Gray] (dot6) at (7,.2) {};
					
					\draw[foo] (dot3) to (dot1);
					\draw[foo] (dot1) to (dot0);
					
					\draw[foo,color=RedOrange] (dot3) to (dot4);
					\draw[foo,color=RedOrange] (dot4) to[out=140,in=40] (dot3);
					\draw[foo,color=RedOrange] (dot3) to[out=140,in=40] (dot1);
					\draw[foo,color=RedOrange] (dot1) to[out=140,in=40] (dot0);
					
					\draw[foo,color=Gray] (dot3) to[out=-30,in=-150] (dot4);
					\draw[foo,color=Gray] (dot4) to (dot5);
					\draw[foo,color=Gray] (dot5) to node (Xmark) {\Xmark} (dot6);
					\draw[foo,color=Gray] (dot6) to[out=100,in=80] (dot5);
					\draw[foo,color=Gray] (dot5) to[out=100,in=80] (dot4);
					\draw[foo,color=Gray] (dot4) to[out=100,in=80] (dot3);
					\draw[foo,color=Gray] (dot3) to[out=100,in=80] (dot1);
					\draw[foo,color=Gray] (dot1) to[out=100,in=80] (dot0);
					

				\end{tikzpicture}
			}
		\end{adjustbox}
		\caption{Effect of guard strengthening: In $\cc_{\textnormal{1dbrw}}$, all the three program traces (distinguished by colors) initiating from $\green{\bullet}$ \emph{terminate} at $\blue{\bullet}$. In $\cc^{\maroon{M}}_{\textnormal{1dbrw}}$, however, the gray trace crossing the \enquote{barrier} $\maroon{M}$ is no longer possible.}
		\label{fig:1-D-RW}
	\end{minipage}
	\hfill
	\begin{minipage}[b]{.49\linewidth}
		\centering
		\begin{adjustbox}{max width=.75\linewidth}
			\scalebox{1.4}{
				\begin{tikzpicture}[>=stealth',label distance=-.5mm,dot/.style={draw,circle,minimum size=.7mm,inner sep=0pt,outer sep=.8pt,fill=black},
					foo/.style={%
						->,
						shorten >=1pt,
						shorten <=1pt,
						decorate,
						decoration={%
							snake,
							segment length=1.64mm,
							amplitude=0.15mm,
							pre length=2pt,
							post length=2pt,
						}
					},
					foobold/.style={%
						-,
						shorten >=1pt,
						shorten <=1pt,
						decorate,
						decoration={%
							snake,
							segment length=1.64mm,
							amplitude=0.15mm,
							pre length=2pt,
							post length=2pt,
						}
					},
					]
					\node [inner sep=.6pt,outer sep=.6pt] (sigma) at (.8,1.6) {\scriptsize \green{$\pstate_0$}};
					\node [dot,label={[xshift=-.3cm,yshift=-.2cm]{\scriptsize $\nblue{\cc}$}}] (prog) at (1,.8) {};
					\node [inner sep=0pt,outer sep=.3pt] (ghost) at (2.4,0.4) {\scriptsize \rotatebox{45}{\maroon{\ldots}}};
					\node [dot,draw=blue,fill=blue,label=below:{{\scriptsize $\orange{f}(\blue{\pstate_{\scaleto{01}{3pt}}})$}}] (term1) at (0,0) {};
					\node [dot,draw=blue,fill=blue,label=below:{{\scriptsize $\orange{f}(\blue{\pstate_{\scaleto{02}{3pt}}})$}}] (term2) at (1,0) {};
					\node (term4) at (1.5,0) {\scriptsize \blue{\ldots}};
					\node [dot,draw=blue,fill=blue,label=below:{{\scriptsize $\orange{f}(\blue{\pstate_{\scaleto{0m}{3pt}}})$}}] (term3) at (2,0) {};
					
					\node (ghostwp) at (-.68,-.20) {};
					\node (wp) at (.89,-.23) {{\scriptsize $\EE\;$}$\big[$\qquad \qquad \qquad \quad \quad$\big]$};
					
					\draw[foo] (sigma) -- (term1);
					\draw[foobold] (sigma) -- (ghost);
					\draw[foo] (sigma) -- (prog);
					\draw[foo] (prog) -- (term2);
					\draw[foo] (prog) -- (term3);
					
					\draw (sigma) edge[|->,bend right=45,above left] node {{\scriptsize $\wp{\nblue{\cc}}{\orange{f}}$}} (ghostwp);
				\end{tikzpicture}
			}
		\end{adjustbox}
		\caption{Illustration of $\wpsymbol$: 
			$\wp{\nblue{\cc}}{\orange{f}}(\green{\pstate_0})$ determines the expected value of $\orange{f}$ evaluated in the final states $\blue{\bullet}$ reached after termination of $\nblue{\cc}$ on input $\green{\pstate_0}$; \rotatebox{45}{\maroon{\ldots}}\! indicates nonterminating (aka, divergent) path of $\nblue{\cc}$.}
		\label{fig:illustration-wp}
	\end{minipage}
\end{figure}

$\cc^M_{\textnormal{1dbrw}}$ terminates by \enquote{hitting} $n \leq 0$ or $n \geq M$. The key observation is that \emph{the probability of $\cc^M_{\textnormal{1dbrw}}$ terminating at $n \leq 0$ is smaller than the termination probability of $\cc_{\textnormal{1dbrw}}$}, since some terminating program traces of $\cc_{\textnormal{1dbrw}}$ -- contributing to its termination probability -- are no longer possible in $\cc^M_{\textnormal{1dbrw}}$ due to the artificial \enquote{barrier} $M$; see \cref{fig:1-D-RW} for an illustration.
Meanwhile, underapproximating the probability that $\cc^M_{\textnormal{1dbrw}}$ terminates at $n \leq 0$ -- thereby yielding a lower bound on the termination probability of $\cc_{\textnormal{1dbrw}}$ -- can be addressed by existing lower induction rules. In fact, since $\cc^M_{\textnormal{1dbrw}}$ has a finite state space for any fixed $M \in \NN$, its \emph{exact} termination probability at $n \leq 0$ can be obtained by probabilistic model checking. Moreover, if we push the \enquote{barrier} further to the right by increasing $M$, then we obtain \emph{increasingly tighter} lower bounds. See \cref{ex:biased_RW} for a detailed analysis.
%
\section{Weakest Preexpectation Reasoning}\label{sec:wp-reasoning}
%
\subsection{The Probabilistic Guarded Command Language}
We consider probabilistic programs described by the simple yet Turing-complete, imperative \emph{probabilistic guarded command language} ($\pgcl$)~\cite{DBLP:series/mcs/McIverM05} which augments Dijkstra's $\gcl$~\cite{DBLP:books/ph/Dijkstra76} with probabilistic choices and random assignments.

\paragraph*{Syntax.}
The syntax of a $\pgcl$ program $\cc$ adheres to the grammar%
%
\begin{align*}
	\cc \quad \Coloneqq \quad & \SKIP \mmid \ASSIGN{x}{\ee} \mmid \PASSIGN{x}{\mu} \mmid \COMPOSE{\cc}{\cc} \mmid \!{\PCHOICE{\cc}{p}{\cc}} \!\mmid\\
	& {\ITE{\guard}{\cc}{\cc}} \!\mmid \WHILEDO{\guard}{\cc}
\end{align*}%
where $x$ is a program variable taken from a countable set $\vars$, $\ee$ is an arithmetic expression over program variables, $\guard$ is a quantifier-free first-order predicate over program variables, and $\mu$ denotes a discrete or continuous distribution. We do not specify the syntax of expressions $e$ and predicates $\guard$ -- they can be arbitrary as long as the corresponding evaluation functions are measurable, as is in~\cite{DBLP:conf/setss/SzymczakK19}. 
The semantics of most program constructs -- including $\SKIP$, (deterministic) assignments, sequential composition, conditional, and (nested) loops -- is standard. The \emph{probabilistic choice} $\PCHOICE{\cc_1}{p}{\cc_2}$ flips a coin with bias $p \in [0, 1]$
and executes $\cc_1$ in case the coin yields heads, and $\cc_2$ otherwise. The \emph{random assignment} $\PASSIGN{x}{\mu}$ draws a sample from the distribution $\mu$ 
-- either discrete or continuous -- and assigns it to the program variable $x$.

\paragraph*{Program States.}
A \emph{program state} $\pstate$ maps every variable in $\vars$ to its value, 
i.e., a real number in $\Reals$.
We denote the (possibly uncountable) set of program states by
\[
	\States \ddefeq \!\left\{\, \pstate\mid \pstate\colon \vars\to \Reals \,\right\}~.
\]%
The evaluation of expressions $\ee$ and guards $\guard$ under a state $\pstate$, denoted by $\ee(\pstate)$ and $\guard(\pstate)$ respectively, is standard. For instance, the evaluation of arithmetic addition is
\[
	(\ee_1 + \ee_2)(\pstate) \ddefeq \ee_1(\pstate) + \ee_2(\pstate) \eeq \ee_1\subst{x}{\pstate(x)} + \ee_2\subst{x}{\pstate(x)}\ \ \text{for all}\ \ x \in \vars
\]%
where $\ee\subst{x}{\pstate(x)}$ denotes the substitution of variable $x$ by its value $\pstate(x)$ in $\ee$.

\paragraph*{Predicates.}
We interpret guards in $\pgcl$ programs as predicates. A \emph{predicate} $\guard$ represents a subset of program states $\States$.
We write $\pstate \models \guard$, reading \enquote{$\pstate$ satisfies $\guard$}, to indicate that state $\pstate$ is in the set represented by predicate $\guard$, i.e., $\guard(\pstate) = \TRUE$; and $\pstate \not\models \guard$ otherwise. We write $\guard_1 \!\implies\! \guard_2$, reading 
\enquote{$\guard_1$ strengthens $\guard_2$}, to indicate that 
under every state $\pstate \in \States$, if $\guard_1(\pstate) = \TRUE$ then $\guard_2(\pstate) = \TRUE$.

\subsection{The Weakest Preexpectation Calculus}

To reason about quantitative properties of probabilistic programs, in particular, to lower-bound expected values of certain probabilistic quantities, we view $\pgcl$ programs as \emph{expectation transformers}~\cite{DBLP:series/mcs/McIverM05,DBLP:phd/dnb/Kaminski19,DBLP:journals/jcss/Kozen85} -- a quantitative extension of the predicate-transformer calculus for non-probabilistic programs of \citet{DBLP:journals/cacm/Dijkstra75,DBLP:books/ph/Dijkstra76}.

An expectation transformer acts on real-valued functions called \emph{expectations}, which map program states to non-negative reals (extended by infinity)\footnote{\label{footnote:non-negative_expectation}For simplicity, we consider the standard case of \emph{non-negative} expectations. An arithmetic expression is thus a well-defined expectation if and only if it takes non-negative values over \emph{all reachable} program states. See \cite{DBLP:conf/lics/KaminskiK17} for more involved techniques addressing \emph{mixed-sign} expectations mapping to the \emph{full} extended reals.}.
Note the distinction between expectations and expected values: instead of an expected value, one can think of an expectation as a \emph{random variable}.%
%
\begin{definition}[Expectations~\textnormal{\cite{DBLP:phd/dnb/Kaminski19}}]
The set of \emph{expectations}, denoted by $\Expectations$, is defined as
\[
	\Expectations \ddefeq \!\left\{\, f\mid f\colon \States\to \PosRealsInf \,\right\}~.
\]%
An expectation $f \in \Expectations$ is \emph{finite}, written as $f \pprec  \infty$, if $f(\pstate) < \infty$ for all $\pstate \in \States$; $f \in \Expectations$ is \emph{bounded}, if there exists $b \in \PosReals$ such that $f(\pstate) \leq b$ for all $\pstate \in \States$.
\end{definition}%
\noindent{}%
For simplicity, a constant expectation $\lambda \pstate \mydot r$ which evaluates to $r \in \PosRealsInf$ for every state is denoted by $r$. Similarly, given an arithmetic expression $\ee$, we denote by $\ee$ the expectation $\lambda \pstate \mydot \ee(\pstate)$.

A partial order $\preceq$ on $\Expectations$ is obtained by point-wise lifting the canonical ordering $\leq$ on $\PosRealsInf$, i.e.,
\[
f_1 \ppreceq f_2 \qiff \forall \pstate\in\States\colon\ \ f_1(\pstate) \lleq f_2(\pstate)~.
\]%
$(\Expectations, {\preceq})$ forms a complete lattice with least element 
$0$ and greatest element 
$\infty$.

A $\pgcl$ program $\cc$  is interpreted as an expectation transformer which pushes a \emph{postexpectation} $f \in \Expectations$ (evaluated in the final states) backward through $\cc$ and gives a \emph{preexpectation} $g \in \Expectations$ (evaluated in the initial states).
%
In particular, as illustrated in \cref{fig:illustration-wp}, the \emph{weakest preexpectation of} $\cc$ \emph{w.r.t.}~$f$ is a \emph{function} $g\colon \States \to \PosRealsInf$ mapping each \emph{initial state} $\pstate_0$ of $\cc$ to the corresponding expected value of $f$ evaluated in the final states reached after termination of $\cc$ on input $\pstate_0$:
%
\begin{definition}[Weakest Preexpectations~\textnormal{\cite{DBLP:series/mcs/McIverM05,DBLP:phd/dnb/Kaminski19,DBLP:journals/jcss/Kozen85}}]
\label{def:wp}
Given probabilistic program $\cc\!$ and initial state $\pstate_0 \in \States$. 
Let $\measure{\pstate_0}{\cc}\!$ be the (sub)probability measure\footnote{\label{footnote:distribution explanation}
$\measure{\pstate_0}{\cc}(\pstate) \in [0,1]$ is the probability that, on input $\pstate_0$, $\cc$ terminates in the final state $\pstate$. Note that $\measure{\pstate_0}{\cc}(\States) \leq 1$, where the \enquote{missing} probability mass is the probability of \emph{nontermination} of $\cc$ on $\pstate_0$. A formal definition of $\measure{\pstate_0}{\cc}$ requires an (operational) semantic model of $\pgcl$, which is out of our scope; we refer interested readers to \cite{dahlqvist_silva_kozen_2020}.} over final states reached after termination of $\,\cc$ on input $\pstate_0$. Given postexpectation $f \in \Expectations$ which is measurable w.r.t.\ $\measure{\pstate_0}{\cc}$, the \emph{weakest preexpectation of} $\cc$ \emph{w.r.t.}~$f$ maps any initial state $\pstate_0$ to the expected value of $\,f$ evaluated in the final states reached after termination of $\,\cc$ on $\pstate_0$, i.e.,\footnote{In case of a countable state space $\States$, the integral can be written as a countable sum $\sum_{\pstate \in \States} \measure{\pstate_0}{\cc}(\pstate) \cdot f(\pstate)$.}
\[
	\wp{\cc}{f}(\pstate_0) \ddefeq \int_{\,\States}~f \dif\,\left(\measure{\pstate_0}{\cc}\right)~.
\]%
\end{definition}%
\noindent
It is known that
\begin{enumerate*}[label=(\roman*)]
	\item for every measurable $f \in \Expectations$, $\wp{\cc}{f}$ is measurable (cf.\ \cite[Lem.\ 3.2]{DBLP:conf/setss/SzymczakK19}), and
	\item the set of measurable expectations also forms a complete lattice under the partial order $\preceq$ (cf.\ \cite[Lem.\ 2]{DBLP:conf/setss/SzymczakK19}).
\end{enumerate*}
Hence, for simplicity, we abuse the notation $\Expectations$ to stand for the set of \emph{measurable} expectations throughout the rest of the paper.

Weakest preexpectations can be determined in a \emph{backward}, \emph{compositional} manner; see \cref{app:compositional-wp}. In fact,
the $\wpsymbol$-\emph{transformer} for all $\pgcl$ constructs can be codified by structural induction: 
\begin{theorem}[\textnormal{$\boldwpsymbol$}-Transformer~\textnormal{\cite{DBLP:series/mcs/McIverM05}}]
	\label{defQQQwp}
	Let $\pgcl$ be the set of programs in the probabilistic guarded command language.
	The \emph{weakest preexpectation transformer}%
	\[
		\wpsymbol\colon\, \pgcl \to \Expectations \to \Expectations
	\]%
	adhering to the rules in \textnormal{\cref{table:wp}} is well-defined; 
	\rev{in fact}, \textnormal{\cref{table:wp}} coincides with \textnormal{\cref{def:wp}}.
%
\end{theorem}
\noindent
A proof of \cref{defQQQwp} can be found in \cite[Sect.\ 5]{DBLP:conf/setss/SzymczakK19}, which extends the well-definedness for discrete probabilistic programs \cite[Thm.\ 4.11]{DBLP:phd/dnb/Kaminski19}.
The function $\charwp{\guard}{\cc}{f}$ in \cref{table:wp} is called the \emph{characteristic function of} $\WHILEDO{\guard}{\cc}\!$ \emph{w.r.t.}~$f$. For simplicity, we omit $\wpsymbol$, $\guard$, $\cc$, or $f$ from $\Phi$ whenever they are clear from the context. $\Phi$ is in fact a \emph{(Scott-)continuous} -- and thus \emph{monotonic} -- operator, i.e., $\Phi(\sup\{g_1 \preceq g_2 \preceq \ldots\}) = \sup\Phi(\{g_1 \preceq g_2 \preceq \ldots\})$; see \textnormal{\cite[Lem.\ 3.1]{DBLP:conf/setss/SzymczakK19}}. Thus by the Kleene fixed point theorem~\textnormal{\cite{LASSEZ1982112}}, its \emph{least fixed point} $\lfp \Phi = \sup_{n \in \NN}\Phi^n(0) = \lim_{n \rightarrow \omega}\Phi^n(0)$ and \emph{greatest fixed point} $\gfp \Phi = \inf_{n \in \NN}\Phi^n(\infty) = \lim_{n \rightarrow \omega}\Phi^n(\infty)$ exist over the partial order $\preceq$ on $\Expectations$. 

The rules for the $\wpsymbol$-transformer in \cref{table:wp} are compositional and, mostly, purely \emph{syntactic}, thus providing the machinery for automating the weakest preexpectation calculus; see \cref{app:compositional-wp} for an example. One exception, however, is the transformation rule for $\texttt{while}$-loops: It amounts to determining the quantitative least fixed point which is often difficult or even impossible to compute~\cite{DBLP:journals/acta/KaminskiKM19}; it is thus desirable to \emph{bound} them from above and/or from below. There are in principle two challenges (cf.\ \cite{DBLP:journals/pacmpl/HarkKGK20}):
\begin{enumerate*}[label=(\roman*)]
\item finding a candidate bound, and
\item verifying that the candidate is indeed an upper or lower bound.
\end{enumerate*}
In this paper, we aim to \emph{verify candidate lower bounds} on $\wp{\cc}{f}$ where $\cc$ is a (possibly nested) $\texttt{while}$-loop that may not terminate almost-surely. 
The termination probability of $\cc$ is captured by $\wp{\cc}{1}(\pstate_0)$:
\begin{table}[t]
	\caption{Rules for the $\wpsymbol$-transformer. $\iverson{\guard}$ denotes the \emph{Iverson-bracket} of $\guard$, i.e., $\iverson{\guard}(\pstate)$ evaluates to $1$ if $\pstate \models \guard$ and to $0$ otherwise. For any variable $x \in \vars$ and any expression $\ee$, $f\subst{x}{\ee}$ denotes the expectation with $f\subst{x}{\ee}(\pstate) = f(\pstate\subst{x}{\ee})$ for any $\pstate \in \States$, where $\pstate\subst{x}{\ee}(x) = \ee(\pstate)$ and $\pstate\subst{x}{\ee}(y) = \pstate(y)$ for all $y \in \vars \setminus\{x\}$.}
	\label{table:wp}
	\renewcommand{\arraystretch}{1.2}
	\begin{tabular}{@{\hspace{.5em}}l@{\hspace{1em}}l@{\hspace{.1em}}}
		\toprule
		$\boldsymbol{\cc}$        & $\boldwp{\cc}{f}$                                                                \\
		\midrule
		$\SKIP$                 & $f$                                                                            \\
		$\ASSIGN{x}{\ee}$                & $f\subst{x}{\ee}$           \\
		$\PASSIGN{x}{\mu}$                & $\int_{\,\Reals}~f\subst{x}{\nu} \dif\,\mu(\nu)$                                                                \\
		$\COMPOSE{\cc_1}{\cc_2}$           & $\wp{\cc_1}{\wp{\cc_2}{f}}$                                                        \\    
		$\PCHOICE{\cc_1}{p}{\cc_2}$        & $p \cdot \wp{\cc_1}{f} + (1 - p) \cdot \wp{\cc_2}{f}$                              \\                                             
		$\ITE{\guard}{\cc_1}{\cc_2}$       & $\iverson{\guard} \cdot \wp{\cc_1}{f} + \iverson{\neg \guard} \cdot \wp{\cc_2}{f}$ \\
		$\WHILEDO{\guard}{\primed{\cc}}$ & $\lfp \charwp{\guard}{\primed{\cc}}{f}$                                          \\[.65em]
		\midrule                                                                                                          
		\multicolumn{2}{c}{$\charwp{\guard}{\primed{\cc}}{f}\colon \Expectations \to \Expectations, \quad h \mapsto \iverson{\neg \guard}\cdot f + \iverson{\guard} \cdot \wp{\primed{\cc}}{h}\qquad
			\begin{array}{c}
				\textnormal{\footnotesize characteristic} \\[-.7em]
				\textnormal{\footnotesize function}
			\end{array}
			$}                                                                                                              \\
		\bottomrule  
	\end{tabular}
\end{table}
\begin{definition}[Almost-Sure Termination and Divergence]
	\label{def:ast-divergence}
	Let $\cc$ be a $\pgcl$ program and let $\pstate_0\in\States$ be an initial program state. Then $\cc$ \emph{terminates almost-surely on input} $\pstate_0$ iff
	\[
		\wp{\cc}{1}(\pstate_0) \eeq 1~.
	\]%
	$\cc$ \emph{terminates almost-surely (AST)} iff $\,\cc$ terminates almost-surely on all inputs, i.e.,
	\[
		\wp{\cc}{1}\! \eeq 1~.
	\]
	$\cc$ \emph{diverges on input} $\pstate_0$ iff $\,\wp{\cc}{1}(\pstate_0) < 1$. $\cc$ \emph{diverges} iff\, $\cc$ diverges on some input $\pstate_0$.
\end{definition}
%
%
\section{Reasoning about Lower Bounds}\label{sec:problem}

\rev{This section formulates our problem of proving lower bounds on $\lfp \!\charwp{\guard}{\cc}{f}$, i.e., on the $\wpsymbol$ of a (possibly divergent) $\texttt{while}$-loop $\cloop = \WHILEDO{\guard}{\cc}$ w.r.t.\ postexpectation $f$. We then give a high-level description of our approach in position to existing proof rules employing induction.}

\subsection{Problem Statement}

The problem concerned in this paper can be formulated as follows.
%
\begin{tcolorbox}[boxrule=1pt,colback=white,colframe=black!75]
	Given a \emph{possibly divergent} probabilistic loop $\cloop= \WHILEDO{\guard}{\cc}$, a \emph{possibly unbounded} postexpectation $f \in \Expectations$, and a \emph{possibly unbounded} candidate lower bound $l \in \Expectations$, verify that
	\begin{equation}\label{eq:problem-stat}
		l \ppreceq \wp{\cloop}{f}~.
	\end{equation}
\end{tcolorbox}
\noindent
We present a new proof rule to address this problem: Our rule does not employ induction, rather, it \emph{reduces} the verification of \cref{eq:problem-stat} with possibly divergent $\cloop$ and possibly unbounded $f,l \in \Expectations$ to
\begin{equation}\label{eq:problem-reduced}
	l \ppreceq \wp{\WHILEDO{\guard'}{\cc}}{[\neg \guard] \cdot f} \qwith \guard' \implies \guard~.
\end{equation}%
Namely, we forge a new loop $\pcloop = \WHILEDO{\guard'}{\cc}$ out of $\cloop$ by \emph{strengthening} its loop guard $\guard$ to $\guard'$. Such guard strengthening restricts the (reachable) state space and, consequently, 
\begin{enumerate*}[label=(\roman*)]
	\item the modified loop $\pcloop$ features a \emph{stronger} termination property (e.g., becoming AST), and
	\item both the uniform integrability of $l$ and the boundedness of expectations are \emph{easier} to verify.
\end{enumerate*}

Our proof rule asserts -- by exploiting the difference between $\wp{\pcloop}{f}$ and $\wp{\cloop}{f}$ -- that a lower bound $l$ w.r.t.\ $\pcloop$ satisfying \cref{eq:problem-reduced} also suffices as a lower bound w.r.t.\ $\cloop$ satisfying \cref{eq:problem-stat}. The former, due to guard strengthening, can often be obtained by applying 
\rev{existing lower induction rules (see \cref{subsec:lower-induction-rules} below)} or -- in case $\pcloop$ has a finite state space -- probabilistic model checking.

\subsection{Induction Rules for Upper Bounds}
%

The \emph{Park induction} principle~\cite{park1969fixpoint} for least fixed points establishes an elegant mechanism for verifying upper bounds on weakest preexpectations:%
\begin{theorem}[Park Induction for Upper Bounds~\textnormal{\cite{DBLP:journals/jcss/Kozen85,DBLP:phd/dnb/Kaminski19}}]
\label{thm:upper_bound_rule}
	Let $\charfunaked{f}$ be the characteristic function of\, $\cloop = \WHILEDO{\guard}{\cc}$ w.r.t.\ postexpectation $f \in \Expectations$ and let $u \in \Expectations$. Then
	\begin{equation}
		\label{eq:upper_bound_rule}
		\charfunaked{f} (u) \ppreceq u \qimplies \wp{\cloop}{f} \ppreceq u~.
	\end{equation}
\end{theorem}%
\noindent{}%
We call $u \in \Expectations$ satisfying $\charfunaked{f} (u) \preceq u$ a \emph{superinvariant}. As pointed out by \citet{DBLP:journals/pacmpl/HarkKGK20}, the striking power of Park induction lies in its \emph{simplicity}: Once an appropriate candidate $u$ is found (which, however, is usually not an easy task), all we have to do is to push $u$ through $\charfunaked{f}$ \emph{once} and check whether it becomes smaller in terms of $\preceq$. If this is the case, we have verified that $u$ is indeed an upper bound on $\lfp \charfunaked{f}$ and thus on the weakest preexpectation.

The \emph{soundness} of Park induction is illustrated by the left (descending) chain in \cref{fig:induction-rules}. We refer the readers to~\cite{DBLP:journals/pacmpl/HarkKGK20} for a formal soundness argument leveraging the Tarski-Kantorovitch principle (cf.\ \cite{jachymski_gajek_pokarowski_2000}). 
See also~\cite{DBLP:conf/cav/BatzCKKMS21} for a strictly more general proof rule via (latticed) $k$-induction for establishing upper bounds on least fixed points.

\subsection{Induction Rules for Lower Bounds}\label{subsec:lower-induction-rules}

A \enquote{dual} version of Park induction -- by flipping $\preceq$ in \cref{eq:upper_bound_rule} -- works for verifying lower bounds on the \emph{greatest} fixed point $\gfp \charfunaked{f}$, but not on $\lfp \charfunaked{f}$. More precisely, for $l \in \Expectations$, the rule
\begin{align*}
	l \ppreceq \charfunaked{f} (l) \qimplies l \ppreceq \wp{\cloop}{f}~, \tag*{\Large\lightning}
\end{align*}%
is \emph{unsound} in general. 
We call $l \in \Expectations$ satisfying $l \preceq \charfunaked{f} (l)$ a \emph{subinvariant} and the above unsound rule \emph{simple lower induction}. 
The \emph{unsoundness} of simple lower induction is illustrated by the right (ascending) chain in \cref{fig:induction-rules}, together with a counterexample below. We refer the readers to~\cite{DBLP:journals/pacmpl/HarkKGK20} for 
a formal argument again using the Tarski-Kantorovitch principle.

\begin{figure}[t]
	\centering
	\begin{adjustbox}{max width=.75\linewidth}
		\scalebox{1.4}{
			\begin{tikzpicture}[>=stealth',node distance=.4cm and .8cm, dot/.style={draw,circle,minimum size=.7mm,inner sep=0pt,outer sep=.8pt,fill=black},
				foo/.style={%
					->,
					shorten >=2pt,
					shorten <=2pt
				}
				]
				\node[dot,label=left:{\scriptsize $\green{u}$}] (u) at (0,0) {};
				\node[dot,below right=of u,label={[label distance=-.08cm]below left:{\scriptsize $\charfunaked{\orange{f}} \left(\green{u}\right)$}}] (phiu) {};
				\node[dot,below right=of phiu,label={[label distance=-.08cm]below left:{\scriptsize $\fp \charfunaked{\orange{f}}$}}] (fpphiu) {};
				\node[dot,below right=of fpphiu,label=below:{\scriptsize $\qquad \qquad \qquad \qquad \!\! \lfp \charfunaked{\orange{f}} = \wp{\nblue{\cloop}}{\orange{f}}$}] (lfpphiu) {};
				
				\node[dot,above right=of lfpphiu,label={[label distance=-.06cm]below right:{\scriptsize $\maroon{l}$}}] (l) {};
				\node[dot,above right=of l,label={[label distance=-.06cm]below right:{\scriptsize $\charfunaked{\orange{f}} \left(\maroon{l}\right)$}}] (phil) {};
				\node[dot,above right=of phil,label=right:{\scriptsize $\fp \charfunaked{\orange{f}}$}] (fpphil) {};
				
				\draw[foo] (phiu) to (u);
				\draw[foo,dashed] (fpphiu) to (phiu);
				\draw[foo] (lfpphiu) to (fpphiu);
				
				\path(lfpphiu) to[out=110,in=0] node (Cmark) {\Cmark} (u);
				\draw[foo] (lfpphiu) to[out=110,in=0] (u);
				
				\draw[foo] (l) to (phil);
				\draw[foo,dashed] (phil) to (fpphil);
				
				\path(l) to node (NAmark) {\NAmark} (lfpphiu);
				\draw[foo] (l) to (lfpphiu);
			\end{tikzpicture}
		}
	\end{adjustbox}
	\caption{Intuition of the soundness of Park induction (left branch) and the unsoundness of simple lower induction (right branch). An arrow from $g_1 \in \Expectations$ to $g_2 \in \Expectations$ indicates $g_1 \preceq g_2$. For Park induction, the iteration of $\charfunaked{\orange{f}}$ on $\green{u}$ converges downwards to a fixed point of $\charfunaked{\orange{f}}$ which is -- by the Knaster-Tarski theorem~\cite{knaster1928theoreme,tarski1955lattice,LASSEZ1982112} -- necessarily above $\lfp \charfunaked{\orange{f}}$, thus proving $\wp{\nblue{\cloop}}{\orange{f}} \preceq \green{u}$. 
		For simple lower induction, however, the ascending chain $\maroon{l} \preceq \charfunaked{\orange{f}} (\maroon{l}) \preceq \ldots$ converges to a fixed point of $\charfunaked{\orange{f}}$ which is necessarily below the \emph{greatest} fixed point $\gfp \charfunaked{\orange{f}}$, but we do not know how $\maroon{l}$ compares to $\lfp \charfunaked{\orange{f}}$.}
	\label{fig:induction-rules}
\end{figure}

\begin{example}[Unsoundness of Simple Lower Induction]
\rev{
Reconsider the loop $\cc_{\textnormal{1dbrw}}$ in \cref{sec:overview} with postexpectation $f = 1$; its characteristic function is $\charfunaked{f} (h) = \iverson{n \leq 0} + \iverson{n >0}\cdot (\nicefrac{1}{3}\cdot h(n-1) + \nicefrac{2}{3}\cdot h(n+1))$. Observe that the constant expectation $g = 1$ is a superinvariant, since $\charfunaked{f} (g) = 1 \preceq g$. This implies that the termination probability of $\cc_{\textnormal{1dbrw}}$ is (trivially) upper-bounded by $1$ (cf.\ \cref{thm:upper_bound_rule}). Meanwhile, $g$ is also a subinvariant as $g \preceq 1 = \charfunaked{f} (g)$, which however does \emph{not} suffice to certify $1$ as a lower bound on the termination probability (recall that $\cc_{\textnormal{1dbrw}}$ is non-AST; cf.\ \cref{sec:overview}).
}
\qedT
\end{example}

To retrieve soundness of lower induction, \citet{DBLP:journals/pacmpl/HarkKGK20} propose \emph{side conditions} relying on notions of almost-sure termination (i.e., $\wp{\cloop}{1} = 1$, cf.~\cref{def:ast-divergence}) and \emph{uniform integrability} from the realm of stochastic processes. To formulate the latter, we denote by $X_n$ the random variable representing the program state after the $n$-th iteration of the loop body $\cc$, by $T^{\neg \guard} \defeq \inf \{n \in \NN \mid X_n \models \neg \guard\}$ the stopping time (aka, \emph{looping time}) indicating the first time that $X_n$ hits $\neg \guard$,\footnote{The looping time $T^{\neg \guard}$ does not take into account the runtime of the loop body $\cc$.} and by $\left\{X_n^{T^{\neg \guard}}\right\}_{n\in \NN}$ the corresponding stopped stochastic process, i.e., $X_n^{T^{\neg \guard}}\! = X_n$ if $n \leq T^{\neg \guard}$ and $X_n^{T^{\neg \guard}}\! = X_{T^{\neg \guard}}$ otherwise; see \cref{app:preliminary} for formal definitions of stopping times and stopped processes.

\begin{definition}[Uniform Integrability]\label{def:ui}
	A stochastic process $\{X_n\}_{n \in \NN}$ on a probability space $(\Omega, \mathcal{F}, P)$ is \emph{uniformly integrable} (\emph{u.i.}, for short), if 
	\begin{equation}\label{eq:ui-def}
		\lim_{R\to \infty}\,\sup_{n \in \NN} \EE\left[\abs{X_n} \cdot \vmathbb{1}_{\abs{X_n} \geq R}\right] \eeq 0
	\end{equation}
	where $\vmathbb{1}_{\abs{X_n} \geq R}$ is the indicator function, i.e., $\vmathbb{1}_{\abs{X_n} \geq R} (\omega) = 1$ if $\omega \in \{\omega \in \Omega \mid \abs{X_n(\omega)} \geq R\}$ and $0$ otherwise. Given $\cloop = \WHILEDO{\guard}{\cc}$, an expectation $h \in \Expectations$ is \emph{uniformly integrable for $\cloop$} if $\left\{h\left(X_n^{T^{\neg \guard}}\right)\right\}_{n\in \NN}$ is uniformly integrable on the probability space induced by $\cloop$ (cf.~\textnormal{\cref{sec:diff_wp_loop}}).
\end{definition}
\noindent
Intuitively, \cref{eq:ui-def} asserts that the tail expected values of $X_n$ are \emph{uniformly} (indicated by the supremum) small in terms of the $L^1$-norm. A u.i.\ process $X_n$ thus satisfies $\EE(\lim_{n \rightarrow \infty} X_n) = \lim_{n \rightarrow \infty} \EE(X_n)$ if $\lim_{n \rightarrow \infty} X_n$ exists almost-surely. 
%
Now, the sound induction rule for lower bounds reads as follows.
\begin{theorem}[Hark et al.'s Induction for Lower Bounds~\textnormal{\cite{DBLP:journals/pacmpl/HarkKGK20}}]\label{thm:lower_bound_hark}
	Let $\charfunaked{f}$ be the characteristic function of\, $\cloop = \WHILEDO{\guard}{\cc}$ w.r.t.\ postexpectation $f \in \Expectations$ and let $l \in \Expectations$. Then
	\begin{align}\label{eq:hark-rule}
		\underbrace{l \ppreceq \charfunaked{f} (l)}_{\textnormal{subinvariance}} \aand \underbrace{\wp{\cloop}{1} = 1 \aand l\ \textnormal{is u.i.\ for}\ \cloop}_{\textnormal{side conditions}} \qimplies l \ppreceq \wp{\cloop}{f}~.
	\end{align}%
\end{theorem}
\noindent
Unlike Park induction for upper bounds, the side conditions in \cref{eq:hark-rule} for establishing lower bounds require extra efforts in proving almost-sure termination and uniform integrability, both of which are computationally \emph{intractable} in general, see, e.g., \cite{DBLP:journals/acta/KaminskiKM19}. Various techniques and tools have been developed in the literature, e.g., \cite{chatterjeeFOPP20,mciver2017new,DBLP:conf/fm/MoosbruggerBKK21}, to prove almost-sure termination of (\emph{subclasses} of) probabilistic programs. For showing uniform integrability, \citet{DBLP:journals/pacmpl/HarkKGK20} propose \emph{sufficient} conditions based on the well-known optional stopping time theorem~\cite[Chap.~10]{williams1991probability}:

\begin{theorem}[Sufficient Criteria for Uniform Integrability~\textnormal{\cite{DBLP:journals/pacmpl/HarkKGK20}}]\label{thm:sufficient_conditions_uniform}
	Given $\cloop = \WHILEDO{\guard}{\cc}$, let $\probability{\pstate_0}$ be the (sub)probability measure induced by $\cloop$ on initial state $\pstate_0 \in \States$.\footnote{The measure $\probability{\pstate_0}$ will be formally defined in \cref{sec:diff_wp_loop}. Note that $\probability{\pstate_0}(T^{\neg\guard} < \infty) = 1$ iff $\wp{\cloop}{1}(\pstate_0) = 1$.} Then, an expectation $h \pprec \infty$ is uniformly integrable for\, $\cloop$ if one of the following conditions holds:
	\begin{enumerate}[label=\textnormal{(\alph*)}]
		\item\label{item:ui-1} The looping time $T^{\neg\guard}$ is almost-surely bounded, i.e., for any initial state $\pstate_0 \in \States$, there exists $N\in \NN$ such that $\probability{\pstate_0}(T^{\neg\guard} \leq N) = 1$, and $\wpsymbol\llbracket \cc\rrbracket^n(h) \pprec \infty$ for any $n \in \NN$.
		\item\label{item:ui-2} The expected looping time is finite and $h$ is \emph{conditionally difference bounded}, i.e., 
		$\expectv{\pstate_0}[T^{\neg\guard}] < \infty$ for any $\pstate_0 \in \States$,
		and there exists $c \in \PosReals$ such that 
		$\wp{\cc}{\abs{h - h(\pstate)}} \leq c$
		for any $\pstate \models \guard$.
		\item\label{item:ui-3} $h$ is bounded, i.e., there exists $c \in \PosReals$ such that $h(\pstate) \leq c$ for any $\pstate \in \States$.
	\end{enumerate}
\end{theorem}%
\noindent
In summary, Hark et al.'s sound lower induction rule in \cref{thm:lower_bound_hark} does not apply to \emph{divergent} programs, and even for AST ones, it requires extra proof efforts in, e.g., looking for supermartingales~\cite{chatterjeeFOPP20} witnessing AST and reasoning about the looping time (\cref{item:ui-1,item:ui-2} in \cref{thm:sufficient_conditions_uniform}) or establishing bounds on expectations (\cref{item:ui-2,item:ui-3} in \cref{thm:sufficient_conditions_uniform}) to achieve u.i..

There is an orthogonal induction rule by~\citet{DBLP:series/mcs/McIverM05} for \emph{bounded} expectations:
\begin{theorem}[McIver \& Morgan's Induction for Lower Bounds~\textnormal{\cite{DBLP:series/mcs/McIverM05}}]\label{thm:MM-induction}
	Let $\charfunaked{f}$ be the characteristic function of\, $\cloop = \WHILEDO{\guard}{\cc}$ w.r.t.\ a \underline{bounded} postexpectation $f \in \Expectations$, $l \in \Expectations$ be a \underline{bounded} expectation such that $l \preceq \charfunaked{f} (l)$ and $[\neg \guard] \cdot l = [\neg \guard] \cdot f$, and $p = \wp{\cloop}{1}$ be the termination probability of\, $\cloop$. Then
	\begin{enumerate}[label=\textnormal{(\alph*)}]
		\item If $l = [G]$ for some predicate $G$, then $p \cdot l \preceq \wp{\cloop}{f}$.
		\item If $[G] \preceq p$ for some predicate $G$, then $[G] \cdot l \preceq \wp{\cloop}{f}$.
		\item If $\varepsilon \cdot l \preceq p$ for some $\varepsilon \in \RR_{>0}$, then $l \preceq \wp{\cloop}{f}$.
	\end{enumerate}
\end{theorem}%
\noindent
McIver and Morgan's lower induction rule applies to divergent programs (with termination probability $< 1$); however, it is confined to \emph{bounded} expectations and requires a priori knowledge on the \emph{termination probability} $p$ of a $\texttt{while}$-loop which is difficult to infer in general (which may in turn ask for lower bounds on $p$). In fact, Hark et al.'s induction rule generalizes McIver and Morgan's in case $\cloop$ is AST~\cite[Thm.~41]{DBLP:journals/pacmpl/HarkKGK20}. These two proof rules, to the best of our knowledge, are the only existing (induction) rules for verifying lower bounds on weakest preexpectations.

%

%

%

\section{Differences of Weakest Preexpectations}\label{sec:diff_wp_loop}

\rev{Our lower-bound proof rule reduces the verification of a probabilistic loop to that of its strengthened counterpart. To justify such a reduction, we need to \emph{quantitatively} relate two probabilistic loops in terms of weakest preexpectations.} In this section, we show how to quantify the difference of the weakest preexpectations of two $\texttt{while}$-loops, which differ \emph{only} in loop guards, with respect to the same postexpectation $f\in\Expectations$, namely,
\begin{equation}\label{eq:wp-diff}
	\wp{\underbrace{\WHILEDO{\guard}{\cc}}_{\cloop}}{f} \ -\  \wp{\underbrace{\WHILEDO{\guard'}{\cc}}_{\pcloop}}{f}
\end{equation}%
where $\guard$ and $\guard'$ are arbitrary predicates representing subsets of $\States$ (here, $\guard'$ does \emph{not} necessarily strengthen $\guard$)\footnote{\rev{Assuming $\guard' \!\implies\! \guard$ suffices to justify our proof rule. However, we are interested in a more general result on $\wpsymbol$-difference with \emph{unrelated} $\guard$ and $\guard'$ due to 
\begin{enumerate*}[label=(\roman*)]
	\item symmetry in $\guard$ and $\guard'$; and
	\item the potential of such a general result for addressing problems beyond verifying lower bounds, e.g., sensitivity analysis \cite{DBLP:journals/pacmpl/0001BHKKM21,DBLP:journals/pacmpl/WangFCDX20,DBLP:journals/pacmpl/BartheEGHS18} and model repair \cite{DBLP:conf/tacas/BartocciGKRS11} for probabilistic programs, as one of the interesting future directions.
\end{enumerate*}%
}} and the shared loop body $\cc$ itself can contain further nested $\texttt{while}$-loops. 
Such a quantification on the $\wpsymbol$-difference forms the basis of our new proof rule 
(cf.~\cref{sec:proof_rule_lower_bound}).

Recall that the weakest preexpectation of a loop w.r.t.\ $f \in \Expectations$ is defined (cf.\ \cref{def:wp}) as the integral of $f$ over the (sub)probability measure over final states reached after termination of the loop, where the termination behavior is determined by the loop guard and the loop body. Thus, to connect the two weakest preexpectations in \cref{eq:wp-diff}, we first abstract away the loop guards $\guard$ and $\guard'$ and thereby obtain a certainly divergent loop $\tcloop = \WHILEDO{\TRUE}{\cc}$.
We then construct a probability space over the set $\SS$ of \emph{infinite traces} (i.e., sequences of program states) of $\tcloop$, formally,
%
\begin{equation*}
    \SS \,\ddefeq \left\{\, \pstate_0 \pstate_1 \cdots \pstate_i \cdots \mid \pstate_0 \in \States, \ \forall i \geq 1\colon \pstate_i \in \States \cup \{\sink\} \land \left( \pstate_i = \sink \!\implies\! \pstate_{i+1} = \sink \right) \,\right\}
\end{equation*}%
where $\pstate_i \neq \sink$ denotes the state in which the loop body $\cc$ \emph{terminates} after its $i$-th iteration and $\sink$ denotes the \emph{sink} state where $\cc$ diverges. See \cref{fig:InfSeq} for an illustration of two types of infinite traces (with or without sink states). 
Given a finite prefix $\prefix$ of an infinite trace, the \emph{cylinder set} of $\prefix$ is 
$\cylinder(\prefix) \defeq \{\trace \in \SS \mid \exists t \in (\States \cup \{\sink\})^\omega \colon s = \prefix t\}$, i.e., the set of infinite traces that have $\prefix$ as a prefix.

\begin{figure}[t]
	\centering
	\begin{minipage}[b]{0.35\textwidth}
		\centering
		\begin{adjustbox}{max width=1\linewidth}
			\scalebox{1.4}{
				\begin{tikzpicture}[>=stealth',label distance=-.5mm,dot/.style={draw,circle,minimum size=.7mm,inner sep=0pt,outer sep=.8pt,fill=black},
					foo/.style={%
						->,
						shorten >=1pt,
						shorten <=1pt,
						decorate,
						decoration={%
							snake,
							segment length=1.64mm,
							amplitude=0.15mm,
							pre length=2pt,
							post length=2pt,
						}
					},
					foobold/.style={%
						-,
						shorten >=1pt,
						shorten <=1pt,
						decorate,
						decoration={%
							snake,
							segment length=1.64mm,
							amplitude=0.15mm,
							pre length=2pt,
							post length=2pt,
						}
					},
					fooboldloop/.style={%
						-,
						shorten >=1pt,
						shorten <=1pt,
						decorate,
						decoration={%
							snake,
							segment length=1mm,
							amplitude=0.1mm,
							pre length=2pt,
							post length=2pt,
						}
					},
					pics/random circle/.style={code={
							\tikzset{random circle/.cd,#1}
							\def\pv##1{\pgfkeysvalueof{/tikz/random circle/##1}}%
							\pgfmathsetmacro{\nextt}{360/\pv{samples}}
							\pgfmathsetmacro{\lastt}{360-\nextt}
							\draw plot[smooth cycle,variable=\t,samples at={0,\nextt,...,\lastt}] 
							(\t:\pv{radius}+rnd*\pv{amplitude});
					}},random circle/.cd,radius/.initial=1,amplitude/.initial=0.1,
					samples/.initial=8
					]
					
					\path (0,0) pic[thick]{random circle={radius=1.75,amplitude=0.1,samples=15}};
					\node at (0,1.55) {\scriptsize $\States \cup \{\textcolor{Maroon}{\sink}\}$};
					
					\node [dot,draw=ForestGreen,fill=ForestGreen,label=below:{{\scriptsize $\green{\pstate_{\scaleto{0}{3pt}}}$}}] (s0) at (-1.5,.6) {};
					\node [dot,label=above:{{\scriptsize $\pstate_{\scaleto{1}{3pt}}$}}] (s1) at (-0.75,.45) {};
					\node [dot,label=below:{{\scriptsize $\pstate_{\scaleto{2}{3pt}}$}}] (s2) at (0,.9) {};
					\node [dot,label={[xshift=.07cm,yshift=-.04cm]{\scriptsize $\pstate_{\scaleto{3}{3pt}}$}}] (s3) at (0.75,.65) {};
					\node [inner sep=.1pt,outer sep=.3pt] (sdots) at (1.5,.25) {\scriptsize \ldots};
					
					\node [dot,draw=ForestGreen,fill=ForestGreen,label=below:{{\scriptsize $\green{\pstate'_{\scaleto{0}{3pt}}}$}}] (sp0) at (-1.45,-.25) {};
					\node [dot,label=above:{{\scriptsize $\pstate'_{\scaleto{1}{3pt}}$}}] (sp1) at (-0.5,-.45) {};
					\node [dot,label=above:{{\scriptsize $\pstate'_{\scaleto{2}{3pt}}$}}] (sp2) at (.25,-.25) {};
					\node [dot,label=above:{{\scriptsize $\pstate'_{\scaleto{3}{3pt}}$}}] (sp3) at (1.05,-.75) {};
					\node [dot,draw=Maroon,fill=none,label={[xshift=.69cm,yshift=-.47cm]{\scriptsize $\scaleto{\textcolor{Maroon}{\sink}\!=\! \pstate'_{\scaleto{4}{3pt}}\!=\! \pstate'_{\scaleto{5}{3pt}} \!=\! \textcolor{Maroon}{\ldots}}{7.1pt}$}}] (sp4) at (-.25,-1.05) {};
					
					\draw[foo] (s0) -- (s1);
					\draw[foo] (s1) -- (s2);
					\draw[foo] (s2) -- (s3);
					\draw[foo,shorten >=2pt] (s3) -- (sdots);
					
					\draw[foo] (sp0) -- (sp1);
					\draw[foo] (sp1) -- (sp2);
					\draw[foo] (sp2) -- (sp3);
					\draw[foobold,shorten >=1pt] (sp3) -- (sp4);
					\draw[fooboldloop] (sp4) to [out=140,in=-140,loop,looseness=15] (sp4);
				\end{tikzpicture}
			}
		\end{adjustbox}
		\caption{Infinite traces in $\SS$.}
		\label{fig:InfSeq}
	\end{minipage}
	$\qquad$
	\begin{minipage}[b]{0.5\textwidth}
		\centering
		\begin{adjustbox}{max width=1\linewidth}
			\scalebox{1.4}{
				\begin{tikzpicture}[>=stealth',label distance=-.5mm,dot/.style={draw,circle,minimum size=.7mm,inner sep=0pt,outer sep=.8pt,fill=black},
					foo/.style={%
						->,
						shorten >=1pt,
						shorten <=1pt,
						decorate,
						decoration={%
							snake,
							segment length=1.64mm,
							amplitude=0.15mm,
							pre length=2pt,
							post length=2pt,
						}
					},
					foobold/.style={%
						-,
						shorten >=1pt,
						shorten <=1pt,
						decorate,
						decoration={%
							snake,
							segment length=1.64mm,
							amplitude=0.15mm,
							pre length=2pt,
							post length=2pt,
						}
					},
					pics/random circle/.style={code={
							\tikzset{random circle/.cd,#1}
							\def\pv##1{\pgfkeysvalueof{/tikz/random circle/##1}}%
							\pgfmathsetmacro{\nextt}{360/\pv{samples}}
							\pgfmathsetmacro{\lastt}{360-\nextt}
							\draw plot[smooth cycle,variable=\t,samples at={0,\nextt,...,\lastt}] 
							(\t:\pv{radius}+rnd*\pv{amplitude});
					}},random circle/.cd,radius/.initial=1,amplitude/.initial=0.1,
					samples/.initial=8
					]
					
					\path (-.85,0) pic[thick]{random circle={radius=1.75,amplitude=0.1,samples=15}};
					\path (.85,0) pic[thick]{random circle={radius=1.75,amplitude=0.1,samples=15}};
					\node at (-1,1.57) {\textcolor{blue}{\scriptsize $\guard$}};
					\node at (1,1.57) {\textcolor{Red}{\scriptsize $\guard'$}};
									
					\node [dot,draw=ForestGreen,fill=ForestGreen,label=below:{{\scriptsize $\green{\bar{\pstate}_{\scaleto{0}{3pt}}}$}}] (sp0) at (-1.25,.7) {};
					\node [dot] (sp1) at (-0.45,.85) {};
					\node [inner sep=.1pt,outer sep=.4pt] (spdots2) at (.32,.75) {\scriptsize \ldots};
					\node [dot,draw=blue,fill=blue,label=right:{\textcolor{blue}{\scriptsize $\bar{\pstate}_{\scaleto{A_2}{3.6pt}}$}}] (sp3) at (1.25,1.15) {};
					\node [inner sep=.1pt,outer sep=.4pt] (spdots1) at (-1.8,.95) {\scriptsize \ldots};
					\node [dot,draw=blue,fill=blue,label=above:{\textcolor{blue}{\scriptsize $\bar{\pstate}_{\scaleto{A_1}{3.6pt}}$}}] (sp4) at (-2.4,1.4) {};
					
					\draw[foo] (sp0) -- (sp1);
					\draw[foo,shorten >=2pt] (sp1) -- (spdots2);
					\draw[foo,shorten <=2pt] (spdots2) -- (sp3);
					\draw[foo,shorten >=2pt] (sp0) -- (spdots1);
					\draw[foo,shorten <=2pt] (spdots1) -- (sp4);
					
					\node [dot,draw=ForestGreen,fill=ForestGreen,label={[label distance=-.06cm]below:{{\scriptsize $\green{\tilde{\pstate}_{\scaleto{0}{3pt}}}$}}}] (spp0) at (1.25,.7) {};
					\node [dot] (spp1) at (.35,1) {};
					\node [inner sep=.1pt,outer sep=.4pt] (sppdots2) at (-0.15,1.23) {\scriptsize \ldots};
					\node [dot,draw=Red,fill=Red,label={[label distance=-.1cm]left:{\textcolor{Red}{\scriptsize $\tilde{\pstate}_{\scaleto{B_2}{3.6pt}}$}}}] (spp3) at (-.9,1.15) {};
					\node [inner sep=.1pt,outer sep=.4pt] (sppdots1) at (1.85,.85) {\scriptsize \ldots};
					\node [dot,draw=Red,fill=Red,label=above:{\textcolor{Red}{\scriptsize $\tilde{\pstate}_{\scaleto{B_1}{3.6pt}}$}}] (spp4) at (2.4,1.4) {};
					
					\draw[foo] (spp0) -- (spp1);
					\draw[foo,shorten >=2pt] (spp1) -- (sppdots2);
					\draw[foo,shorten <=2pt] (sppdots2) -- (spp3);
					\draw[foo,shorten >=2pt] (spp0) -- (sppdots1);
					\draw[foo,shorten <=2pt] (sppdots1) -- (spp4);
					
					\node [dot,draw=ForestGreen,fill=ForestGreen,label={[label distance=-.13cm]above right:{{\scriptsize $\green{\pstate_{\scaleto{0}{3pt}}}$}}}] (s0) at (0,-.2) {};
					\node [dot] (s11) at (-0.35,.15) {};
					\node [inner sep=.1pt,outer sep=.4pt] (sdots1) at (-.7,-.2) {\scriptsize \ldots};
					\node [dot,draw=Red,fill=Red,label={[label distance=-.06cm]left:{\textcolor{Red}{\scriptsize $\pstate_{\scaleto{D\left(\neg\guard',\guard\right)}{4.2pt}}$}}}] (s1end) at (-1.45,0) {};
					
					\draw[foo] (s0) -- (s11);
					\draw[foo,shorten >=2pt] (s11) -- (sdots1);
					\draw[foo,shorten <=2pt] (sdots1) -- (s1end);
					
					\node [inner sep=.1pt,outer sep=.4pt] (sdots2) at (.6,-.2) {\scriptsize \ldots};
					\node [dot] (s22) at (0.7,.3) {};
					\node [dot,draw=blue,fill=blue,label=right:{\textcolor{blue}{\scriptsize $\pstate_{\scaleto{D\left(\neg\guard,\guard'\right)}{4.2pt}}$}}] (s2end) at (1.45,.15) {};
					
					\draw[foo,shorten >=2pt] (s0) -- (sdots2);
					\draw[foo,shorten <=1pt] (sdots2) -- (s22);
					\draw[foo] (s22) -- (s2end);
					
					\node [dot] (s31) at (.45,-.6) {};
					\node [dot]
					 (s32) at (1.1,-.75) {};
					\node [inner sep=.1pt,outer sep=.4pt] (sdots3) at (1.55,-1.3) {\scriptsize \ldots};
					\node [dot,draw=Red,fill=Red,label=below:{\textcolor{Red}{\scriptsize $\pstate_{\scaleto{B}{2.5pt}}$}}] (s3end) at (2.4,-1.4) {};
					
					\draw[foo] (s0) -- (s31);
					\draw[foo] (s31) -- (s32);
					\draw[foo,shorten >=2pt] (s32) -- (sdots3);
					\draw[foo,shorten <=1pt] (sdots3) -- (s3end);
					
					\node [dot] (s41) at (-.5,-.7) {};
					\node [dot] (s42) at (-1.25,-.9) {};
					\node [inner sep=.1pt,outer sep=.4pt] (sdots4) at (-1.55,-1.35) {\scriptsize \ldots};
					\node [dot,draw=blue,fill=blue,label=below:{\textcolor{blue}{\scriptsize $\pstate_{\scaleto{A}{2.5pt}}$}}] (s4end) at (-2.4,-1.4) {};
					
					\draw[foo] (s0) -- (s41);
					\draw[foo] (s41) -- (s42);
					\draw[foo,shorten >=2pt] (s42) -- (sdots4);
					\draw[foo,shorten <=2pt] (sdots4) -- (s4end);
					
					\node [dot] (s51) at (-.07,-.65) {};
					\node [inner sep=.1pt,outer sep=.4pt] (sdots5) at (0,-1.2) {\scriptsize \ldots};
					\node [dot,draw=Violet,fill=Violet,label={[label distance=-.1cm]right:{\textcolor{Violet}{\scriptsize $\pstate_{\scaleto{C}{2.5pt}}$}}}] (s5end) at (-0.04,-1.78) {};
					
					\draw[foo] (s0) -- (s51);
					\draw[foo,shorten >=2pt] (s51) -- (sdots5);
					\draw[foo,shorten >=0pt] (sdots5) -- (s5end);

%
%
				\end{tikzpicture}
			}
		\end{adjustbox}
		\caption{Illustration of \cref{thm:exact_diff} (cf.~\cref{eq:proof-decomp}).}
		\label{fig:exact_diff}
	\end{minipage}
\end{figure}
%

The operational semantics~\cite{dahlqvist_silva_kozen_2020} of $\tcloop$ induces a family of (sub)probability measures -- denoted by $\probability{\pstate}$ for any $\pstate\in \States$ -- over $\SS$ with the smallest $\pstate$-algebra containing all cylinder sets. That is, for any finite prefix $\prefix = \pstate_0 \pstate_1 \cdots \pstate_n$, 
\begin{equation*}
	\probability{\pstate}\left(\cylinder(\prefix)\right) =
	\begin{cases}
		[\pstate = \pstate_0] \cdot \measure{\pstate_0}{\cc}(\pstate_1) \cdot \ldots \cdot \measure{\pstate_{i-1}}{\cc}(\pstate_i) \cdot \ldots \cdot \measure{\pstate_{n-1}}{\cc}(\pstate_n) &\text{if}\,\ \forall i \leq n \colon \pstate_i \neq \sink~,\\
		[\pstate = \pstate_0] \cdot \measure{\pstate_0}{\cc}(\pstate_1) \cdot 
		\ldots \cdot \measure{\pstate_{i-1}}{\cc}(\pstate_{i}) \cdot (1 - \measure{\pstate_{i}}{\cc}(\States)) &\text{if}\,\ \exists i < n \colon \pstate_{i} \neq \sink \land \pstate_{i+1} = \sink~, 
	\end{cases}%
\end{equation*}%
where $\measure{\pstate}{\cc}$ represents the (sub)probability measure over final states reached after termination of the loop body $\cc$ on input $\pstate$. The random variable $X_n\colon \SS \to \States$ representing the program state after the $n$-th iteration of the loop body $\cc$ is 
$X_n(\pstate_0\pstate_1\cdots) = \pstate_n$. Given $\cloop = \WHILEDO{\guard}{\cc}$, the looping time $T^{\neg\guard}\colon \SS\to \NN$ of $\cloop$ is defined as $T^{\neg\guard}(\trace)=\inf\{n \mid X_n(\trace) \models \neg\guard\}$. For any random variable $X$ over $\SS$ and any predicate $\phi$ over $X$, we abbreviate $\probability{\pstate}(\{\trace \in \SS \mid \phi(X(s))\})$ as $\probability{\pstate}(\phi(X))$, e.g., $\probability{\pstate}(T^{\neg\guard} \geq n)$ is a shorthand for $\probability{\pstate}(\{\trace \in \SS \mid T^{\neg\guard}(s) \geq n\})$.
Based on the measure $\probability{\pstate}$ defined above, the difference between $\wp{\cloop}{f}$ and $\wp{\pcloop}{f}$ as in \cref{eq:wp-diff} can be quantified as follows.
\begin{theorem}[\textnormal{$\boldwpsymbol$}-Difference]\label{thm:exact_diff}
Given loops\, $\cloop = \WHILEDO{\guard}{\cc}$ and\, $\pcloop = \WHILEDO{\guard'}{\cc}$, then, for any postexpectation $f\in\Expectations$,\footnote{For better understandability, \cref{eq:diff} is formulated in the form of \emph{difference} between weakest expectations. This formulation may raise the issue of \enquote{$\infty - \infty$}, however, one can avoid this issue by shifting all the negative terms in \cref{eq:diff} to the other side of the equation. The same reformulation tactic applies to the proof of \cref{thm:exact_diff}.}
\begin{multline}\label{eq:diff}
\wp{\cloop}{f} - \wp{\pcloop}{f} \eeq \\
\begin{aligned}
&\wp{\WHILEDO{\guard\wedge\guard'}{\cc}}{\left[\neg\guard\wedge \guard'\right] \cdot f} + \lambda \pstate \mydot \int_{A}~\fsub{\cloop} \dif\,\left(\probability{\pstate}\right) -\\
&\wp{\WHILEDO{\guard\wedge\guard'}{\cc}}{\left[\guard\wedge\neg \guard'\right] \cdot f} - \lambda \pstate \mydot \int_{B}~\fsub{\pcloop} \dif\,\left(\probability{\pstate}\right)~,
\end{aligned}%
\end{multline}%
where $A \subseteq \SS$ is the set of infinite traces that hit $\neg \guard'$ before hitting $\neg \guard$, and $\fsub{\cloop}$\! is a partial function mapping a trace $\trace \in \SS$ to $f(\pstate_n)$ if \,$\trace$ hits $\neg \guard$ for the first time at $\pstate_n$, namely,
\begin{equation}\label{eq:A-f_C_loop}
\begin{gathered}
    A \,\defeq\, \!\left\{\, \pstate_0 \pstate_1 \cdots \pstate_i \cdots \in \SS \mid 
    \exists n \in \NN \colon 
    \left(\pstate_n \models \neg\guard\right) \wedge \left(\forall i < n \colon \pstate_i \models \guard\right) \wedge \left(\exists k < n \colon \pstate_k \models \neg\guard'\right) \,\right\}~,\\
	\fsub{\cloop} \colon \SS \rightharpoonup \PosRealsInf \qquad
	\pstate_0 \pstate_1 \cdots \pstate_i \cdots \mapsto f(\pstate_n) \quad \textnormal{if}~\left(\pstate_n \models \neg\guard\right) \land \left(\forall i < n \colon \pstate_i \models \guard\right)~;
\end{gathered}%
\end{equation}%
dually, $B \subseteq \SS$ is the set of infinite traces that hit $\neg \guard$ before hitting $\neg \guard'$, and $\fsub{\pcloop}$\! maps a trace $\trace \in \SS$ to $f(\pstate_n)$ if \,$\trace$ hits $\neg \guard'$ for the first time at $\pstate_n$ (the definitions of\, $B$ and $\fsub{\pcloop}$\! are analogous to \textnormal{\cref{eq:A-f_C_loop}}).
%
\end{theorem}
\begin{proof}
	We prove the theorem by exploring different types of traces in $\SS$, as depicted in \cref{fig:exact_diff}.
	Given a predicate, $\phi$ 
	let $\event \phi$ be the set of all traces that \emph{eventually} hit $\phi$, i.e., 
	\begin{equation*}
		\event \phi \ddefeq \!\left\{\, \pstate_0 \pstate_1 \cdots \pstate_i\cdots \in \SS \mid \exists n \in \NN \colon \pstate_n \models \phi \,\right\}~.
	\end{equation*}%
	By definition of the (sub)probability measure $\probability{\pstate}$, we have
	\begin{equation}\label{eq:proof-event}
	\begin{alignedat}{2}
		\wp{\cloop}{f} &\eeq \lambda \pstate\mydot \int_{\States}~f \dif \,\left(\measure{\pstate}{\cloop}\right) & &\eeq \lambda \pstate\mydot \int_{\event(\neg\guard)}~\fsub{\cloop} \dif\,\left(\probability{\pstate}\right)~,\\
		\wp{\pcloop}{f} &\eeq \lambda \pstate\mydot \int_{\States}~f \dif \,\left(\measure{\pstate}{\pcloop}\right) & &\eeq \lambda \pstate\mydot \int_{\event(\neg\guard')}~\fsub{\pcloop} \dif\,\left(\probability{\pstate}\right)~.
	\end{alignedat}%
	\end{equation}%
	To quantify the $\wpsymbol$-difference, we first decompose the set of traces in $\event(\neg \guard)$ and $\event(\neg \guard')$ into disjoint parts, respectively. To this end, let
	\begin{equation*}
		C \ddefeq \!\left\{\, \pstate_0 \pstate_1 \cdots \pstate_i \cdots \in \SS \mid \exists n \in \NN \colon 
		\left(\pstate_n \models \neg\guard \wedge \neg \guard' \right) \wedge \left(\forall i < n \colon \pstate_i \models \guard \wedge \guard'\right) \,\right\}~,
	\end{equation*}%
	and for any predicates $\phi_1, \phi_2$, let
	\begin{equation*}
		D\left(\phi_1,\phi_2\right)\! \ddefeq \!\left\{\, \pstate_0 \pstate_1 \cdots \pstate_i \cdots \in \SS \mid \exists n \in \NN \colon 
		\left(\pstate_n \models \phi_1 \right) \wedge \left(\forall i \leq n \colon \pstate_i \models \phi_2\right) \,\right\}~.
	\end{equation*}%
	It follows that
	\begin{equation}\label{eq:proof-decomp}
	\begin{aligned}
		\event\left(\neg \guard\right) &\eeq \stackunder{A}{\overbrace{\substack{\text{hitting}~\neg\guard'~\text{before hitting}~\neg\guard\\\text{e.g.,}~\green{\pstate_0}\cdots\blue{\pstate_A}\cdots,\, \green{\bar{\pstate}_0}\cdots~\text{in \cref{fig:exact_diff}}}}} \uuplus
		\stackunder{C}{\overbrace{\substack{\text{hitting}~\neg\guard',\, \neg\guard~\text{simultaneously}\\\text{e.g.,}~\green{\pstate_0}\cdots\violet{\pstate_C}\cdots~\text{in \cref{fig:exact_diff}}}}} \uuplus
		\stackunder{D\left(\neg\guard,\guard'\right)}{\overbrace{\substack{\text{hitting}~\neg\guard~\text{while satisfying}~\guard'\\\text{e.g.,}~\green{\pstate_0}\cdots\blue{\pstate_{D(\neg\guard,\guard')}}\cdots~\text{in \cref{fig:exact_diff}}}}}~,\\
		\event\left(\neg \guard'\right) &\eeq \stackunder{B}{\overbrace{\substack{\text{hitting}~\neg\guard~\text{before hitting}~\neg\guard'\\\text{e.g.,}~\green{\pstate_0}\cdots\red{\pstate_B}\cdots,\, \green{\tilde{\pstate}_0}\cdots~\text{in \cref{fig:exact_diff}}}}} \uuplus
		\stackunder{C}{\overbrace{\substack{\text{hitting}~\neg\guard,\, \neg\guard'~\text{simultaneously}\\\text{e.g.,}~\green{\pstate_0}\cdots\violet{\pstate_C}\cdots~\text{in \cref{fig:exact_diff}}}}} \uuplus
		\stackunder{D\left(\neg\guard',\guard\right)}{\overbrace{\substack{\text{hitting}~\neg\guard'~\text{while satisfying}~\guard\\\text{e.g.,}~\green{\pstate_0}\cdots\red{\pstate_{D(\neg\guard',\guard)}}\cdots~\text{in \cref{fig:exact_diff}}}}}~.
	\end{aligned}%
	\end{equation}%
	%
	It is evident that, by definition, $\fsub{\cloop}$\! and $\fsub{\pcloop}$\! coincide on $C \subseteq \SS$, i.e., 
	\begin{equation}\label{eq:proof-C}
		\forall s \in C\colon\ \ \fsub{\cloop}\left(s\right) \eeq \fsub{\pcloop}\left(s\right)~.
	\end{equation}%
	For $D(\cdot,\cdot)$, we have (see \cref{lem:equality_for_D} in \cref{app:proof_equ12} for a more detailed proof of \cref{eq:proof-D} below)
	\begin{equation}\label{eq:proof-D}
	\begin{aligned}
		\lambda \pstate \mydot \int_{D(\neg\guard,\guard')}~\fsub{\cloop} \dif\,\left(\probability{\pstate}\right) &\eeq \underbrace{\wp{\WHILEDO{\guard\wedge\guard'}{\cc}}{\left[\neg\guard\wedge \guard'\right] \cdot f}}_{\text{evaluating}~f~\text{over, e.g.,}~\blue{\pstate_{D(\neg\guard,\guard')}}~\text{in \cref{fig:exact_diff}}}~,\\
		\lambda \pstate \mydot \int_{D(\neg\guard',\guard)}~\fsub{\pcloop} \dif\,\left(\probability{\pstate}\right) &\eeq \underbrace{\wp{\WHILEDO{\guard\wedge\guard'}{\cc}}{\left[\guard\wedge\neg \guard'\right] \cdot f}}_{\text{evaluating}~f~\text{over, e.g.,}~\red{\pstate_{D(\neg\guard',\guard)}}~\text{in \cref{fig:exact_diff}}}~.
	\end{aligned}%
	\end{equation}%
	By combining the facts above, we have
	\begingroup
	\allowdisplaybreaks
	\begin{align*}
		&\phantom{=\ \,}\wp{\cloop}{f} - \wp{\pcloop}{f} = \lambda \pstate\mydot\! \int_{\event(\neg\guard)}\fsub{\cloop}\! \dif\,\left(\probability{\pstate}\right) -
		\lambda \pstate\mydot\! \int_{\event(\neg\guard')}\fsub{\pcloop}\! \dif\,\left(\probability{\pstate}\right)\TAG{by \cref{eq:proof-event}}\\
		&
		\begin{aligned}[b]
			=\ \,&\lambda \pstate \mydot\! \int_{A}\fsub{\cloop}\! \dif\,\left(\probability{\pstate}\right) + \lambda \pstate \mydot\! \int_{C}\fsub{\cloop}\! \dif\,\left(\probability{\pstate}\right) + \lambda \pstate \mydot\! \int_{D(\neg\guard,\guard')}\fsub{\cloop}\! \dif\,\left(\probability{\pstate}\right) -\\
			&\lambda \pstate \mydot\! \int_{B}\fsub{\pcloop}\! \dif\,\left(\probability{\pstate}\right) - \lambda \pstate \mydot\! \int_{C}\fsub{\pcloop}\! \dif\,\left(\probability{\pstate}\right) - \lambda \pstate \mydot\! \int_{D(\neg\guard',\guard)}\fsub{\pcloop}\! \dif\,\left(\probability{\pstate}\right)
		\end{aligned}\TAG{by linearity of $\int$, \cref{eq:proof-decomp}}\\
		&
		\begin{aligned}[b]
			=\ \,&\wp{\WHILEDO{\guard\wedge\guard'}{\cc}}{\left[\neg\guard\wedge \guard'\right] \cdot f} + \lambda \pstate \mydot\! \int_{A}\fsub{\cloop}\! \dif\,\left(\probability{\pstate}\right) -\\
			&\wp{\WHILEDO{\guard\wedge\guard'}{\cc}}{\left[\guard\wedge\neg \guard'\right] \cdot f} - \lambda \pstate \mydot\! \int_{B}\fsub{\pcloop}\! \dif\,\left(\probability{\pstate}\right)~.
		\end{aligned}\TAG{by \cref{eq:proof-C,eq:proof-D}}%
	\end{align*}%
	\endgroup%
	This completes the proof.
\end{proof}

\begin{example}[\textnormal{$\boldwpsymbol$}-Difference]
\rev{
	Consider two loops with postexpectation $f = [0\leq n\leq 11]$:
	\begin{align*}
		\cloop \colon
		\quad &\WHILEDO{0<n<10}{\ASSIGN{n}{n-1}~[\nicefrac{1}{2}]~{\ASSIGN{n}{n+1}}}~,\\
		\pcloop \colon
		\quad &\WHILEDO{1<n<11}{\ASSIGN{n}{n-1}~[\nicefrac{1}{2}]~{\ASSIGN{n}{n+1}}}~.
	\end{align*}%
	Let $\cc_{\wedge\!}$ be the loop $\WHILESYMBOL(1<n<10)\{\ASSIGN{n}{n-1}\,[\nicefrac{1}{2}]\,\ASSIGN{n}{n+1}\}$. To illustrate \cref{thm:exact_diff}, we show
	\begin{multline}\label{eq:wp-difference-exmp}
		\wp{\cloop}{[0\leq n\leq 11]} - \wp{\pcloop}{[0\leq n\leq 11]} \eeq \\
		\begin{aligned}
			&\wp{\cc_{\wedge}}{\left[n=10\right]} + \lambda \pstate \mydot \int_{A}~\fsub{\cloop} \dif\,\left(\probability{\pstate}\right) -\\
			&\wp{\cc_{\wedge}}{\left[ n=1 \right]} - \lambda \pstate \mydot \int_{B}~\fsub{\pcloop} \dif\,\left(\probability{\pstate}\right)~.
		\end{aligned}%
	\end{multline}%
	Observe that 
	\begin{align*}
		\wp{\cloop}{[0\leq n\leq 11]} \eeq \wp{\pcloop}{[0\leq n\leq 11]} \eeq [0\leq n\leq 11]~,
	\end{align*}%
	due to the key fact that $\cloop$ and $\pcloop$ both terminate with probability $1$. Thus, the left-hand side of \cref{eq:wp-difference-exmp} equals $0$.
	For the right-hand side of \cref{eq:wp-difference-exmp}, by applying Hark et al.'s induction for lower bounds and Park induction for upper bounds (details omitted), one can show that
	\begin{align*}
	\wp{\cc_{\wedge}}{[n=10]} \eeq [1\leq n\leq 10] \cdot \frac{n-1}{9}~,\qquad
	\wp{\cc_{\wedge}}{[n=1]} \eeq [1\leq n\leq 10] \cdot \frac{10-n}{9}~.
	\end{align*}%
	Moreover, notice that $\fsub{\cloop} = 1$ over $A$ and $\fsub{\pcloop} = 1$ over $B$, we have
	\begin{align*}
		\int_{A}~\fsub{\cloop} \dif\,\left(\probability{\pstate}\right) \eeq \left[1\leq n <10\right]\cdot \frac{10-n}{9}~,\qquad 
		\int_{B}~\fsub{\pcloop} \dif\,\left(\probability{\pstate}\right) \eeq \left[1 < n \leq 10\right]\cdot \frac{n-1}{9}~.
	\end{align*}%
	It follows that the right-hand side of \cref{eq:wp-difference-exmp} also equals $0$.
	}
\qedT
\end{example}

The intuition behind \cref{thm:exact_diff}
is to decompose the weakest preexpectations into disjoint parts covering different types of traces in $\SS$ -- the type of a trace is determined by its temporal behavior of hitting $\neg\guard$ and/or hitting $\neg\guard'$, see \cref{eq:proof-decomp} and \cref{fig:exact_diff} -- thus yielding the \emph{exact} difference between $\wp{\cloop}{f}$ and $\wp{\pcloop}{f}$ by eliminating their common parts, cf.~\cref{eq:proof-C}. The integrals in~\cref{eq:diff} for their exclusive parts remain hard to resolve, however, it suffices to obtain a lower bound on $\wp{\cloop}{f}$ if $\guard'$ strengthens $\guard$
(tightness of the so-obtained lower bound is shown in \cref{sec:tightness}):
\begin{restatable}{corollary}{restateApproxDiffLower}
\label{thm:approx_diff_lower}
Given loops\, $\cloop = \WHILEDO{\guard}{\cc}$ and\, $\pcloop = \WHILEDO{\guard'}{\cc}$, suppose $\guard' \!\implies\! \guard$, then, for any postexpectation $f\in\Expectations$, 
\begin{equation*}
	\begin{aligned}
		\wp{\cloop}{f} \ssucceq \wp{\pcloop}{[\neg\guard] \cdot f}~.
	\end{aligned}%
\end{equation*}%
\end{restatable}
%
%
\noindent
The intuition of \cref{thm:approx_diff_lower} is as follows: Recall that $\wp{\cloop}{f}$ maps any initial state $\pstate_0$ to the expected value of $f$ evaluated in the final states reached after termination of $\cloop$ on $\pstate_0$, i.e., upon violating the loop guard $\guard$. In order to obtain a sound lower bound on $\wp{\cloop}{f}$ via the modified loop $\pcloop$, we have to restrict the postexpectation $f$ to $[\neg\guard] \cdot f$ such that $f$ is evaluated only in states violating the original guard $\guard$ (and hence also violating the strengthened guard $\guard'$).
The proof of \cref{thm:approx_diff_lower} leverages the fact that, as $\guard' \!\implies\! \guard$, we have $\guard \wedge \guard' = \guard'$, $[\neg\guard \wedge \guard'] = 0$, and $B = \emptyset$ (i.e., no trace can ever hit $\neg\guard$ before hitting $\neg\guard'$, cf.\ \cref{eq:proof-decomp}); see the detailed proof in \cref{app:proof_restateApproxDiffLower}.

\section{Proof Rule for Lower Bounds}\label{sec:proof_rule_lower_bound}

In this section, we present our new proof rule for verifying lower bounds on weakest preexpectations -- termed the \emph{guard-strengthening rule} -- based on the $\wpsymbol$-difference and the guard-strengthening technique in \cref{sec:diff_wp_loop}. We then showcase the usefulness of this proof rule in several aspects, in particular, for reasoning about possibly divergent probabilistic programs.
\begin{restatable}[Guard Strengthening for Lower Bounds]{theorem}{restateInferenceRule}
\label{thm:inference_rule}
Given loops\, $\cloop = \WHILEDO{\guard}{\cc}$, $\pcloop = \WHILEDO{\guard'}{\cc}$, and postexpectation $f \in \Expectations$, let $l \in \Expectations$, then the following inference rule holds:
\begin{equation}\label{eq:inference_rule}
\inference{
	\guard' \!\implies\! \guard ~&~
	l \ppreceq \wp{\pcloop}{[\neg \guard] \cdot f}
}{l \ppreceq \wp{\cloop}{f}}\ \ \makecell{\textnormal{\textsc{\footnotesize Guard-}}\\[-.15cm]\textnormal{\textsc{\footnotesize Strengthening}}}~.
\end{equation}%
\end{restatable}
\begin{proof}
The (soundness of the) proof rule in \cref{eq:inference_rule} is an immediate consequence of \cref{thm:approx_diff_lower}. We provide in \cref{app:proof_trace_agnostic} an alternative proof which is \emph{trace-agnostic} and thus simpler, yet does not contribute to our argument on the tightness of the proof rule in \cref{sec:tightness}.
\end{proof}
\noindent
Our guard-strengthening rule asserts that a lower bound $l$ on $\wp{\pcloop}{[\neg \guard] \cdot f}$ suffices as a lower bound on $\wp{\cloop}{f}$ provided that $\guard' \!\implies\! \guard$. Such guard strengthening restricts the (reachable) state space and, consequently, 
\begin{enumerate*}[label=(\roman*)]
	\item the modified loop $\pcloop$ features a \emph{stronger} termination property (e.g., becoming AST), and
	\item both the uniform integrability of $l$ and the boundedness of expectations are \emph{easier} to verify.
\end{enumerate*}
We will show that, due to these effects, our proof rule is 
\begin{enumerate}[label=(\alph*)]
	\item \emph{general} (\cref{label:subsec_generality,subsec:alternative-ui}): it is applicable to divergent $\cloop$ and unbounded $f,l \in \Expectations$ where existing rules, e.g., \cref{thm:lower_bound_hark,thm:MM-induction}, do not apply; even when $\cloop$ is AST, it is capable of certifying a lower bound $l$ which is \emph{not} uniformly integrable for $\cloop$ (thus beyond the scope of Hark et al.'s lower induction rule~\cite{DBLP:journals/pacmpl/HarkKGK20}); moreover, it admits an orthogonal sufficient criterion for proving uniform integrability.
	\item \emph{tight} (\cref{sec:tightness}): the error incurred by underapproximating $\wp{\cloop}{f}$ with $\wpsymbol\llbracket \pcloop\rrbracket\allowbreak\left([\neg \guard] \cdot f\right)$ approaches zero when $\guard'$ approaches $\guard$ appropriately; and
	\item \emph{automatable} (\cref{subsec:automation}): it is amenable to automation -- modulo an appropriate strengthening -- via probabilistic model checking in case $\pcloop$ has a finite state space.
\end{enumerate}
\begin{remark}
	The \emph{completeness} of the guard-strengthening rule in \cref{eq:inference_rule} is evident: If $l$ is a lower bound on $\wp{\cloop}{f}$, then there always exists $\guard'$ strengthening $\guard$ such that $l$ is a lower bound of $\wpsymbol\llbracket\pcloop\rrbracket\allowbreak\left([\neg \guard] \cdot f\right)$, as one can always choose $\guard' = \guard$ yielding $\wp{\pcloop}{[\neg \guard] \cdot f} = \wp{\cloop}{f}$. This completeness argument, however, shall not be viewed as a characterization of the generality of our proof rule in \emph{absolute} terms, as choosing $\guard' = \guard$ does not turn a non-AST loop into an AST one. We remark that characterizing the absolute generality of our proof rule is non-trivial as it depends highly on the \enquote{quality} of guard strengthening which may in turn rely on expert knowledge. Nonetheless, we provide useful heuristics in \cref{subsec:automation} for finding a \enquote{good} strengthening.
	\qedT
\end{remark}
\begin{remark}
	Our guard-strengthening rule in \cref{eq:inference_rule} applies also to \emph{non-probabilistic} loops $\cloop$. In this case, the weakest preexpectation transformer degenerates to the weakest precondition transformer of \citet{DBLP:journals/cacm/Dijkstra75,DBLP:books/ph/Dijkstra76}, $l$ and $f$ become predicates partially ordered by $\!\implies\!$, and $\cdot$ becomes $\wedge$. Our proof rule is then capable of verifying an underapproximation $l$ of the \emph{set of initial states} on which $\cloop$ certainly terminates in states satisfying $f$.
\end{remark}

\subsection{Application to Possibly Divergent Programs}\label{label:subsec_generality}


Our guard-strengthening rule reduces the verification of $l \preceq \wp{\cloop}{f}$ -- with possibly divergent $\cloop$ and possibly unbounded $f,l \in \Expectations$ -- to that of $l \preceq \wp{\pcloop}{[\neg \guard] \cdot f}$. In order to apply existing lower induction rules to the latter, one has to prove almost-sure termination of the modified loop $\pcloop$ and/or boundedness of expectations. To show AST, one may apply established techniques based on \emph{supermartingales}, e.g., \cite{chatterjeeFOPP20,mciver2017new,DBLP:conf/fm/MoosbruggerBKK21}. However, due to our guard-strengthening technique, a novel, simpler argument that does not rely on the discovery of fine-tuned supermartingales often suffices to prove AST of $\pcloop$:
%
\begin{restatable}[AST Witness]{lemma}{restateASTWitness}
	\label{lem:exponential_decrease}
	Given $\pcloop = \WHILEDO{\guard'}{\cc}$, let $T^{\neg\guard'}$\! be the looping time of\, $\pcloop$. If there exist $N \in \NN$ and $p \in (0,1)$ such that, for any initial state $\pstate \in \States$,
	\begin{equation}\label{eq:ast-witness}
		\probability{\pstate}\left(T^{\neg\guard'} > N \right)\! \lleq p~,
	\end{equation}%
	then\, $\pcloop$ terminates almost-surely.
\end{restatable}%
\begin{proof}[Proof (sketch)]
	It suffices to show that the \emph{tail probability} $\probability{\pstate}(T^{\neg\guard'} > n)$ (see, e.g., \textnormal{\cite{SankaranarayananFOPP2020}}), i.e., the probability that $\pcloop$ does not terminate within $n \in \NN$ steps, decreases exponentially in $n$. Then, by pushing $n$ ad infinitum, we have
		$\probability{\pstate}(T^{\neg\guard'} < \infty) = 1 - \lim_{n \to \infty} \probability{\pstate}(T^{\neg\guard'} > n) \geq 1 - 0 = 1$.
	Hence, $\pcloop$ terminates almost-surely. See an elaborated proof in \cref{app:proof_AST}.
\end{proof}

\begin{remark}
	How to specify the class of loops $\cloop$ for which there exist strengthened counterparts $\pcloop$ fulfilling the AST criterion in \cref{lem:exponential_decrease} is left as an interesting open question. This may also contribute to the characterization of the absolute generality of our proof rule as discussed above.
	\qedT
\end{remark}

There are (semi-)automated approaches, e.g., \cite{DBLP:conf/cav/ChatterjeeFG16}, tailored for proving exponentially decreasing nontermination probabilities via \emph{ranking-supermartingales}. 
In our setting, however,
\cref{lem:exponential_decrease} is arguably \emph{simpler} to verify since after strengthening the guard (by, e.g., bounding the variables, cf.\ \cref{subsec:automation}), one can usually \emph{read off} the constants $N$ and $p$ in \cref{eq:ast-witness} directly from the modified program with (certain) bounded variables. In fact, \cref{lem:exponential_decrease} suffices to witness AST of the modified loops in \emph{all} the examples presented in this paper; moreover, it enables an additional sufficient condition for establishing uniform integrability which is orthogonal to \cref{thm:sufficient_conditions_uniform} (cf.\ \cref{subsec:alternative-ui}).



\begin{example}[1-D Biased Random Walk on $\Ints$~\textnormal{\cite{mciver2017new}}]\label{ex:biased_RW}
Recall the example in \cref{sec:overview}:
\begin{align*}
	\cc_{\textnormal{1dbrw}}\colon
	\quad
	\WHILEDO{n > 0}{\ASSIGN{n}{n-1}~[\nicefrac{1}{3}]~{\ASSIGN{n}{n+1}}}~.
\end{align*}%
This loop terminates as soon as it reaches a state $\pstate$ with $\pstate(n) \leq 0$. We are interested in establishing non-trivial lower bounds on the termination probability of $\cc_{\textnormal{1dbrw}}$, i.e., on $\wp{\cc_{\textnormal{1dbrw}}}{f}$ with $f = 1$. However, \emph{none of the existing proof rules applies}: 
\begin{enumerate*}[label=(\roman*)]
\item $\cc_{\textnormal{1dbrw}}$ is not AST 
due to its biased nature (see, e.g., \cite{mciver2017new}), and thus Hark et al.'s lower induction \cite{DBLP:journals/pacmpl/HarkKGK20} does not apply; moreover,
\item it makes no sense to apply McIver and Morgan's lower induction \cite{DBLP:series/mcs/McIverM05} as having (a lower bound on) the termination probability -- the quantity we aim to infer -- is in turn one of the prerequisites for applying the rule (cf.\ \cref{thm:MM-induction}).
\end{enumerate*}

We now show how our guard-strengthening rule can be used to verify non-trivial lower bounds on $\wp{\cc_{\textnormal{1dbrw}}}{1}$. To this end, we strengthen the guard $(n > 0)$ of $\cc_{\textnormal{1dbrw}}$ to $\guard^M = (0 < n < M)$ for any fixed $M \in \NN$, and hence obtain a modified loop which differs from $\cc_{\textnormal{1dbrw}}$ only in the guard:
\begin{align*}
	\cc^M_{\textnormal{1dbrw}}\colon
	\quad
	\WHILEDO{0 < n < M}{\ASSIGN{n}{n-1}~[\nicefrac{1}{3}]~{\ASSIGN{n}{n+1}}}~.
\end{align*}%
Our proof rule asserts that, for any fixed $M \in \NN$, a lower bound on $\wp{\cc^M_{\textnormal{1dbrw}}}{[n \leq 0]\cdot 1}$ suffices as a lower bound on $\wp{\cc_{\textnormal{1dbrw}}}{1}$. Since $\cc^M_{\textnormal{1dbrw}}$ terminates after at most $M-1$ steps of consecutively moving to the right with probability $(\nicefrac{2}{3})^{M-1}$, we have $\probability{\pstate}\big(T^{\neg\guard^M}\! > M\big) < 1 - (\nicefrac{2}{3})^{M-1}$ for any initial state $\pstate(n) \in \Ints$. Thus, by \cref{lem:exponential_decrease}, $\cc^M_{\textnormal{1dbrw}}$ terminates almost-surely. We can therefore try to apply Hark et al.'s lower induction for a lower bound on $\wp{\cc^M_{\textnormal{1dbrw}}}{[n \leq 0]\cdot 1}$. Let
\begin{equation*}
	l_M \eeq 
	[n < 0] +
	[0 \leq n \leq M] \cdot \left(\left(\nicefrac{1}{2}\right)^n - \left(\nicefrac{1}{2}\right)^M\right)~.
\end{equation*}%
It is straightforward to check (see \cref{app:check_1dbrw}) that, for any fixed $M \in \NN$, $l_M$ is a subinvariant, i.e.,
\begin{equation*}
	l_M \ppreceq \charfunaked{g}^M\left(l_M\right)
\end{equation*}%
where $g = [n \leq 0]\cdot 1$ is the restricted postexpectation, and $\charfunaked{g}^M$ is the characteristic function of $\cc^M_{\textnormal{1dbrw}}$ with respect to $g$. Furthermore, $l_M$ is bounded from above by $1$, and thus is uniformly integrable for $\cc^M_{\textnormal{1dbrw}}$ due to condition \cref{item:ui-3} in \cref{thm:sufficient_conditions_uniform}. Consequently, we have
\begin{align*}
	\qquad\qquad l_M &\ppreceq \wp{\cc^M_{\textnormal{1dbrw}}}{[n \leq 0]\cdot 1}\TAG{by Hark et al.'s induction, cf.\ \cref{thm:lower_bound_hark}}\\
	&\ppreceq \wp{\cc_{\textnormal{1dbrw}}}{1}~.\TAG{by guard-strengthening rule, cf.\ \cref{thm:inference_rule}}
\end{align*}%
We therefore conclude that, for any fixed $M \in \NN$, $l_M$ is a lower bound on $\wp{\cc_{\textnormal{1dbrw}}}{1}$. In particular, by pushing $M$ ad infinitum -- to obtain the \emph{tightest} possible bound; this step is however \emph{not necessary} to establish a \emph{valid} lower bound --
we have
\begin{align*}
	\wp{\cc_{\textnormal{1dbrw}}}{1} \ssucceq \lim_{M\to \infty} l_M \eeq [n < 0] + [n \geq 0] \cdot \left(\nicefrac{1}{2}\right)^n~,
\end{align*}
namely, $l_\infty = [n < 0] + [n \geq 0] \cdot (\nicefrac{1}{2})^n$ is a lower bound on $\wp{\cc_{\textnormal{1dbrw}}}{1}$.
In fact, for this specific example, one can verify (see \cref{app:check_1dbrw}) that $l_\infty$ is a superinvariant, i.e., $\charfunaked{f} (l_\infty) \preceq l_\infty$, where $\charfunaked{f}$ is the characteristic function of $\cc_{\textnormal{1dbrw}}$ w.r.t.\ postexpectation $f = 1$. Thus, by Park induction (cf.\ \cref{thm:upper_bound_rule}), $l_\infty$ is also an upper bound on $\wp{\cc_{\textnormal{1dbrw}}}{1}$. We can thus claim that the \emph{exact} least fixed point of $\charfunaked{f}$, i.e., the \emph{exact termination probability of\, $\cc_{\textnormal{1dbrw}}$ starting from initial position $n$}, is
\begin{align*}
	\wp{\cc_{\textnormal{1dbrw}}}{1} \eeq l_\infty \eeq [n < 0] + [n \geq 0] \cdot \left(\nicefrac{1}{2}\right)^n~.
\end{align*}%
Finally, we remark that the reasoning process in this example applies to a more general form of one-dimensional random walk with \emph{parametrized} biased probability $p \in [0, \nicefrac{1}{2})$, whose exact termination probability can be shown to be $[n < 0] + [n \geq 0] \cdot (\nicefrac{p}{1-p})^n$.
\qedT
\end{example}


Next, we show via the following example that even when the targeted loop $\cloop$ almost-surely terminates, a genuine lower bound on $\wp{\cloop}{f}$ may \emph{not} be uniformly integrable for $\cloop$, thereby staying out-of-reach by Hark et al.'s lower induction rule; our guard-strengthening rule, however, is capable of certifying such a lower bound.

\begin{example}[Bernoulli's St.\ Petersburg Paradox~\textnormal{\cite{10.2307/1909829}}]\label{ex:pd}
Consider the following loop $\cc_{\textnormal{pd}}$ modelling a lottery game via fair-coin tosses: The stake $b \in \Ints_{>0}$ is doubled every time heads appears; the first time tails occurs, the game ends and the player wins all the stake in the pot:
\begin{align*}
	\cc_{\textnormal{pd}}\colon
	\quad
	\WHILEDO{a = 1}{\ASSIGN{a}{0}~[\nicefrac{1}{2}]~{\ASSIGN{b}{2 \cdot b}}}~.
\end{align*}%
We are interested in establishing non-trivial lower bounds on the expected payoff of the lottery, i.e., lower bounds on $\wp{\cc_{\textnormal{pd}}}{f}$ with $f = b$. Since $\cc_{\textnormal{pd}}$ terminates by setting $a$ to 0 with probability $\nicefrac{1}{2}$, we have $\probability{\pstate}(T^{a \neq 1} > 1) < \nicefrac{1}{2}$ for any initial state $\pstate$. Thus, by \cref{lem:exponential_decrease}, $\cc_{\textnormal{pd}}$ terminates almost-surely. However, the genuine lower bound $l_\infty = ([a \neq 1]\cdot b+[a=1]\cdot \infty) \npprec \infty$ -- which we ultimately aim to verify -- is \emph{not} uniformly integrable for $\cc_{\textnormal{pd}}$, and thus cannot be addressed by Hark et al.'s lower induction rule (see a detailed argument in \cite[Appx.~A]{DBLP:journals/corr/abs-1904-01117}).

To apply our new proof rule, we strengthen the guard $(a = 1)$ of $\cc_{\textnormal{pd}}$ to $\guard^M = (a = 1 \wedge b < M)$ for any fixed $M \in \Ints_{>0}$, and thereby obtain the modified loop:
\begin{align*}
	\cc^M_{\textnormal{pd}}\colon
	\quad
	\WHILEDO{a = 1 \wedge b < M}{\ASSIGN{a}{0}~[\nicefrac{1}{2}]~{\ASSIGN{b}{2 \cdot b}}}~.
\end{align*}%
Our proof rule asserts that, for any fixed $M \in \Ints_{>0}$, a lower bound on $\wp{\cc^M_{\textnormal{pd}}}{[a \neq 1]\cdot b}$ suffices as a lower bound on $\wp{\cc_{\textnormal{pd}}}{b}$. Program $\cc^M_{\textnormal{pd}}$ is clearly AST (by the same argument as for $\cc_{\textnormal{pd}}$); in fact, its looping time $T^{\neg \guard^M}$\! is certainly bounded by $\lceil \log_2 M \rceil$, and thus by condition \cref{item:ui-1} in \cref{thm:sufficient_conditions_uniform}, any expectation $h \pprec \infty$ is uniformly integrable for $\cc^M_{\textnormal{pd}}$ (as $\wpsymbol\llbracket \cc\rrbracket^n(h) \pprec \infty$ for any $n \in \NN$ is vacuously satisfied by $h \pprec \infty$ and loop-free body $\cc$, which is the case for $\cc^M_{\textnormal{pd}}$).
Let 
\begin{align*}
	l_M \eeq [a\neq 1] \cdot b + \sum\nolimits_{i = 1}^{\lceil \log_2 M \rceil}\left[a =1 \wedge \left(\nicefrac{M}{2^i} \leq b < \nicefrac{M}{2^{i-1}}\right)\right] \cdot i \cdot \nicefrac{b}{2}~.
\end{align*}%
It is straightforward to check (see \cref{app:check_pd}) that, for any fixed $M \in \Ints_{>0}$, $l_M$ is a subinvariant, i.e.,
\begin{equation*}
	l_M \ppreceq \charfunaked{g}^M\left(l_M\right)
\end{equation*}%
where $g = [a \neq 1]\cdot b$ is the restricted postexpectation, and $\charfunaked{g}^M$ is the characteristic function of $\cc^M_{\textnormal{pd}}$ with respect to $g$. Since $l_M \pprec \infty$ is uniformly integrable for $\cc^M_{\textnormal{pd}}$, by \cref{thm:lower_bound_hark,thm:inference_rule}, we have 
\begin{align*}
	l_M \ppreceq \wp{C^M_{\textnormal{pd}}}{[a\neq 1] \cdot b} \ppreceq \wp{C_{\textnormal{pd}}}{b}
\end{align*}%
for any fixed $M \in \Ints_{>0}$. It follows that, when $M$ tends to infinity,
\begin{align*}
	\wp{C_{\textnormal{pd}}}{b} \ssucceq \lim_{M\to \infty} l_M \eeq [a \neq 1]\cdot b+[a=1]\cdot \infty \eeq l_\infty~.
\end{align*}%
In analogy to \cref{ex:biased_RW}, one can further show that $l_\infty$ suffices as a superinvariant and ultimately conclude that $\wp{C_{\textnormal{pd}}}{b} = l_\infty$, that is, the \emph{exact} expected value of $b$ upon termination.
\qedT
\end{example}

\subsection{Orthogonal Sufficient Condition for Uniform Integrability}\label{subsec:alternative-ui}

%

\Cref{ex:biased_RW,ex:pd} demonstrate how we can certify lower bounds on $\wp{\cloop}{f}$ leveraging existing sufficient criteria for uniform integrability (cf.\ \cref{thm:sufficient_conditions_uniform}) for the modified loop $\pcloop$. The fact that $\pcloop$ often features a strong and easily checkable termination property (cf.\ \cref{lem:exponential_decrease}) enables an alternative sufficient condition for establishing uniform integrability (see proof in \cref{app:proof_addui}):
\begin{restatable}[Orthogonal Sufficient Criterion for Uniform Integrability]{theorem}{restateSuffUI}
	\label{thm:additional_conditions_uniform}
	Suppose $\pcloop = \WHILEDO{\guard'}{\cc}$ satisfies the AST condition in \textnormal{\cref{lem:exponential_decrease}}. Then, an expectation $h \pprec \infty$ is uniformly integrable for\, $\pcloop$ if\,
	\begin{enumerate*}[label=(\roman*)]
		\item \,$\pcloop$ has the \emph{bounded update property}\footnote{The bounded update property has been used to analyze expected costs for nondeterministic probabilistic programs \cite{DBLP:conf/pldi/Wang0GCQS19}; it subsumes the class of constant probability programs \cite{DBLP:conf/cade/GieslGH19}.}, i.e., there exists $c \in \PosReals$ such that for any $\pstate\in \States$, 
		$\probability{\pstate}(\forall n \in \Nats\colon \abs{X_{n+1} - X_{n}} \leq c) = 1$, and
		\item\label{item-polynomial-bound} \!$h$ is bounded by a polynomial expectation\footnote{One can replace \cref{item-polynomial-bound} with (ii') $h$ is bounded by a \emph{piece-wise} polynomial expectation, and the theorem still holds.}.
	\end{enumerate*}
\end{restatable}
\noindent
The sufficient criterion for uniform integrability in \cref{thm:additional_conditions_uniform} is orthogonal to those in \cref{thm:sufficient_conditions_uniform}, in particular, to condition \cref{item:ui-2} thereof: Our new sufficient condition leverages the fact that $\pcloop$ with a strengthened guard often features a strong and easily checkable termination property (cf.\ \cref{lem:exponential_decrease}); it is confined to loops with the bounded update property, yet relaxes the conditional difference boundedness of $h$ to the polynomial boundedness of $h$. Our new sufficient criterion is usually \emph{easier} to check than condition \cref{item:ui-2} in \cref{thm:sufficient_conditions_uniform} in the sense that
\begin{enumerate*}[label=(\roman*)]
	\item in case $\pcloop$ has a loop-free body, one can simply read off the bounded update property from the syntax of $\pcloop$, and
	\item any (piece-wise) polynomial expectation $h$ fulfils the polynomial boundedness condition.
\end{enumerate*}
The following example demonstrates a scenario where \emph{none} of the conditions in \cref{thm:sufficient_conditions_uniform} suffices to show uniform integrability, but our (orthogonal) sufficient condition in \cref{thm:additional_conditions_uniform} does.
\begin{example}[1-D Symmetric Random Walk with Bounded Updates]\label{ex:additional-ui}
Consider the following loop $\cc_{\textnormal{bu}}$ modelling a symmetrized 1-D random walk on $\Ints$, where we additionally increase (resp.\ decrease) the value of $b \in \PosReals$ by 2 along every left (resp.\ right) move with probability $\nicefrac{1}{2}$:
\begin{align*}
	\cc_{\textnormal{bu}}\colon
	\quad
	\WHILEDO{n > 0}{\{\COMPOSE{\ASSIGN{n}{n-1}}{\ASSIGN{b}{b+2}}\}~[\nicefrac{1}{2}]~\{\COMPOSE{\ASSIGN{n}{n+1}}{\ASSIGN{b}{b-2}}\}}~.
\end{align*}%
We are interested in certifying non-trivial lower bounds on the expected value of $b$ upon termination, i.e., lower bounds on $\wp{\cc_{\textnormal{bu}}}{f}$ with $f = b$. Notice that $\cc_{\textnormal{bu}}$ exhibits exactly the same termination behavior as the standard 1-D symmetric random walk \cite{mciver2017new}, namely, $\cc_{\textnormal{bu}}$ terminates almost-surely yet with an \emph{infinite} expected looping time. Moreover, the candidate lower bound $l_\infty = [n<0]\cdot b + [n\geq 0]\cdot (b+2 \cdot n)$ -- which we ultimately aim to verify -- is \emph{unbounded}, thereby out-of-reach by Hark et al.'s lower induction (as \cref{thm:sufficient_conditions_uniform} does not suffice to show u.i.\ of $l_\infty$).

To apply our new proof rule, we strengthen the guard $(n > 0)$ of $\cc_{\textnormal{bu}}$ to $\guard^M = (0 < n < M)$ for any fixed $M \in \Ints_{>0}$, and thereby obtain the modified loop:
\begin{align*}
	\cc^M_{\textnormal{bu}}\colon
	\quad
	\WHILEDO{0 < n < M}{\{\COMPOSE{\ASSIGN{n}{n-1}}{\ASSIGN{b}{b+2}}\}~[\nicefrac{1}{2}]~\{\COMPOSE{\ASSIGN{n}{n+1}}{\ASSIGN{b}{b-2}}\}}~.
\end{align*}%
Our proof rule asserts that, for any fixed $M \in \Ints_{>0}$, a lower bound on $\wp{\cc^M_{\textnormal{bu}}}{[n\leq 0]\cdot b}$ suffices as a lower bound on $\wp{\cc_{\textnormal{bu}}}{b}$. Let 
\begin{align*}
	l_M \eeq [n<0] \cdot b + [0 \leq n \leq M] \cdot \left(b+2 \cdot n\right) \cdot \left(1 - \nicefrac{n}{M}\right)~.
\end{align*}%
One can readily verify that, for any fixed $M \in \Ints_{>0}$, $l_M$ is a subinvariant (see \cref{app:check_bu}).
Since $\cc^M_{\textnormal{bu}}$ terminates after at most $M-1$ steps of consecutively increasing $n$ with probability $(\nicefrac{1}{2})^{M-1}$, we have $\probability{\pstate}\big(T^{\neg\guard^M}\! > M\big) < 1 - (\nicefrac{1}{2})^{M-1}$ for any initial state $\pstate(n) \in \Ints$. Thus, by \cref{lem:exponential_decrease}, $\cc^M_{\textnormal{bu}}$ terminates almost-surely. 
Nevertheless, still \emph{none of the conditions in \textnormal{\cref{thm:sufficient_conditions_uniform}} suffices to show that $l_M$ is uniformly integrable for $\cc^M_{\textnormal{bu}}$}: the looping time $T^{\neg\guard^M}$\! is not almost-surely bounded, $l_M$ is not bounded due to unbounded variable $b$ thereof, and $l_M$ is not conditionally difference bounded (see \cref{app:check_bu}).

However, it is evident that our (orthogonal) sufficient criterion in \cref{thm:additional_conditions_uniform} is fulfilled, thereby witnessing uniform integrability of $l_M$ for $\cc^M_{\textnormal{bu}}$: $\cc^M_{\textnormal{bu}}$ has the bounded update property, as it has a loop-free body with only bounded updates; moreover, $l_M$ meets the polynomial boundedness condition as, for any fixed $M \in \Ints_{>0}$, $l_M$ is a piece-wise polynomial in $n$ and $b$. Consequently, by \cref{thm:lower_bound_hark,thm:inference_rule}, we have, for any fixed $M \in \Ints_{>0}$,
\begin{align*}
	l_M \ppreceq \wp{C^M_{\textnormal{bu}}}{[n \leq  0] \cdot b} \ppreceq \wp{C_{\textnormal{bu}}}{b}~.
\end{align*}%
We can eventually certify the lower bound $l_\infty$ by pushing $M$ ad infinitum:

\vspace{\abovedisplayskip}
\hfill $\displaystyle \wp{C_{\textnormal{bu}}}{b} \ssucceq \lim_{M\to \infty} l_M \eeq [n<0] \cdot b + [n\geq 0]\cdot \left(b+2 \cdot n\right) \eeq l_\infty~.$
\qedT
\end{example}

\subsection{Tightness of the Guard-Strengthening Rule}\label{sec:tightness}

Next, we discuss the error incurred by our guard-strengthening rule, i.e., the difference between $\wp{\cloop}{f}$ and the established lower bound $l$ as per \cref{thm:inference_rule}. To this end, observe that
%
%
\begin{align}\label{eq:error-terms}
    \wp{\cloop}{f} - l \ =\ \underbrace{\left(\wp{\cloop}{f} - \wp{\pcloop}{[\neg\guard] \cdot f}\right)}_{\textnormal{error incurred by guard strengthening, $\succeq 0$}} + \underbrace{\left(\wp{\pcloop}{[\neg\guard] \cdot f} - l\right)}_{\mathclap{\textnormal{error incurred in the reduced problem, $\succeq 0$}}}
\end{align}%
where, on the right-hand side, the first summand encodes the error inherently caused by our guard-strengthening technique, whereas the second summand  accounts for the error in underapproximating $\wp{\pcloop}{[\neg\guard] \cdot f}$ by $l$, which is independent of the proof rule itself. In what follows, we quantify the first summand, i.e., the error incurred by strengthening the guard.
%
\begin{restatable}[]{lemma}{restateApproxUpper}
	\label{thm:approx_diff_upper}
	Given loops\, $\cloop = \WHILEDO{\guard}{\cc}$ and\, $\pcloop = \WHILEDO{\guard'}{\cc}$, suppose $\guard' \!\implies\! \guard$, then, for any postexpectation $f\in\Expectations$,
	\begin{align*}
	\wp{\cloop}{f} - \wp{\pcloop}{[\neg\guard] \cdot f} \ppreceq \wp{\pcloop}{[\guard]} \cdot \sup_{\pstate \models \neg\guard' \wedge \guard} \wp{\cloop}{f}\left(\pstate\right)~.
	\end{align*}%
\end{restatable}

\begin{proof}[Proof (sketch)]
	As shown in the proof of \cref{thm:approx_diff_lower} (cf.\ \cref{app:proof_restateApproxDiffLower}), we have 
	\begin{align*}
		\wp{\cloop}{f} -  \wp{\pcloop}{[\neg\guard] \cdot f} \eeq \lambda \pstate \mydot \int_{A}~\fsub{\cloop} \dif\,\left(\probability{\pstate}\right)~.
	\end{align*}%
	It thus suffices to bound the integral on the right-hand side, which can be achieved by examining different fragments (witnessed by guard violations) of traces in $A$. See details in \cref{app:proof_diff_upper}.
\end{proof}
An immediate consequence of \cref{thm:approx_diff_upper} yields the \emph{tightness} of our guard-strengthening rule, that is, the error incurred by underapproximating $\wp{\cloop}{f}$ by $\wp{\pcloop}{[\neg \guard] \cdot f}$ approaches zero when $\guard'$ approaches $\guard$ in an appropriate manner:
\begin{theorem}[Tightness of Guard Strengthening]\label{cor:tightness_rule}
Given $\cloop = \WHILEDO{\guard}{\cc}$, suppose there exists a sequence of guards $\left\{\guard^{m}\right\}_{m \in \NN}$ such that $\forall m \in \NN\colon \guard^{m} \!\implies\! \guard$. Let\, $\cc^{m}_{\textnormal{loop}} =  \WHILEDO{\guard^{m}}{\cc}$.
If one of the following two conditions holds:
\begin{enumerate}[label=\textnormal{(\alph*)}]
	\item\label{item:tightness-1} $\lim_{m \to \infty} \sup_{\pstate \models \neg\guard^{m} \wedge \guard} \wp{\cloop}{f}\left(\pstate\right) < \infty \aand \lim_{m \to \infty} \wp{\cc^{m}_{\textnormal{loop}}}{[\guard]}  = 0$,
	\item\label{item:tightness-2} $\lim_{m \to \infty} \sup_{\pstate \models \neg\guard^{m} \wedge \guard} \wp{\cloop}{f}\left(\pstate\right) = 0$,\footnote{Note that $\wp{\cc^{m}_{\textnormal{loop}}}{[\guard]}$ maps a state to a \emph{probability}, and thus $\lim_{m \to \infty} \wp{\cc^{m}_{\textnormal{loop}}}{[\guard]} \preceq 1 \pprec \infty$.}
\end{enumerate}%
then, the error incurred by strengthening $\guard$ with $\guard^{m}$ converges to zero as $m$ tends to $\infty$, i.e.,
\begin{align*}
    \wp{\cloop}{f} \eeq \lim_{m \to \infty} \wp{\cc^{m}_{\textnormal{loop}}}{[\neg\guard] \cdot f}~.
\end{align*}
\end{theorem}
%
%
%
%
%
%

\begin{remark}
	\rev{There are cases where both conditions \cref{item:tightness-1,item:tightness-2} in \cref{cor:tightness_rule} do \emph{not} hold: Recall the 1-D random walk given in \cref{sec:overview} and consider, for instance, the sequence of guards $\left\{\guard^{m}\right\}_{m \in \NN}$ with $\guard^{m} = (0 < n < 9+\nicefrac{(2^m-1)}{2^m})$. We have that $\lim_{m \to \infty} \guard^{m} = (0 < n < 10)$ and $\forall m \in \NN\colon \guard^{m} \!\implies\! \guard$. However, one can verify that neither \cref{item:tightness-1} nor \cref{item:tightness-2} holds (as $\left\{\guard^{m}\right\}_{m \in \NN}$ do not approach $\guard$ at all).}
	\qedT
\end{remark}

Conditions \cref{item:tightness-1,item:tightness-2} in \cref{cor:tightness_rule} can be checked by certifying upper bounds on the weakest preexpectations therein via, e.g., Park induction, as demonstrated by the following example.
%
\begin{example}\label{ex:biased_RW-tightness}
Reconsider the probabilistic loops $\cc_{\textnormal{1dbrw}}$ and $\cc^M_{\textnormal{1dbrw}}$ with postexpectation $f = 1$ in \cref{ex:biased_RW}. We have shown -- with the aid of Park induction -- that $l_\infty = [n < 0] + [n \geq 0] \cdot (\nicefrac{1}{2})^n$ coincides with the exact termination probability of $\cc_{\textnormal{1dbrw}}$ starting from initial position $n$, i.e., $\wp{\cc_{\textnormal{1dbrw}}}{1} = l_\infty \pprec \infty$. Thus, by \cref{eq:error-terms}, one can already conclude the tightness, namely,
\begin{align*}
	\wp{\cc_{\textnormal{1dbrw}}}{1} \eeq \lim_{M \to \infty} \wp{\cc^M_{\textnormal{1dbrw}}}{[n \leq 0] \cdot 1}~.
\end{align*}%
We now show that the same conclusion can be drawn via condition \cref{item:tightness-2} in \cref{cor:tightness_rule}. Observe that
\begingroup
\allowdisplaybreaks
\begin{align*}
	\lim_{M \to \infty} \sup_{\pstate \models \neg\guard^{M} \wedge \guard} \wp{\cc_{\textnormal{1dbrw}}}{1}\left(\pstate\right)
	&\eeq \lim_{M \to \infty} \sup_{\pstate \models n \geq M} \wp{\cc_{\textnormal{1dbrw}}}{1}\left(\pstate\right)\\
	&\lleq \lim_{M \to \infty} \sup_{\pstate \models n \geq M} \left([n < 0] + [n \geq 0] \cdot \left(\nicefrac{1}{2}\right)^n\right)\left(\pstate\right)\TAG{by~$\wp{\cc_{\textnormal{1dbrw}}}{1} \preceq l_\infty$}\\
	&\eeq \lim_{M \to \infty} \sup_{\pstate \models n \geq M} \left(\left(\nicefrac{1}{2}\right)^n\right)\left(\pstate\right)\\
	&\eeq \lim_{M \to \infty} \left(\nicefrac{1}{2}\right)^M \eeq 0~.
\end{align*}%
\endgroup%
Thus, by condition \cref{item:tightness-2} in \cref{cor:tightness_rule}, the error incurred by strengthening $\guard$ with $\guard^{M}$ converges to zero as $M$ tends to $\infty$.
\qedT
\end{example}

\subsection{On Automation of the Guard-Strengthening Rule}\label{subsec:automation}

%
We briefly discuss the potential to \emph{automate} our guard-strengthening rule for \emph{finding} -- rather than just verifying -- lower bounds on weakest preexpectations. Such automation needs to address two questions:
\begin{enumerate*}[label=(\roman*)]
	\item how to find a \enquote{good} guard strengthening, i.e., $\guard' \!\implies\! \guard$?
	\item how to generate a non-trivial lower bound for the modified loop $\pcloop$?
\end{enumerate*}

\paragraph*{How to Find a \enquote{Good} Strengthening?}
In terms of logical strength, a good modified guard $\guard'$ cannot be too strong, otherwise only trivial lower bound can be derived (think of $\guard' = \FALSE$); it cannot be too weak either, otherwise the resulting loop $\pcloop$ may not fulfil our conditions on termination (cf.\ \cref{lem:exponential_decrease}) and uniform integrability (cf.\ \cref{thm:sufficient_conditions_uniform,thm:additional_conditions_uniform}). In general, it is hard to find -- or even to characterize-- the \enquote{best} strengthening $\guard' \!\implies\! \guard$, however, a naive strengthening pattern by simply bounding a subset $\mathcal{V} \subseteq \vars$ of program variables 
turns out to be rather effective, i.e.,
\begin{align*}
	\guard' \eeq \guard \wedge \bigwedge\nolimits_{x \in \mathcal{V}} x  \diamond c~,
\end{align*}%
where $c$ is a constant and $\diamond \in \{<, >, \leq, \geq, =\}$. The subset $\mathcal{V} \subseteq \vars$ shall be selected in a way such that $\pcloop$ meets our conditions on termination and uniform integrability. Such a strengthening pattern is indeed heuristic, but it suffices to produce good strengthened guards $\guard'$ (in the sense discussed above) for \emph{all} the examples presented in this paper \rev{-- except for those in \cref{sec:limitations} where we address limitations of the naive strengthening pattern.} Investigating a potentially more advanced and clever guard-strengthening strategy is subject to future work.

\paragraph*{How to Generate a Non-trivial Lower Bound for $\pcloop$?}
Due to the nice properties of $\pcloop$ on, e.g., termination, 
the desired lower bounds on $\wp{\pcloop}{[\neg \guard] \cdot f}$ can be discovered by leveraging various (semi-)automated techniques for synthesizing lower bounds on least fixed points. These include -- in the probabilistic setting -- techniques based on
\emph{constraint solving} \cite{DBLP:conf/sas/KatoenMMM10,DBLP:conf/atva/FengZJZX17,DBLP:conf/cav/ChenHWZ15} and \emph{invariant learning} \cite{DBLP:conf/cav/BaoTPHR22} for generating lower bounds on weakest preexpectations of probabilistic programs, \emph{recurrence solving} \cite{DBLP:conf/atva/BartocciKS19} for synthesizing moment-based invariants of (solvable) probabilistic loops, 
and various forms of \emph{value iteration} \cite{DBLP:conf/cav/Baier0L0W17,DBLP:conf/cav/QuatmannK18,DBLP:conf/cav/HartmannsK20} for determining reachability probabilities in Markov models.
In particular, we identify a special case where computing the \emph{exact} least fixed point can be done by \emph{probabilistic model checking}:
%
%
\begin{theorem}[Reduction to Probabilistic Model Checking~\textnormal{\cite[Chap.~10]{baier2008principles}}]\label{thm:pbmc}
	Given $\pcloop = \WHILEDO{\guard'}{\cc}$ and postexpectation $g \in \Expectations$, if\, $\pcloop$ has a \underline{finite} set of reachable states,  
	then $\wp{\pcloop}{g}$ can be computed exactly by solving a system of linear equations.
\end{theorem}
\noindent
\Cref{thm:pbmc} holds since 
a finite-state loop 
$\pcloop$ can be unrolled into a finite-state Markov chain $\mathcal{M}$, and $\wp{\pcloop}{g}$ coincides with the unique solution to a linear-equation system 
induced by $\mathcal{M}$ with rewards modelling $g$ \cite[Chap.~10]{baier2008principles}, which can be computed symbolically by probabilistic model checkers, e.g., \toolname{Storm} \cite{DBLP:conf/cav/DehnertJK017}, \toolname{PRISM} \cite{kwiatkowska2002prism}.

\emph{The combination of our guard-strengthening rule with \textnormal{\Cref{thm:pbmc}} facilitates inferring quantitative properties of infinite-state probabilistic programs by model checking finite-state probabilistic models:} 

\begin{example}[3-D Symmetric Random Walk on $\Ints^3$~\textnormal{\cite{mccrea1940xxii,montroll1956random}}]\label{ex:3drw}
Consider the loop $\cc_{\textnormal{3dsrw}}$ modelling a symmetric random walk on the 3-D lattice over $\Ints^3$:
\begin{align*}
	\cc_{\textnormal{3dsrw}}\colon
	\quad
	&\WHILESYMBOL\left(\,x\neq 0\vee y\neq 0\vee z\neq 0\,\right)\left\{\,\right.\\
	&\quad \left.\ASSIGN{x}{x-1}~\oplus~\ASSIGN{x}{x+1}~\oplus~\ASSIGN{y}{y-1}~\oplus~\ASSIGN{y}{y+1}~\oplus~\ASSIGN{z}{z-1}~\oplus~\ASSIGN{z}{z+1}\,\right\}
\end{align*}%
where iterated $\oplus$ is shorthand for discrete uniform choice (in this case, with probability $\nicefrac{1}{6}$ each).

The random nature underneath $\cc_{\textnormal{3dsrw}}$ is fundamentally \emph{different} from its 1- and 2-D counterparts: \citet{Polya1921} proved that the probability $\mathcal{P}$ that such a random walk returns to the origin at $(0, 0, 0)$ is strictly less than 1, indicating that $\cc_{\textnormal{3dsrw}}$ does \emph{not} terminate almost-surely. More precisely, the termination probability of $\cc_{\textnormal{3dsrw}}$ starting from any neighbor location of the origin is
\begin{align*}\label{eq:3d-return-prob}
	\mathcal{P} \eeq 1 - \left( \frac{3}{(2\pi)^3} \int^\pi_{-\pi} \int^\pi_{-\pi} \int^\pi_{-\pi} \frac{\dif\, x \dif\, y \dif\, z}{3 - \cos{x} - \cos{y} - \cos{z}}\right)^{-1}\! \eeq 0.3405373296\ldots
\end{align*}%
which is known as one of P\'{o}lya's random walk constants. To the best of our knowledge, existing techniques cannot cope with the verification of $\cc_{\textnormal{3dsrw}}$ due to its complex nature of divergence.

We now show how our guard-strengthening rule can be used to \emph{automatically} derive lower bounds on the termination probability $\mathcal{P}$ of $\cc_{\textnormal{3dsrw}}$ (i.e., with postexpectation $f = 1$) by leveraging probabilistic model checking. To this end, we first bound the possible positions $(x, y, z)$ with a cube of side-length $2 \cdot M$ with fixed $M \in \Ints_{>0}$, and thereby obtain the modified loop:
\begin{align*}
	\cc^M_{\textnormal{3dsrw}}\colon
	\quad
	&\WHILESYMBOL\left(\,(x\neq 0\vee y\neq 0\vee z\neq 0) \wedge \absd{x} < M \wedge \absd{y} < M \wedge \absd{z} < M\,\right)\left\{\,\right.\\
	&\quad \left.\ASSIGN{x}{x-1}~\oplus~\ASSIGN{x}{x+1}~\oplus~\ASSIGN{y}{y-1}~\oplus~\ASSIGN{y}{y+1}~\oplus~\ASSIGN{z}{z-1}~\oplus~\ASSIGN{z}{z+1}\,\right\}~.
\end{align*}%
%
{\makeatletter
	\let\par\@@par
	\par\parshape0
	\everypar{}
\begin{wrapfigure}{r}{0.43\textwidth}
		\begin{tikzpicture}
			\draw (0, 0) node[inner sep=0] {\includegraphics[width=1\linewidth]{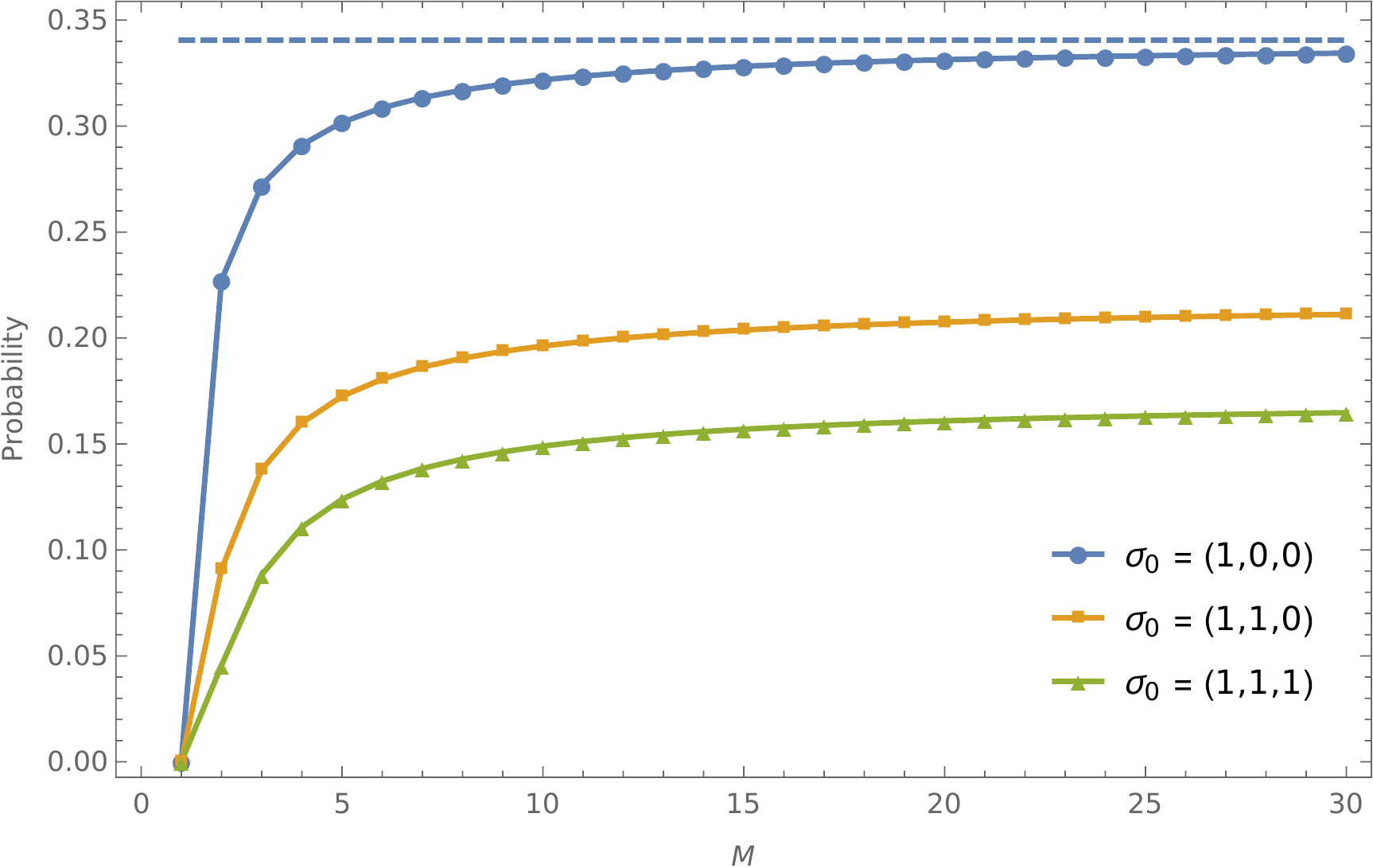}};
			\draw 
			(-2.61, 1.67) node {\tiny \textcolor{MidnightBlue}{$\mathcal{P}$}};
		\end{tikzpicture}
	\vspace*{-5.5mm}
	\caption{Lower bounds produced by \toolname{PRISM} on the termination probability of $\cc_{\textnormal{3dsrw}}$ in \cref{ex:3drw} on initial position $\pstate_0$. $\textcolor{MidnightBlue}{\mathcal{P}} \approx 0.34$ marks the true termination probability when starting from a neighbor location of the origin, e.g., $\pstate_0 = (1,0,0)$; the true termination probabilities on other initial positions are unknown.}
	\label{fig:3drw}
\end{wrapfigure}
\noindent
Observe that, starting from any initial position in the lattice, the modified random walk $\cc^M_{\textnormal{3dsrw}}$ can reach only a \emph{finite} number of positions, i.e., it either
\begin{enumerate*}[label=(\roman*)]
\item terminates by escaping the cube or returning to the origin, or 
\item keeps strolling within the cube.
\end{enumerate*}
Thus, by the connection in \cref{thm:pbmc}, the reduced weakest preexpectation $\wp{\cc^M_{\textnormal{3dsrw}}}{[x = 0 \wedge y = 0 \wedge z = 0] \cdot 1}$ -- which underapproximates the termination probability $\mathcal{P}$ of $\cc_{\textnormal{3dsrw}}$ -- can be computed symbolically by off-the-shelf probabilistic model checkers. For instance, \cref{fig:3drw} depicts such underapproximations produced by \toolname{PRISM} \cite{kwiatkowska2002prism} under different choices of $M$ (cf.\ the $X$-axis). In particular, \cref{fig:3drw} confirms our previous claim on tightness (cf.\ \cref{cor:tightness_rule})\footnote{A rigorous proof of tightness in this example requires finding an appropriate superinvariant -- like we did in \cref{ex:biased_RW-tightness}, yet now for the 3-D random walk -- which is extremely hard and beyond the scope of this paper.}: as $M$ gets larger, the strengthening gets weaker, thus yielding tighter lower bounds on the actual termination probability $\mathcal{P}$. 
\par}%
\noindent 
\cref{fig:3drw} also confirms our intuition that the further the walker is away from the origin, the lower probability returns the walker back to the origin.
\qedT
\end{example}

\begin{remark}
	In the aforementioned special case where our problem is reducible to model checking finite-state discrete-time Markov chains, our guard-strengthening rule acts analogously (yet on the source-code level) to the so-called \emph{partial exploration} technique \cite{DBLP:conf/atva/BrazdilCCFKKPU14} established in probabilistic model checking. The latter constructs a sequence of increasingly precise approximations (lower and upper bounds on reachability probabilities) by (partially) exploring an incremental subset of the state space. This procedure yields a similar effect as we gradually weaken our strengthening to obtain tighter and tighter lower bounds. Since we always infer safe underapproximations, our technique may be used to generate \emph{critical subsystems} \cite{DBLP:conf/sfm/AbrahamBDJKW14} serving as counterexamples to quantitative properties in probabilistic verification.
\end{remark}

%

%
\section{Case Studies}\label{sec:case-studies}

In this section, we present a few non-trivial examples to demonstrate the effectiveness of our guard-strengthening rule in establishing lower bounds on weakest preexpectations. These examples include probabilistic programs with \emph{continuous distributions}, \emph{nested loops}, and \emph{state-dependent probabilities}, as well as a real-world \emph{randomized networking protocol}.
These are all \emph{infinite-state} programs, for some of which our proof rule can be used to prove almost-sure termination by certifying 1 as a lower bound on the termination probability.

\begin{example}[Continuous Sampling]\label{ex:continuous}
	Consider the following loop $\cc_{\textnormal{cs}}$ modelling a 1-D random walk where the branching probability is drawn form a continuous uniform distribution over $(0,1)$.
	\begin{align*}
		\cc_{\textnormal{cs}}\colon
		\quad
		&\WHILESYMBOL\left(\,n > 0\,\right)\{\\
		&\quad\PASSIGNUNIFORM{p}{0}{1}{\,\fatsemi}~\\
		&\quad  \{\,\COMPOSE{\ASSIGN{n}{n-1}}{\ASSIGN{b}{b+\UNIFORM{0}{1}}}\,\}~[p]~\{\,\COMPOSE{\ASSIGN{n}{n+1}}{\ASSIGN{b}{b-\UNIFORM{0}{1}}}\,\}\,\}~.
	\end{align*}%
	We aim to verify $l_{\infty} = [n<0]\cdot b + [n\geq 0]\cdot (b+ \nicefrac{1}{2} \cdot n)$ as a lower bound on the expected value of $b$ upon termination, i.e., $l_\infty\preceq \wp{\cc_{\textnormal{cs}}}{b}$. Observe that $l_\infty$ is unbounded; moreover, $\cc_{\textnormal{cs}}$ behaves -- in expectation -- as the standard 1-D symmetric random walk \cite{mciver2017new} and thus terminates almost-surely yet with an \emph{infinite} expected looping time. Existing proof rules hence do not apply.
	
	We now strengthen the guard $(n > 0)$ of $\cc_{\textnormal{cs}}$ to $\guard^M = (0 < n < M)$ for any fixed $M \in \Ints_{>0}$, and denote the so-obtained loop by $\cc_{\textnormal{cs}}^M$. 
	It can be shown (cf.\ \cref{app:check_continuous}) that
	\begin{align*}
		l_M \eeq [n<0] \cdot b + [0 \leq n \leq M] \cdot \left(b+\nicefrac{1}{2} \cdot n\right) \cdot \left(1 - \nicefrac{n}{M}\right)
	\end{align*}%
	is a subinvariant, i.e., $l_M \preceq \charfunaked{g}^M(l_M)$, where $g = [n \leq 0]\cdot b$ is the restricted postexpectation, and $\charfunaked{g}^M$ is the characteristic function of $\cc^M_{\textnormal{cs}}$ with respect to $g$. 
	Furthermore, $\cc^M_{\textnormal{cs}}$ terminates after at most $M-1$ steps of consecutively increasing $n$ with probability at least $(\nicefrac{1}{2}\cdot\nicefrac{1}{2})^{M-1}$ (i.e., by first drawing $p \in (\nicefrac{1}{2}, 1)$ and then picking the right branch), we have $\probability{\pstate}\big(T^{\neg\guard^M}\! > M\big) < 1 - (\nicefrac{1}{4})^{M-1}$ for any initial state $\pstate(n) \in \Ints$. Thus, by \cref{lem:exponential_decrease}, $\cc^M_{\textnormal{cs}}$ terminates almost-surely. Moreover, $\cc^M_{\textnormal{cs}}$ is uniformly integrable for $\cc^M_{\textnormal{cs}}$ as witnessed by our new sufficient condition in \cref{thm:additional_conditions_uniform}. It follows that $l_M\preceq \wp{\cc^M_{\textnormal{cs}}}{[n\leq 0]\cdot b}\preceq \wp{\cc_{\textnormal{cs}}}{b}$ for any fixed $M \in \Ints_{>0}$. We can eventually certify the lower bound $l_\infty$ by pushing $M$ ad infinitum:
	
	\vspace{\abovedisplayskip}
	\hfill$
	\wp{\cc_{\textnormal{cs}}}{b} \ssucceq \lim_{M\rightarrow\infty} l_M \eeq  [n<0]\cdot b + [n\geq 0]\cdot (b+ \nicefrac{1}{2} \cdot n) \eeq l_\infty~.
	$
	\qedT
\end{example}

\begin{example}[Nested Loops]\label{ex:nested_loop}
Consider the program $\cc_{\textnormal{nl}}$ with two nested loops (the outer loop alters $n\in\Ints$ based on the outcome of the inner loop; $a,b\in \Ints_{>0}$ are parameters satisfying $a>b$\footnote{The restriction $a>b$ makes $\cc_{\textnormal{nl}}$ a non-AST program.}):
\begin{align*}
\cc_{\textnormal{nl}}\colon
\quad
&\WHILESYMBOL\left(\,n > 0\,\right)\{\\
&\quad	\ASSIGN{k}{n}{\,\fatsemi}~\\
&\quad \WHILEDO{-a< k-n < b}{\ASSIGN{k}{k-1}~[\nicefrac{1}{2}]~\ASSIGN{k}{k+1}}{\,\fatsemi}~\\
&\quad	\ITE{k<n}{\ASSIGN{n}{n-1}}{\ASSIGN{n}{n+1}}\, \}~.
\end{align*}%
Suppose we are interested in proving a non-trivial lower bound on the termination probability $\wp{\cc_{\textnormal{nl}}}{1}$. This problem is out-of-reach by existing proof rules as $\cc_{\textnormal{nl}}$ is not AST. To apply our proof rule, we strengthen the guard of the outer loop to $\guard^M = (0 < n < M)$ for any fixed $M \in \Ints_{>0}$. The modified program $\cc^M_{\textnormal{nl}}$ is AST due to a similar argument per \cref{lem:exponential_decrease}. Let 
\begin{align*}
	l_M \eeq [n<0] + [0 \leq n\leq M]\cdot \left(\left(\nicefrac{b}{a}\right)^n - \left(\nicefrac{b}{a}\right)^M \right)~.
\end{align*}%
Since $l_M$ is bounded, we know by \cref{thm:lower_bound_hark,thm:inference_rule} that $l_M$ underapproximates the termination probability of $\cc_{\textnormal{nl}}$ if $l_M$ is a subinvariant of $\cc_{\textnormal{nl}}^M$. In that case, we have proved the lower bound $l_\infty$:
\begin{align*}
	\wp{\cc_{\textnormal{nl}}}{1} \ssucceq \lim_{M\rightarrow\infty} l_M \eeq [n<0] + [n\geq 0]\cdot \left(\nicefrac{b}{a}\right)^n \ddefeq l_\infty~.
\end{align*}%

We now show that $l_M$ is indeed a subinvariant of $\cc_{\textnormal{nl}}^M$, i.e., $l_M \preceq \charfunaked{g}^M(l_M)$, where $g = [n \leq 0] \cdot 1$ is the restricted postexpectation, and $\charfunaked{g}^M$ is the characteristic function of $\cc^M_{\textnormal{nl}}$ w.r.t.\ $g$. Checking subinvariance in this case is more involved due to the nested feature of $\cc_{\textnormal{nl}}^M$: one also has to reason about \emph{lower bounds for the inner loop} w.r.t.\ a specific postexpectation depending on $l_M$. More precisely, let $\cc^M_{\textnormal{body}}$ and $\cc^M_{\textnormal{inner}}$ denote the loop body and the inner loop of $\cc^M_{\textnormal{nl}}$, respectively. We have
\begin{align*}
\charfunaked{g}^M\left(l_M\right) &\eeq [n\leq 0\vee n\geq M]\cdot [n\leq 0] + [0<n<M]\cdot \wp{\cc^M_{\textnormal{body}}}{l_M}\\
&\eeq
\begin{aligned}[t]
	 &[n\leq 0] + [0<n<M]\,\cdot\\[-1mm]
&\wp{\ASSIGN{k}{n}}{\wp{{\cc^M_{\textnormal{inner}}}}{[k<n] \cdot l_M\subst{n}{n-1} + [k\geq n]\cdot l_M\subst{n}{n+1} }}~.
\end{aligned}
\end{align*}%
By combining the fact (see \cref{app:check_nl} for a detailed proof) that, for any bounded postexpectation $h$,
\begin{align*}
	\wp{{\cc^M_{\textnormal{inner}}}}{h} \ssucceq &[\left(k-n< -a\right)\vee \left(k-n> b\right)]\cdot h~+ \\
	&[-a\leq k-n\leq b]\cdot \left( \frac{b-(k-n)}{b+a} \cdot h\subst{k}{n-a} +  \frac{(k-n)+a}{b+a} \cdot h\subst{k}{n+b} \right)~,
\end{align*}%
we can conclude the subinvariance of $l_M$, i.e., $l_M \preceq \charfunaked{g}^M(l_M)$ (see details in \cref{app:check_nl}). This completes our verification of the lower bound $l_\infty$.
\qedT
\end{example}
%

\begin{example}[Fair-in-the-Limit Random Walk~\textnormal{\cite{mciver2017new}}]\label{ex:fair-limit-rw}
Consider the following loop $\cc_{\textnormal{flrw}}$ modelling a biased random walk on $\Ints$, where in particular, the probability of taking the next move depends on the current position of the walker:\footnote{The loop remains AST if we replace the biased probability $\nicefrac{n+1}{2n+1}$ by $\nicefrac{n}{2n+1}$. However, in this case, one needs a more advanced subinvariant in order to certify AST of the loop. 
See, e.g., a proof 
using harmonic numbers in \cite{mciver2017new}.}
\begin{align*}
	\cc_{\textnormal{flrw}}\colon
	\quad
	\WHILEDO{n > 0}{\ASSIGN{n}{n-1}~[\nicefrac{n+1}{2n+1}]~\ASSIGN{n}{n+1}}~.
\end{align*}%
The further the walker moves to the right, the more fair becomes the random walk as $\nicefrac{n+1}{2\cdot n+1}$ approaches $\nicefrac{1}{2}$ asymptotically.
We show how our guard-strengthening rule can be used to prove almost-surely termination of $\cc_{\textnormal{flrw}}$ by certifying $1$ as a lower bound on $\wp{\cc_{\textnormal{flrw}}}{1}$. Similarly, we strengthen the guard $(n > 0)$ to $\guard^M = (0 < n < M)$ for any fixed $M \in \Ints_{>0}$ and denote the so-obtained loop by $\cc_{\textnormal{flrw}}^M$.
Since $\cc_{\textnormal{flrw}}^M$ terminates after at most $M -1$ steps of consecutively moving to the left with probability at least $\left(\nicefrac{1}{2}\right)^{M-1}$, $\cc_{\textnormal{flrw}}^M$ terminates almost-surely by \cref{lem:exponential_decrease}. Let
\begin{align*}
l_M \eeq [n < 0] + [0\leq n\leq M] \cdot \left(1-\nicefrac{n}{M}\right)~.
\end{align*}%
$l_M$ is bounded from above by 1, thus is uniformly integrable for $\cc_{\textnormal{flrw}}^M$ by condition \cref{item:ui-3} in \cref{thm:sufficient_conditions_uniform}. Furthermore, $l_M$ is a subinvariant since
\begingroup
\allowdisplaybreaks
\begin{align*}
           \charfunaked{g}(l_M) & = [n\leq 0\vee n\geq M]\!\cdot\! [n\leq 0] + [0< n<M]\cdot \left(\frac{n+1}{2n+1}\cdot l_M\subst{n}{n-1} +  \frac{n}{2n+1}\cdot l_M\subst{n}{n+1}\right)\\
           &=  [n\leq 0\vee n\geq M]\!\cdot\! [n\leq 0] +[0< n<M]\cdot \left(\frac{n+1}{2n+1}\cdot\frac{M-n+1}{M} + \frac{n}{2n+1}\cdot\frac{M-n-1}{M}\right)\\
           &= [n\leq 0] +[0< n<M]\cdot \left(1 - \frac{n}{M} + \frac{1}{(2n+1)\cdot M}\right) \ssucceq l_M~,
\end{align*}%
\endgroup%
where $g = [n \leq 0]\cdot 1$ is the restricted postexpectation, and $\charfunaked{g}^M$ is the characteristic function of $\cc^M_{\textnormal{flrw}}$ with respect to $g$.
This implies $l_M \preceq \wp{\cc_{\textnormal{flrw}}}{1}$, i.e., $l_M$ is a lower bound on the termination probability of $\cc_{\textnormal{flrw}}$ for any fixed $M \in \Ints_{>0}$. In particular, when $M$ approaches infinity, 
we have
\begin{align*}
    \wp{\cc_{\textnormal{flrw}}}{1} \ssucceq \lim_{M\rightarrow\infty} l_M \eeq 1~.
\end{align*}
We can thus conclude that $\cc_{\textnormal{flrw}}$ terminates almost-surely.
\qedT
\end{example}

\begin{example}[Zeroconf Protocol~\textnormal{\cite{DBLP:conf/dsn/BohnenkampSHV03}}]\label{ex:zc}
	Consider the loop $\cc_{\textnormal{zc}}$ encoding the randomized IPv4 Zeroconf protocol (parameterized in $N$) for self-configuring IP network interfaces:
	\begin{align*}
		\cc_{\textnormal{zc}}\colon
		\quad
		&\COMPOSE{\textit{start}=1}{\COMPOSE{\textit{established}=0}{\textit{probe}=0}}\,\fatsemi\\
		&\WHILESYMBOL\left(\,\textit{start} \le 1 \wedge \textit{established} \le 0 \wedge \textit{probe} < N \wedge  N \geq \rev{4}\, \right)\{\\ %
		&\quad \IFSYMBOL\left(\, \textit{start} = 1 \,\right)
		\{\,\\
		&\quad \quad \{\,\ASSIGN{\textit{start}}{0}\,\}~[0.5]~\{\,\COMPOSE{\ASSIGN{\textit{start}}{0}}{\ASSIGN{\textit{established}}{1}}\,\}
		\, \}\\
		&\quad \ELSESYMBOL\,\{\,
		\{\,\ASSIGN{\textit{probe}}{\textit{probe} + 1}\,\}~[\rev{0.001}]~\{\,\COMPOSE{\ASSIGN{\textit{start}}{1}}{\ASSIGN{\textit{probe}}{0}}\,\}
		\, \}\, \}~.
	\end{align*}%
	\rev{This program is intended for providing a convenient way to analyze the probability that an unused IP address is successfully assigned to a host that is newly connected to a network of $m$ existing devices: Once being connected, the host first randomly selects an IP address from a pool of $n$ available addresses. With probability $\nicefrac{m}{n}$ (instantiated as $0.5$), this address is already in use; with probability $1-\nicefrac{m}{n}$, the chosen address is unused and thus the IP connection can be established (signified by terminating with $\textit{established} = 1$). For the former case, the host broadcasts a \enquote{probe} message to other devices in the network asking whether the chosen IP address is already taken; If the probe is received by a device that already uses the address, it replies with a message indicating so. After receiving this message, the host to be configured restarts.\footnote{\rev{The program $\cc_{\textnormal{zc}}$ is an abstract model of the actual protocol in the sense that $\cc_{\textnormal{zc}}$ abstracts away the need for broadcasting probe messages when the right branch of the $\IFSYMBOL$-statement is executed (in this case, no reply message will be sent anyway). Such an abstraction does not affect the analysis of the probability of terminating with $\textit{established} = 1$.}} The probability $0.001$ in  the $\ELSESYMBOL$-statement encodes the probability of message loss (for probe or reply). To increase reliability, the host is required to send multiple probes (upper-bounded by $N$). In fact, $\cc_{\textnormal{zc}}$ encodes an \emph{infinite family} of Markov chains (cf.\ \cite[Exmp.\ 10.5]{baier2008principles}) by parameterizing $N$, and thus cannot be handled by probabilistic model checking.}

	We are interested in verifying a non-trivial lower bound \rev{$0.99999999999$} on the probability $\mathcal{P}$ that, starting from a state satisfying the loop guard, $\cc_{\textnormal{zc}}$ terminates in a state with $\textit{established} = 1$. 
	We do so by trying to \emph{synthesize} a subinvariant justifying this lower bound using the recently developed tool \toolname{cegispro2} \cite{DBLP:journals/corr/abs-2205-06152} which automates Hark et al.'s sound lower induction rule \cite{DBLP:journals/pacmpl/HarkKGK20}. Unfortunately, \toolname{cegispro2} failed in this case (timeout in 20 minutes). However, by strengthening the loop guard with upper bounds on $N$, e.g., $N \leq \rev{10}$, and applying our proof rule, \toolname{cegispro2} found a piece-wise linear subinvariant (in less than a second; omitted due to limited space) that suffices to certify \rev{$0.99999999999$} as a lower bound on $\mathcal{P}$. 
	Overall, this demonstrates the potential of our proof rule to enable verifying lower bounds via automated techniques for practical randomized algorithms that are otherwise out-of-reach.
	\qedT
\end{example}

\rev{
Our proof rule applies further to a few extra case studies from the Quantitative Verification Benchmark Set \cite{hartmanns2019quantitative}, including the (infinite family of) Bounded Retransmission protocol, the (parametrized) Rabin Mutual Exclusion algorithm, and the Coupon Collectors. 
These case studies are not included because they either do not bring extra insights over the Zeroconf protocol or can already be tackled by existing techniques. 
The main challenges of pursuing 
a wider range of real randomized algorithms are
\begin{enumerate*}[label=(\roman*)]
\item to find good guard strengthening (cf.\ \cref{sec:limitations} below) and, in some cases,
\item to find subinvariants 
certifying the given lower bounds.
\end{enumerate*}
We plan to address these challenges by exploiting the potential for \emph{automating} our proof rule as discussed in \cref{subsec:automation}.
}
%

\section{Limitations of the Guard-Strengthening Rule}\label{sec:limitations}

\rev{There are interesting divergent programs for which the proposed guard-strengthening principle 
	is insufficient or inadequate. We address such limitations by means of two examples.}

\rev{The following example demonstrates the case where the naive strengthening heuristic is insufficient, yet a more advanced strengthening strategy suffices to establish non-trivial lower bounds.}

\begin{example}[2-D Discrete Spiral]\label{ex:2D-spiral}
\rev{
		Consider the probabilistic program
	\begin{align*}
		&\COMPOSE{\ASSIGN{x}{0}}{\COMPOSE{\ASSIGN{y}{0}}{\ASSIGN{\textit{flag}}{0}}}\,\fatsemi\\
		&\WHILESYMBOL\left(\,(x-1)^2 + (y-1)^2 < 4\,\right)\{\\
		&\quad \{\,\COMPOSE{\ASSIGN{x}{-1}}{\ASSIGN{y}{-1}}\,\}\mathrel{\left[\,\nicefrac{\abs{x-1}}{4}\,\right]}\{\\
		&\quad \quad 
		\COMPOSE{\ITE{\textit{flag}}{\ASSIGN{y}{\nicefrac{(3-y)}{2}}}{\ASSIGN{x}{\nicefrac{(3-x)}{2}}}}{\ASSIGN{\textit{flag}}{1-\textit{flag}}} \,\} \,\}~.
	\end{align*}%
	This program terminates only if it visits the left branch of the probabilistic choice; otherwise, it diverges with $(x,y)$ approaching $(1,1)$ in the shape of a 2-D discrete spiral. Note that the program is non-AST and we aim to verify lower bounds on its termination probability. Observe that, in this case, the naive strengthening heuristic $\guard' = \guard \wedge x < c \wedge y < c$ by adding a constant bound $c$ on program variables is \emph{insufficient} for any $c > \nicefrac{3}{2}$ as the resulting program is still non-AST; whereas for any $c \leq \nicefrac{3}{2}$, the resulting program terminates after at most one loop iteration thus only yielding \emph{trivial} lower bounds. In contrast, by applying the more advanced (yet intuitive) strengthening $\guard' = \guard \wedge (x-1)^2 + (y-1)^2 > \epsilon$ with a small constant $\epsilon > 0$ (i.e., to rule out the spiral center $(1,1)$), we reduce the program to the case of a finite state space, which can then be tackled by probabilistic model checking for establishing \emph{non-trivial} lower bounds.
}
	\qedT
\end{example}
\noindent
\rev{As remarked in \cref{subsec:automation}, it is an interesting future direction to investigate potentially more advanced, (semi-)automatable guard-strengthening strategies than the heuristic used in this paper.}

\rev{The following example demonstrates a corner case where no useful strengthening exists.}

\begin{example}[Dummy Swapper]\label{ex:corner-case}
\rev{
	Consider the probabilistic program
	\begin{align*}
		\WHILEDO{x\neq y}{ \ASSIGN{x}{y}~\oplus~\left\{\COMPOSE{\COMPOSE{\ASSIGN{\textit{temp}}{x}}{\ASSIGN{x}{y}}}{\ASSIGN{y}{\textit{temp}}}\right\}~\oplus~\DIVERGE
		}
	\end{align*}%
	where iterated $\oplus$ is shorthand for discrete uniform choice (in this case, with probability $\nicefrac{1}{3}$ each); $\DIVERGE$ is syntactic sugar for $\WHILESYMBOL(\TRUE)\{\SKIP\}$. This program is non-AST and it terminates only by visiting the leftmost branch of the uniform choice; otherwise, it diverges by either keeping swapping the values of $x$ and $y$ (i.e., pacing between two points) or diverges immediately by visiting the rightmost branch. It can be shown that \emph{every} possible strengthening that can turn this program AST necessarily leads to $\guard'=\FALSE$ -- otherwise, at least one loop iteration is executed and the program diverges with probability at least $\nicefrac{1}{3}$ -- and therefore only yield \emph{trivial} lower bounds.
}
	\qedT
\end{example}
%
\section{Related Work}\label{sec:related-work}


\paragraph*{Weakest Preexpectation Reasoning}
As a probabilistic analog to Dijkstra's predicate-transformer calculus \cite{DBLP:journals/cacm/Dijkstra75,DBLP:books/ph/Dijkstra76}, \emph{expectation transformers} have been extensively used 
to reason about \emph{quantitative} properties of probabilistic programs. 
$\wpsymbol$-style reasoning goes back to the seminal work of \citet{DBLP:conf/stoc/Kozen83,DBLP:journals/jcss/Kozen85} on probabilistic propositional dynamic logic. Amongst others, \citet{DBLP:phd/ethos/Jones90}, \citet{DBLP:journals/toplas/MorganMS96}, \citet{DBLP:series/mcs/McIverM05}, and \citet{DBLP:journals/fac/Hehner11} furthered this line of research by considering, e.g., nondeterminism and proof rules for bounding preexpectations in the presence of loops. The classical weakest preexpectation calculus has been significantly advanced in several directions to reason about, e.g., \emph{expected runtimes} of (recursive) probabilistic programs \cite{DBLP:conf/esop/KaminskiKMO16,DBLP:journals/jacm/KaminskiKMO18,DBLP:conf/lics/OlmedoKKM16}, probabilistic programs with \emph{conditioning} \cite{DBLP:journals/toplas/OlmedoGJKKM18,DBLP:conf/setss/SzymczakK19}, \emph{mixed-sign} postexpectations \cite{DBLP:conf/lics/KaminskiK17}, \emph{sensitivity} of probabilistic programs \cite{DBLP:journals/pacmpl/0001BHKKM21}, probabilistic temporal logic \cite{DBLP:journals/igpl/MorganM99}, and quantitative separation logic \cite{DBLP:journals/pacmpl/BatzKKMN19}. A relative completeness result has been recently established by \citet{DBLP:journals/pacmpl/BatzKKM21} for weakest preexpectation reasoning.

\paragraph*{Bounds on Weakest Preexpectation and Fixed Points}
There are a wide spectrum of results on establishing upper and/or lower bounds on loop semantics captured by (least) fixed points. These include (semi-)automated techniques based on \emph{metering functions} \cite{DBLP:journals/toplas/FrohnNBG20,DBLP:conf/cade/FrohnNHBG16} for underapproximating runtimes of non-probabilistic programs, 
\emph{constraint solving} \cite{DBLP:conf/sas/KatoenMMM10,DBLP:conf/atva/FengZJZX17,DBLP:conf/cav/ChenHWZ15} and \emph{invariant learning} \cite{DBLP:conf/cav/BaoTPHR22} for generating lower bounds on weakest preexpectations of probabilistic programs, \emph{recurrence solving} \cite{DBLP:conf/atva/BartocciKS19} for synthesizing moment-based invariants of the so-called prob-solvable probabilistic loops, \emph{bounded model checking} \cite{DBLP:conf/atva/0001DKKW16} for verifying probabilistic programs with nondeterminism and conditioning,
and various forms of \emph{value iteration} \cite{DBLP:conf/cav/Baier0L0W17,DBLP:conf/cav/QuatmannK18,DBLP:conf/cav/HartmannsK20} for determining reachability probabilities in Markov models.

Apart from the proof rules elaborated in this paper, \citet{DBLP:conf/fossacs/BaldanE0P21} recently proposed lattice-theoretic proof rules for verifying lower bounds on least fixed points of \emph{finite-state} stochastic systems. \citet{DBLP:conf/cav/BatzCKKMS21} invented \emph{latticed $k$-induction} for establishing upper bounds on least fixed points of possibly infinite-state probabilistic programs. A closely related concept to our guard-strengthening rule is the notion of \emph{$\omega$-subinvariant} \cite{DBLP:conf/esop/KaminskiKMO16,DBLP:journals/jacm/KaminskiKMO18,DBLP:phd/ethos/Jones90,DBLP:journals/scp/AudebaudP09}, which is a monotonically increasing sequence $\{I_n\}_{n \in \NN}$ of expectations that are subinvariants relative to each other, i.e., $I_0 = 0$ and $I_{n+1} \preceq \Phi(I_n)$. It is known that $\sup_{n \in \NN} I_n$ suffices as a lower bound on the least fixed point of $\Phi$, however, finding such a lower bound amounts to 
\begin{enumerate*}[label=(\roman*)]
	\item learning a closed form of $I_n$ in $n$ by inspecting the \enquote{pattern} through fixed point iterations, and
	\item finding the supremum of the closed form.
\end{enumerate*}
These two steps may barely save efforts compared to just inferring the supremum (limit) of the sequence $\sup_{n \in \NN}\Phi^n(0)$ to obtain the exact least fixed point. In some of our examples, we also reason about limits of $l_M$ w.r.t.\ $M$ -- the constant used to strengthen the guard -- to get tight lower bounds. This step, however, is \emph{not necessary} since $l_M \preceq \Phi(l_M)$ suffices already as a lower bound for any fixed $M$, that is, we only need to \enquote{push $l_M$ through the loop semantics} \emph{once}. Moreover, for discrete-state programs with complex closed form solutions, e.g., the 3-D random walk in \cref{ex:3drw}, our rule is capable of inferring tight lower bounds (via probabilistic model checking) \emph{without} the need for taking limits.

\paragraph*{Space-Restricting in Analyzing Random Walks}
There is an intellectual connection between our guard-strengthening technique and the \emph{space-restricting} tactic in analyzing random walks over infinite, discrete lattices: Restricting an infinite-space random walk \`{a} la \citet[Chap.\ XIV]{feller1950introduction} and \citet{mccrea1940xxii} by setting \enquote{barriers}
and then taking the limit is analogous to our approach. Nonetheless, they both proceed, after the restriction, by \emph{solving the recurrence equation} (aka, difference equation) to obtain a closed-form solution for the restricted random walk in order to take the limit. Although a similar methodology can be embedded in our approach -- as solving recurrences is one (expensive) way to obtain subinvariants -- it is \emph{not necessary}: our proof rule \emph{reduces} the problem and places it under the lens of probabilistic model checking. The rule is thus able to handle examples that are out-of-reach by existing verification techniques, e.g., the 3-D random walk, \emph{without} the need for recurrence solving nor taking limits. Moreover, our proof rule codifies \emph{ad-hoc} proof methods in probability theory as a \emph{syntactic} rule of a program logic, and 
enables the use of
existing 
proof rules 
on verification problems that are otherwise out-of-scope.
	
\paragraph*{Martingale-Based Reasoning}
Probabilistic program analysis via martingales was pioneered by \citet{DBLP:conf/cav/ChakarovS13,DBLP:conf/sas/ChakarovS14} for finding upper bounds on expected runtimes and synthesizing expectation invariants. \citet{DBLP:conf/popl/ChatterjeeFNH16} further extended the martingale-based approach to address nondeterminism. \citet{DBLP:conf/cav/BartheEFH16} focus on synthesizing exact martingale expressions. \citet{DBLP:conf/popl/FioritiH15} develop a type system for uniform integrability in order to prove AST
of probabilistic programs and provide upper bounds on \emph{expected runtimes}. In contrast, \citet{DBLP:conf/vmcai/FuC19} give lower bounds on expected runtimes; \citet{DBLP:conf/cav/ChatterjeeGMZ22} prove lower bounds on \emph{termination probabilities}. \citet{DBLP:journals/lmcs/KobayashiLG20} provide a semi-decision procedure for upper-bounding termination probabilities of probabilistic higher-order recursive programs. \citet{DBLP:conf/pldi/WangS0CG21,DBLP:conf/popl/ChatterjeeNZ17} focus on finding upper and lower bounds on \emph{assertion-violation probabilities}. 
\citet{DBLP:conf/pldi/NgoC018} perform automated template-driven \emph{resource} analysis, yet infer upper bounds only. \citet{DBLP:conf/pldi/Wang0GCQS19} provide sufficient conditions for finding upper and lower bounds on \emph{expected costs}. \citet{DBLP:conf/pldi/Wang0R21} present conditions to derive upper and lower bounds on higher moments of \emph{expected accumulated costs}.
In contrast to our proof rule, most of these martingale-based methods for inferring lower bounds address only AST probabilistic programs.

\paragraph*{Probabilistic Bisimulation and Equivalence}
Our result on $\wpsymbol$-difference falls 
in the general scope of quantifying the \enquote{distance} between two programs. This gives rise to the notions of \emph{probabilistic bisimulation} \cite{DBLP:conf/cav/HongLMR19,DBLP:journals/iandc/LarsenS91} and \emph{probabilistic equivalence} (see \cite{DBLP:conf/concur/MurawskiO05} over \emph{finite} data types and \cite{DBLP:conf/popl/BartheGB09} for \emph{straight-line} programs). \citet{CAV2022Chen} recently proved a decidability result for checking the equivalence of an \emph{infinite-state} discrete, loopy probabilistic program with a loop-free specification program (assuming \emph{AST}). \citet{DBLP:conf/mpc/BartheGB12,DBLP:conf/popl/BartheGB09} developed a \emph{relational Hoare logic} for probabilistic programs, which has been extensively used for proving program equivalence with applications in provable security and side-channel analysis. Our proof rule is dedicated to reasoning about lower bounds on expected values of possibly divergent probabilistic programs, nevertheless, it is interesting to explore its generalization in proving equivalence, bisimulation, and \emph{$\varepsilon$-closeness} of possibly \emph{infinite-state}, \emph{non-AST} probabilistic programs by, e.g., embedding our proof rule in the probabilistic relational Hoare logic.

%
\section{Conclusion}\label{sec:conclusion}

We have presented a new proof rule -- based on $\wpsymbol$-difference and the guard-strengthening technique -- for verifying lower bounds on weakest preexpectations of probabilistic programs. This is the first lower bound rule that admits \emph{divergent} probabilistic loops with possibly \emph{unbounded} expectations. In particular, we unleash existing lower induction rules for general applicability to possibly divergent programs whose strengthened counterparts feature easily provable almost-sure termination and uniform integrability. Moreover, we have shown that
the error incurred by our guard-strengthening technique can be arbitrarily small thereby yielding tight lower bounds, and
in case the modified loop has a finite state space, our proof rule can be automated to generate lower bounds via probabilistic model checking.
The effectiveness of our proof rule has been demonstrated on several examples. In particular, we managed to infer tight lower bounds on the termination probability of the well-known 3-D random walk on a lattice, which has not been addressed yet in the context of verification.

Future directions include 
\begin{enumerate*}[label=(\roman*)]
	\item extending our proof rule to the expected runtime calculus for lower-bounding expected runtimes \cite{DBLP:conf/esop/KaminskiKMO16,DBLP:journals/jacm/KaminskiKMO18}, to probabilistic programs with angelic/demonic nondeterminism \cite{DBLP:journals/tcs/McIverM01a,DBLP:series/mcs/McIverM05} and/or with soft/hard conditioning \cite{DBLP:journals/toplas/OlmedoGJKKM18,DBLP:conf/setss/SzymczakK19};
	\item exploring the applicability of our result on $\wpsymbol$-difference in the context of sensitivity analysis \cite{DBLP:journals/pacmpl/0001BHKKM21,DBLP:journals/pacmpl/WangFCDX20,DBLP:journals/pacmpl/BartheEGHS18} and model repair \cite{DBLP:conf/tacas/BartocciGKRS11} for probabilistic programs; and
	\item investigating (semi-)automated synthesis of (candidate) quantitative (sub)invariants used in our proof rule.
\end{enumerate*}
\ifanonymous
\else
\begin{acks}
%
This work has been partially funded by the National Key R\&D Program of China under grant No.\ 2022YFA1005101, by the NSFC under grant No.\ 62192732 and 62032024, by the CAS Project for Young Scientists in Basic Research under grant No.\ YSBR-040, by the ZJU Education Foundation's Qizhen Talent program, by the ERC Advanced Project FRAPPANT under grant No.~787914, and by the European Union's Horizon 2020 research and innovation programme under the Marie Sk\l{}odowska-Curie grant agreement No.~101008233. The authors would like to thank Kevin Batz, Tim Quatmann, and anonymous reviewers for the insightful discussions on the connection respectively to $\omega$-invariants in weakest preexpectation reasoning, partial exploration in Markov models, and the space-restricting tactic in analyzing random walks.
\end{acks}
\fi
%
\bibliography{references}

\newpage
\appendix
\section*{Appendix}

\section{Basic Concepts in Measure Theory}\label{app:preliminary}
We recapitulate basic concepts in measure theory which form the basis of understanding probabilistic programs. We refer interested readers to~\cite[Sect.~1.2]{dahlqvist_silva_kozen_2020} for a more detailed introduction to measure theory geared toward probabilistic programming.

%

\paragraph*{Measurable Spaces and Functions.}
A \emph{$\sigma$-algebra} $\mathcal{B}$ on a set $S$ is a collection of subsets of $S$ containing the empty set $\emptyset$ that is closed under complementation in $S$ and countable union. Given $\sigma$-algebras $\mathcal{B}_1$ and $\mathcal{B}_2$ on the same set $S$, $\mathcal{B}_1$ is a \emph{sub-$\sigma$-algebra} of $\mathcal{B}_2$ iff $\mathcal{B}_1 \subseteq \mathcal{B}_2$. The \emph{Borel $\sigma$-algebra} on $\RR$ -- consisting of \emph{Borel sets} on $\RR$ --  is the smallest $\sigma$-algebra on $\RR$ that contains the intervals. A pair $(S, \mathcal{B})$, where $S$ is a set and $\mathcal{B}$ is a $\sigma$-algebra on $S$, is called a \emph{measurable space}. Given two measurable spaces $(S, \mathcal{B}_S)$ and $(T, \mathcal{B}_T)$, a function $f\colon S\to T$ is \emph{measurable} if the inverse image $f^{-1}(B) = \{x \in S \mid f(x) \in B\}$ of every measurable subset $B \in \mathcal{B}_T$ is a measurable subset of $S$.

\paragraph*{(Sub)probability Measures.}
A \emph{signed (finite) measure} on a measurable space $(S, \mathcal{B})$ is a countably additive map $\mu \colon \mathcal{B} \to \Reals$ such that $\mu(\emptyset) = 0$. Note that $\mu$ is \emph{countably additive} means that if $\mathcal{A}$ is a countable set of pairwise disjoint events, then $\mu(\cup\mathcal{A}) = \sum_{A \in \mathcal{A}}\mu(A)$. A signed measure on $(S, \mathcal{B})$ is called \emph{positive} if $\mu(B) \geq 0$ for all $B \in \mathcal{B}$. A positive measure on $(S, \mathcal{B})$ is termed a \emph{probability measure} if $\mu(S) = 1$ and a \emph{subprobability measure} if $\mu(S) \leq 1$. A measurable set $B$ such that $\mu(B) = 0$ is called a \emph{$\mu$-nullset}, or simply a \emph{nullset} when it is clear from the context. A property over $S$ is said to hold \emph{($\mu$-)almost-surely} 
if the set of points on which it does not hold is contained in a nullset.
%

\paragraph*{Probability Spaces and Random Variables.}
A probability space $(\Omega, \mathcal{F}, P)$ consists of a sample space $\Omega$, a $\sigma$-algebra $\mathcal{F}$ on $\Omega$, and a probability measure $P\colon \mathcal{F} \to [0, 1]$ on the measurable space $(\Omega, \mathcal{F})$. A \emph{random variable} $X$ defined on the probability space $(\Omega, \mathcal{F}, P)$ is an $\mathcal{F}$-measurable function $X\colon \Omega \to \RR$; its $\emph{expected value}$ (w.r.t.\ $P$) is denoted by $\EE[X]$. Every random variable $X$ induces a (sub)probability measure $\mu_X\colon \mathcal{B} \to [0,1]$ on $\RR$, defined as $\mu_X(B) \defeq P(X^{-1}(B))$ for Borel sets $B$ in the Borel $\sigma$-algebra $\mathcal{B}$ on $\RR$. $\mu_X$ is also called the \emph{distribution of $X$}. $\mu_X$ is \emph{continuous} on the measurable space $(\RR, \mathcal{B})$ if $\mu(\{s\}) = 0$ for all singleton sets $\{s\} \in \mathcal{B}$; $\mu_X$ is \emph{discrete} if it is a countable weighted sum of \emph{Dirac measures} $\delta_s$ with $s \in S$ and $\delta_s(B) = 1$ if $s \in B$, and $0$ otherwise.

\paragraph*{Stochastic Processes and Stopping Times.}
Given a probability space $(\Omega, \mathcal{F}, P)$,
a (discrete-time) \emph{stochastic process} is a parametrized collection $\{X_n\}_{n \in \NN}$ of random variables on $(\Omega, \mathcal{F}, P)$; we sometimes drop the brackets in $\{X_n\}$ when it is clear from the context. A collection $\{\mathcal{F}_n\}_{n \in \NN}$ of sub-$\sigma$-algebras of $\mathcal{F}$ is a \emph{filtration} if $\mathcal{F}_i \subseteq \mathcal{F}_{j}$ for all $i \leq j$; intuitively, $\mathcal{F}_n$ carries the information known to an observer at step $n$. A random variable $\tau\colon \Omega \to \NN$ is called a \emph{stopping time} with respect to some filtration $\{\mathcal{F}_n\}_{n \in \NN}$ of $\mathcal{F}$ if $\{\tau \le n\} \in \mathcal{F}_n$ for all $n \in \NN$. Given a stochastic process $X_n$ and a stopping time $\tau$, the corresponding \emph{stopped process} $\{X^\tau_n\}_{n \in \NN}$ is defined by
\begin{align*}
	X^\tau_n \ddefeq
	\begin{cases}
		X_n &\text{if}~n \leq \tau~,\\
		X_{\tau} &\text{if}~n > \tau~,
	\end{cases}%
\end{align*}%
namely, a process that is forced to keep the same value after a prescribed (random) time $\tau$.
%
%

\begin{figure}[b]
	\begin{center}
		\begin{adjustbox}{max width=.95\linewidth}%
			\scalebox{1.7}{
				\begin{tikzpicture}[wp/.style={inner sep=0pt,outer sep=2pt},
					foo/.style={%
						->,
						shorten >=1pt,
						shorten <=1pt,
						decorate,
						decoration={%
							snake,
							segment length=1.64mm,
							amplitude=0.2mm,
							pre length=2pt,
							post length=2pt,
						}
					}]
					\node[wp] (wpcomp) at (-1.98,0) {\tiny $\wp{\nblue{\nblue{\COMPOSE{\cc_1}{\cc_2}}}}{\orange{f}} =$};
					\node[wp] (wpc1c2) at (0,0) {\tiny $\wpC{\nblue{\cc_{\scaleto{1}{2.5pt}}}}\!\left(\vphantom{\big(}\wpC{\nblue{\cc_{\scaleto{2}{2.5pt}}}}(\orange{f})\right)$};
					\node (c1) at (1.5,0) {\tiny $\nblue{\cc_{\scaleto{1}{2.5pt}}}$};
					\node [wp,right of=c1] (wpc2) {\tiny $\wpC{\nblue{\cc_{\scaleto{2}{2.5pt}}}}(\orange{f})$};
					\node [right of=wpc2] (c2) {\tiny $\nblue{\cc_{\scaleto{2}{2.5pt}}}$};
					\node [wp,inner sep=0pt,outer sep=1pt] (f) at (4.1,0) {\tiny $\orange{f}$};
					
					\draw[foo] (f) to[out=120,in=40] (wpc2);
					\draw[foo] (wpc2) to[out=140,in=30] (wpc1c2);
				\end{tikzpicture}
			}
		\end{adjustbox}
	\end{center}
	\caption{Weakest preexpectation transformer in a backward continuation-passing style~\cite{DBLP:journals/pacmpl/HarkKGK20}.}
	\label{fig:wptrans}
\end{figure}

\section{Compositional Reasoning of weakest Preexpectations}\label{app:compositional-wp}

Weakest preexpectations per \cref{def:wp} can be determined in a \emph{compositional} manner \`{a} la McIver and Morgan's \emph{weakest preexpectation calculus}. 
As explained in \cite{DBLP:journals/pacmpl/HarkKGK20}, this calculus employs \emph{expectation transformers} which move \emph{backward} through the program in a \emph{continuation-passing} style, see an illustration in \cref{fig:wptrans}. If we are interested in the expected value of some postexpectation $f$ after executing the sequential composition $\COMPOSE{\cc_1}{\cc_2}$, then we can first compute the weakest preexpectation of $\cc_2$ w.r.t.\ $f$, i.e., $\wp{\cc_2}{f}$. Thereafter, we can use the intermediate result $\wp{\cc_2}{f}$ as \emph{postexpectation} to determine the weakest preexpectation of $\cc_1$ w.r.t.\ $\wp{\cc_2}{f}$. Overall, we obtain the weakest preexpectation of $\COMPOSE{\cc_1}{\cc_2}$ w.r.t.\ postexpectation $f$. The example below demonstrates the application of the $\wpsymbol$ calculus in such a backward, compositional manner.
\begin{example}[Applying the \textnormal{$\boldwpsymbol$} Calculus] Consider the probabilistic program $\cc$ given by
	\begin{align*}
		\COMPOSE{\PASSIGNUNIFORM{p}{0}{1}}{\,\ASSIGN{n}{n-1}~[p]~\ASSIGN{n}{n+1}}~,
	\end{align*}%
	where $\UNIFORM{0}{1}$ is the continuous uniform distribution between 0 and 1. Suppose we aim to infer the expected value of $n$ upon termination of $\cc$ by determining $\wp{\cc}{n}$. By rules in \cref{table:wp},
	\begin{align*}
		\wp{\cc}{n} &\eeq \wp{\PASSIGNUNIFORM{p}{0}{1}}{\wp{\ASSIGN{n}{n-1}~[p]~\ASSIGN{n}{n+1}}{n}}\\
		&\eeq \wp{\PASSIGNUNIFORM{p}{0}{1}}{p\cdot \left(n-1\right) + \left(1-p\right)\cdot \left(n+1\right)}\\
		&\eeq \wp{\PASSIGNUNIFORM{p}{0}{1}}{n+1-2p}\\
		&\eeq \int_{[0,1]}\left(n+1-2p\right) \dif\,p \eeq n~.
	\end{align*}%
	This tells us that the expected value of $n$ upon termination of $\cc$ is $n$, i.e., $\cc$ does not change the initial value of $n$ in expectation.
	\qedT
\end{example}

\section{Additional Proofs}\label{app:proofs}

\subsection{Proof of~\textnormal{\cref{eq:proof-D}} in~\textnormal{\cref{thm:exact_diff}}}\label{app:proof_equ12}
\begin{lemma}\label{lem:equality_for_D}
Following notations in \textnormal{\cref{thm:exact_diff}}, let $\phi_1$ and $\phi_2$ be arbitrary two predicates, we have
\begin{equation}
	\lambda \pstate\mydot \int_{D(\neg \phi_1,\phi_2)} \fsub{\cloop} \dif\,\left(\probability{\pstate}\right) \eeq \wp{\WHILEDO{\phi_1\wedge\phi_2}{\cc}}{[\neg\phi_1\wedge \phi_2] \cdot f}~.
\end{equation}
\end{lemma}
\begin{proof}
Let $\cloop^{\wedge}$ denote loop $\WHILEDO{\phi_1\wedge\phi_2}{\cc}$, and for any $l\colon \States\to \PosRealsInf$, let $\,l_{\cloop^{\wedge}}$ represent the partial function mapping a trace $s\in \SS$ to $l(\pstate_n)$ if $s$ hits $\neg(\phi_1\wedge\phi_2)$ for the first time at $\pstate_n$, that is
\begin{equation*}
	l_{\cloop^{\wedge}} \colon \SS \rightharpoonup \PosRealsInf \qquad
	\pstate_0 \pstate_1 \cdots \pstate_i \cdots \mapsto g(\pstate_n) \quad \textnormal{if}~\left(\pstate_n \models \neg(\phi_1\wedge\phi_2)\right) \land \left(\forall i < n \colon \pstate_i \models (\phi_1\wedge\phi_2)\right)~.
\end{equation*}
Let $h \defeq [\neg\phi_1\wedge \phi_2] \cdot f$. It is checked without difficulties that $h_{\cloop^{\wedge}}(s) = \fsub{\cloop^{\wedge}}(s)$ for $s\in D(\neg \phi_1,\phi_2)$; $h_{\cloop^{\wedge}}(s) = 0$ for $s\in D(\neg \phi_2,\phi_1)\cup C$. By definition of the (sub)probability measure $\probability{\pstate}$, we have
\begingroup
\allowdisplaybreaks
\begin{align*}
	\wp{\cloop^{\wedge}}{[\neg\phi_1\wedge \phi_2] \cdot f} &\eeq \lambda \pstate\mydot \int_{\event(\neg (\phi_1\wedge\phi_2))} h_{\cloop^\wedge} \dif\,\left(\probability{\pstate}\right)\\
	& \eeq \lambda \pstate\mydot \int_{D(\neg \phi_1,\phi_2) \uplus D(\neg \phi_2,\phi_1)\uplus C} h_{\cloop^\wedge} \dif\,\left(\probability{\pstate}\right) 
	\\ 
	& \eeq \lambda \pstate\mydot \int_{D(\neg \phi_1,\phi_2)} h_{\cloop^\wedge} \dif\,\left(\probability{\pstate}\right) +\lambda \pstate\mydot \int_{ D(\neg \phi_2,\phi_1)\cup C} h_{\cloop^\wedge} \dif\,\left(\probability{\pstate}\right) \TAG{by linearity of $\int$}\\
	& \eeq \lambda \pstate\mydot \int_{D(\neg \phi_1,\phi_2)} f_{\cloop^\wedge} \dif\,\left(\probability{\pstate}\right) \TAG{by definition of $h_{\cloop^{\wedge}}$}~,
\end{align*}%
\endgroup%
where the second equality is due to set decomposition: 
\[
	\event\left(\neg \left(\phi_1\wedge\phi_2\right)\right) \eeq D\left(\neg \phi_1,\phi_2\right)\uuplus D\left(\neg \phi_2,\phi_1\right)\uuplus C~.
\]%
This completes the proof.
\end{proof}

\subsection{Proof of \textnormal{\cref{thm:approx_diff_lower}}}\label{app:proof_restateApproxDiffLower}
\restateApproxDiffLower*

\begin{proof}
	Since $\guard' \!\implies\! \guard$, we have $\guard \wedge \guard' = \guard'$, $[\neg\guard \wedge \guard'] = 0$, and $B = \emptyset$ (i.e., no trace can ever hit $\neg\guard$ before hitting $\neg\guard'$). Thus, by \cref{thm:exact_diff},
	\begin{align}
		\wp{\cloop}{f} - \wp{\pcloop}{f}
		&\eeq \lambda \pstate \mydot \int_{A}~\fsub{\cloop} \dif\,\left(\probability{\pstate}\right) - \wp{\WHILEDO{\guard'}{\cc}}{\left[\guard\wedge\neg \guard'\right] \cdot f}\notag\\
		&\eeq \lambda \pstate \mydot \int_{A}~\fsub{\cloop} \dif\,\left(\probability{\pstate}\right) - \wp{\WHILEDO{\guard'}{\cc}}{\left[\guard\right] \cdot f}\TAG{by definition of $\wpsymbol$-transformer, cf.\ \cref{table:wp}}\\
		&\eeq \lambda \pstate \mydot \int_{A}~\fsub{\cloop} \dif\,\left(\probability{\pstate}\right) - \wp{\pcloop}{\left[\guard\right] \cdot f}~.\notag
	\end{align}%
	It follows that
	\begingroup
	\allowdisplaybreaks
	\begin{align*}
		\wp{\cloop}{f} &\eeq \wp{\pcloop}{f} - \wp{\pcloop}{\left[\guard\right] \cdot f} + \lambda \pstate \mydot \int_{A}~\fsub{\cloop} \dif\,\left(\probability{\pstate}\right)\notag\\
		&\eeq \wp{\pcloop}{\left[\neg\guard\right] \cdot f} + \lambda \pstate \mydot \int_{A}~\fsub{\cloop} \dif\,\left(\probability{\pstate}\right)\TAG{by linearity of $\wpsymbol$-transformer, cf., e.g., \cite[Chap.\ 8.5]{DBLP:phd/dnb/Kaminski19}}\\
		&\ssucceq \wp{\pcloop}{\left[\neg\guard\right] \cdot f}~.\TAG{$\fsub{\cloop} \geq 0$}
	\end{align*}%
	\endgroup%
	This completes the proof.%
\end{proof}%

\subsection{Trace-Agnostic Proof of Our Lower Bound Rule in \cref{thm:inference_rule}}\label{app:proof_trace_agnostic}
\restateInferenceRule*
 
\begin{proof}[Proof (trace-agnostic)]
It suffices to show that $\wp{\pcloop}{[\neg \guard] \cdot f} \preceq \wp{\cloop}{f}$. By definition of the $\wpsymbol$-transformer (see \cref{defQQQwp}), we have 
\begin{align*}
	\wp{\pcloop}{[\neg \guard] \cdot f} &\eeq \lfp\charwp{\guard'}{\cc}{[\neg \guard] \cdot f} \eeq  \lim_{k\to \omega} \left(\charwp{\guard'}{\cc}{[\neg \guard] \cdot f}\right)^k\left(0\right)~,
\end{align*}%
and, analogously,
\begin{align*}
	\wp{\cloop}{f} & \eeq \lfp \charwp{\guard}{\cc}{f} \eeq \lim_{k\to \omega} \left(\charwp{\guard}{\cc}{f}\right)^k\left(0\right)~.
\end{align*}%
We now prove by induction that, for any $k\in \NN$,
\begin{equation}\label{eq:induction_goal}
	\left(\charwp{\guard'}{\cc}{[\neg \guard] \cdot f}\right)^k\left(0\right) \ppreceq \lim_{k\to \omega} \left(\charwp{\guard}{\cc}{f}\right)^k\left(0\right)~.
\end{equation}%
If this is the case, $\wp{\pcloop}{[\neg \guard] \cdot f} \preceq \wp{\cloop}{f}$ follows by taking the limit over $k$ on both sides of \cref{eq:induction_goal}. For the base step of $k=0$, induction goal \cref{eq:induction_goal} trivial holds:
\begin{equation}
	\left(\charwp{\guard'}{\cc}{[\neg \guard] \cdot f}\right)^0\left(0\right) \eeq 0 \eeq \left(\charwp{\guard}{\cc}{f}\right)^0\left(0\right)~.
\end{equation}%
For the induction step, suppose \cref{eq:induction_goal} holds for $k=i$, then for $k=i+1$, we have
\begin{align*}
	\left(\charwp{\guard'}{\cc}{[\neg \guard] \cdot f}\right)^{i+1}\left(0\right) &\eeq [\neg \guard]\cdot [\neg \guard']  \cdot f+ [\guard']\cdot\wp{\cc}{  \left(\charwp{\guard'}{\cc}{[\neg \guard] \cdot f}\right)^{i}\left(0\right)}\\
	&\ppreceq [\neg \guard] \cdot f + [\guard]\cdot \wp{\cc}{\left(\charwp{\guard}{\cc}{f}\right)^{i}\left(0\right)}\\
	& \eeq\left(\charwp{\guard}{\cc}{f}\right)^{i+1}\left(0\right)~.
\end{align*}%
Thus, \cref{eq:induction_goal} holds for any $k\in \NN$. This completes the proof.
\end{proof}

\subsection{Detailed Proof of~\textnormal{\cref{lem:exponential_decrease}}}\label{app:proof_AST}
\restateASTWitness*

\begin{proof}
	We first show that the probability that $\pcloop$ does not terminate within $n \in \NN$ steps decreases exponentially in $n$. To this end, we have, for any $k \in \NN$,
	\begin{align*}
		\probability{\pstate}\left(T^{\neg\guard'} > k N\right) &\eeq \probability{\pstate}\left(\forall i \leq k N\colon X_i \models \guard'\right)\\
		&\lleq \max_{X_0,X_{N},\ldots,X_{(k-1)N} \models \guard'} \probability{X_0}\left(T^{\neg\guard'} > N\right)\cdot \probability{X_N}\left(T^{\neg\guard'} > N\right) \cdots  \probability{X_{(k-1)N}}\left(T^{\neg\guard'} > N\right)\\
		&\lleq p\cdot p \cdots p\TAG{by \cref{eq:ast-witness}}\\
		&\eeq p^k~.
	\end{align*}%
	It follows that, for any $n \in \NN$ and any initial state $\pstate \in \States$,
	\begin{equation*}
		\probability{\pstate}\left(T^{\neg\guard'} > n\right) \lleq p^{\floor{\nicefrac{n}{N}}} \lleq p^{\nicefrac{n}{N} -1} \eeq \nicefrac{1}{p} \cdot \mathrm{e}^{\log(p)\cdot \nicefrac{n}{N}}~.
	\end{equation*}%
	For any fixed $p \in (0,1)$, we have $\log(p) < 0$, and thus $\probability{\pstate}(T^{\neg\guard'} > n)$ decreases exponentially in $n$.
	Let $n$ tend to infinity, we have
	\begin{equation*}
		\probability{\pstate}\left(T^{\neg\guard'} < \infty\right) \eeq 1 - \lim_{n \to \infty} \probability{\pstate}\left(T^{\neg\guard'} > n\right) \ggeq 1 - \lim_{n \to \infty} \nicefrac{1}{p} \cdot \mathrm{e}^{\log(p)\cdot \nicefrac{n}{N}} \eeq 1 - 0 \eeq 1~.
	\end{equation*}%
	Hence, $\pcloop$ terminates almost-surely.
\end{proof}

\subsection{Proof of~\textnormal{\cref{thm:additional_conditions_uniform}}}\label{app:proof_addui}
\restateSuffUI*

\begin{proof}
	Let $\probability{\pstate}$ be the probability measure induced by $\pcloop$ on initial state $\pstate \in \States$, $T^{\neg \guard'}$ be the looping time of $\pcloop$, and $\big\{X_n^{T^{\neg \guard'}}\big\}_{n\in \NN}$ be the corresponding stopped process. By \cref{def:ui}, we need to show that $\big\{h\big(X_n^{T^{\neg \guard'}}\big)\big\}_{n\in \NN}$ is uniformly integrable on $\probability{\pstate}$.
	Due to \cite[Thm.~13.3]{williams1991probability}, it suffices to show that the stochastic process $\big\{h\big(X_n^{T^{\neg \guard'}}\big)\big\}_{n\in \NN}$  is \emph{dominated} by an integrable random variable, i.e., $\big\lvert h\big(X_n^{T^{\neg \guard'}}\big)\big\rvert$ is bounded by a random variable whose expected value is finite.
	
	By the bounded update property of $\pcloop$, there exists $c \in \PosReals$ such that
	\begin{align*}
	\forall n \in \NN\colon\ \ \absd{X_n} \lleq \absd{X_n - X_{n-1}} + \absd{X_{n-1} - X_{n-2}} + \ldots + \absd{X_1- X_{0}} + \absd{X_0} \lleq c \cdot n + \absd{\pstate}
	\end{align*}%
	holds almost-surely. Moreover, since $h$ is bounded by a polynomial expectation, we have
	\begin{align*}
	\absd{h\left(X_{n+1}\right) - h\left(X_n\right)} \lleq \poly\left(n\right)
	\end{align*}%
	for some polynomial $\poly$ in $n$. Furthermore, observe that
	\begin{align}\label{eq:dominance}
	\big\lvert h\big(X_n^{T^{\neg \guard'}}\big)\big\rvert \lleq \absd{h(X_0)} + \sum\nolimits_{k=0}^{\infty} \absd{h\left(X_{k+1}\right) - h\left(X_k\right)} \cdot  \big[T^{\neg \guard'} > k\big]~.
	\end{align}%
	It then suffices to show that the right-hand side of \cref{eq:dominance} has a finite expected value. Observe that
	\begingroup
	\allowdisplaybreaks
	\begin{align*}
	\expectv{\pstate}\left[\absd{h\left(X_{k+1}\right) - h\left(X_k\right)} \cdot  \big[T^{\neg \guard'} > k\big]\right] &\lleq \expectv{\pstate} \left[\poly\left(k\right) \cdot \big[T^{\neg \guard'} > k\big]\right] \\
	&\eeq \poly\left(k\right) \cdot \probability{\pstate}\left(T^{\neg\guard'} > k\right)\\
	&\lleq \poly\left(k\right) \cdot a \cdot \mathrm{e}^{-b \cdot k}~.\TAG{for $a,b \in \StrictPosReals$, cf.\ proof of \cref{lem:exponential_decrease}}
	\end{align*}%
	\endgroup%
	It follows that the right-hand side of \cref{eq:dominance} has a finite expected value, i.e.,
	\begin{align*}
	\expectv{\pstate}\left[\absd{h(X_0)} + \sum\nolimits_{k=0}^{\infty} \absd{h\left(X_{k+1}\right) - h\left(X_k\right)} \cdot \big[T^{\neg \guard'} > k\big]\right] &\lleq \expectv{\pstate}[\absd{h(X_0)}] + \sum\nolimits_{k=0}^{\infty} \poly\left(k\right) \cdot a \cdot \mathrm{e}^{-b \cdot k}\\
	&\LL \infty~.
	\end{align*}%
	This completes the proof.
\end{proof}

\subsection{Detailed Proof of~\textnormal{\cref{thm:approx_diff_upper}}}\label{app:proof_diff_upper}
\restateApproxUpper*

\begin{proof}
	As shown in the proof of \cref{thm:approx_diff_lower}, we have 
	\begin{align*}
		\wp{\cloop}{f} -  \wp{\pcloop}{[\neg\guard] \cdot f} \eeq \lambda \pstate \mydot \int_{A}~\fsub{\cloop} \dif\,\left(\probability{\pstate}\right)~.
	\end{align*}%
	It thus suffices to bound the integral above. Recall that, in case $\guard' \!\implies\! \guard$, $A \subseteq \SS$ is the set of traces that hit $\neg \guard' \wedge \guard$ before hitting $\neg \guard$, and $\fsub{\cloop}$ maps a trace $\trace \in \SS$ to $f(\pstate_n)$ if $\trace$ hits $\neg \guard$ for the first time at $\pstate_n$. Let $X$ map a trace $\trace \in \SS$ to $\pstate_n$ if $\trace$ hits $\neg \guard'$ for the first time at $\pstate_n$, i.e., $X(\pstate_0\pstate_1\cdots\pstate_i\cdots)\defeq \sigma_n$ if $(\pstate_n \models \neg\guard') \land (\forall i < n \colon \pstate_i \models \guard')$, and let $[\guard]_{\pcloop}$ map a trace $\trace \in \SS$ to $[\guard](X(s))$. We have
	\begingroup
	\allowdisplaybreaks
	\begin{align*}
		\lambda \pstate \mydot\int_{A}~\fsub{\cloop} \dif\,\left(\probability{\pstate}\right) &\eeq \lambda \pstate \mydot\int_{\event \neg\guard'}~\left([\guard]_{\pcloop}\cdot \lambda s\mydot\int_{\event \neg\guard}~\fsub{\cloop}\dif\,\left(\probability{X(s)}\right) \right) \dif\,\left(\probability{\pstate}\right)\\
		&\eeq \lambda \pstate \mydot\int_{\event \neg\guard'}~\left([\guard]_{\pcloop}\cdot \lambda s\mydot \wp{\cloop}{f}\left(X\left(s\right)\right) \right) \dif\,\left(\probability{\pstate}\right)\\
		&\ppreceq  \lambda \pstate \mydot\int_{\event \neg\guard'}~\left([\guard]_{\pcloop}\cdot \sup_{\pstate \models \neg\guard' \wedge \guard}\wp{\cloop}{f}\left(\pstate\right) \right) \dif\,\left(\probability{\pstate}\right)\\
		&\eeq  \lambda \pstate \mydot\int_{\event \neg\guard'}~\left([\guard]_{\pcloop} \right) \dif\,\left(\probability{\pstate}\right)\cdot \sup_{\pstate \models \neg\guard' \wedge \guard}\wp{\cloop}{f}\left(\pstate\right)\\
		&\eeq \wp{\pcloop}{[\guard]} \cdot \sup_{\pstate \models \neg\guard' \wedge \guard} \wp{\cloop}{f}\left(\pstate\right)~.
	\end{align*}%
	\endgroup%
	This completes the proof.
\end{proof}

\section{Details on the Examples}\label{app:subinvariance}

\subsection{Checking Sub- and Superinvariance for~\textnormal{\cref{ex:biased_RW}}}\label{app:check_1dbrw}

We present detailed calculations for checking subinvariance of $l_M$ and superinvariance of $l_\infty$. We first check the validity of $l_M\preceq\charfunaked{g}^M(l_M)$. By definition of $\charfunaked{g}^M$, we have
\begingroup
\allowdisplaybreaks
\begin{align*}
	\charfunaked{g}^M\left(l_M\right) &\eeq [n\leq 0\vee n\geq M] \cdot [n\leq 0] + [0 < n < M]\cdot \left(\nicefrac{1}{3}\cdot l_M\subst{n}{n-1} + \nicefrac{2}{3}\cdot l_M\subst{n}{n+1} \right)\\
	& \eeq [n\leq 0] + [0 < n < M]\cdot \left(\nicefrac{1}{3}\cdot \left(\left(\nicefrac{1}{2}\right)^{n-1} - \left(\nicefrac{1}{2}\right)^M\right) + \nicefrac{2}{3}\cdot\left(\left(\nicefrac{1}{2}\right)^{n+1} - \left(\nicefrac{1}{2}\right)^M\right)\right)\\
	& \eeq [n\leq 0] + [0 < n < M]\cdot\left(\left(\nicefrac{1}{2}\right)^n - \left(\nicefrac{1}{2}\right)^M\right)\\
	& \ssucceq [n < 0] +
	[0 \leq n \leq M] \cdot \left(\left(\nicefrac{1}{2}\right)^n - \left(\nicefrac{1}{2}\right)^M\right)\\
	& \eeq l_M~.
\end{align*}%
\endgroup%
For the validity of $\charfunaked{f} (l_\infty) \preceq l_\infty$, we have
\begin{align*}
\charfunaked{f} (l_\infty) & \eeq [n\leq 0]\cdot 1 + [n>0 ]\cdot \left(\nicefrac{1}{3}\cdot l_\infty\subst{n}{n-1} + \nicefrac{2}{3}\cdot l_\infty\subst{n}{n+1} \right)\\
& \eeq [n\leq 0] + [n>0]\cdot\left(\nicefrac{1}{3}\cdot \left(\nicefrac{1}{2}\right)^{n-1}+ \nicefrac{2}{3}\cdot \left(\nicefrac{1}{2}\right)^{n+1}\right)\\
& \eeq [n\leq 0] + [n>0]\cdot\left(\nicefrac{1}{2}\right)^n \\
& \eeq [n< 0] + [n\geq 0]\cdot\left(\nicefrac{1}{2}\right)^n\\
&\eeq l_\infty~.
\end{align*}%

\subsection{Checking Subinvariance for~\textnormal{\cref{ex:pd}}}\label{app:check_pd}

We present detailed calculations for checking subinvariance of $l_M$, i.e., $l_M \preceq \charfunaked{g}^M(l_M)$. Recall
\begin{align*}
	l_M \eeq [a\neq 1] \cdot b + \sum\nolimits_{i = 1}^{\lceil \log_2 M \rceil}\left[a =1 \wedge \left(\nicefrac{M}{2^i} \leq b < \nicefrac{M}{2^{i-1}}\right)\right] \cdot i \cdot \nicefrac{b}{2}~.
\end{align*}%
By definition of $\charfunaked{g}^M$, we have
\begingroup
\allowdisplaybreaks
\begin{align*}
	\charfunaked{g}^M\left(l_M\right) &\eeq [a\neq 1 \vee b\geq M]\cdot [a\neq 1]\cdot b +[a=1\wedge b<M]\cdot \left(\nicefrac{1}{2}\cdot l_\infty\subst{a}{0} + \nicefrac{1}{2}\cdot l_\infty\subst{b}{2b}\right)\\
	&\eeq [a\neq 1] \cdot b + [a=1\wedge b<M]\cdot \left(\nicefrac{1}{2}\cdot b + \nicefrac{1}{2}\cdot \sum\nolimits_{i = 1}^{\lceil \log_2 M \rceil} [\nicefrac{M}{2^i} \leq 2b < \nicefrac{M}{2^{i-1}}] \cdot i\cdot b \right)\\
	&\eeq [a\neq 1] \cdot b + [a=1\wedge b<M]\cdot \left(\nicefrac{1}{2}\cdot b + \nicefrac{1}{2}\cdot \sum\nolimits_{i = 1}^{\lceil \log_2 M \rceil} [\nicefrac{M}{2^{i+1}} \leq b < \nicefrac{M}{2^{i}}] \cdot i\cdot b \right)\\
	&\eeq [a\neq 1] \cdot b + [a=1\wedge b<M]\cdot \left(\sum\nolimits_{i = 0}^{\lceil \log_2 M \rceil} [\nicefrac{M}{2^{i+1}} \leq b < \nicefrac{M}{2^{i}}] \cdot (i+1)\cdot \nicefrac{b}{2} \right)\\
	&\eeq [a\neq 1] \cdot b + [a=1\wedge b<M]\cdot \left(\sum\nolimits_{i = 1}^{\lceil \log_2 M \rceil +1} [\nicefrac{M}{2^{i}} \leq b < \nicefrac{M}{2^{i-1}}] \cdot 1\cdot \nicefrac{b}{2} \right)\\
	&\eeq [a\neq 1] \cdot b + \sum\nolimits_{i = 1}^{\lceil \log_2 M \rceil}\left[a =1 \wedge \left(\nicefrac{M}{2^i} \leq b < \nicefrac{M}{2^{i-1}}\right)\right] \cdot i \cdot \nicefrac{b}{2}\\
	& \eeq  l_M~.
\end{align*}%
\endgroup%

\subsection{Checking Subinvariance and Non-Conditional Difference Boundedness for~\textnormal{\cref{ex:additional-ui}}}\label{app:check_bu}

We present detailed computations for checking subinvariance of $l_M$, i.e., $l_M \preceq \charfunaked{g}^M(l_M)$ and show that $l_M$ is \emph{not} conditionally difference bounded.
Recall
\begin{align*}
	l_M \eeq [n<0] \cdot b + [0 \leq n \leq M] \cdot \left(b+2 \cdot n\right) \cdot \left(1 - \nicefrac{n}{M}\right)~.
\end{align*}%
By definition of $\charfunaked{g}^M(l_M)$, we have
\begingroup
\allowdisplaybreaks
\begin{align*}
	\charfunaked{g}^M\left(l_M\right)&\eeq [n \le 0]\cdot b+[0 < n < M]\cdot\left(\nicefrac{1}{2}\cdot l_M\subst{n}{n-1}\subst{b}{b+2}+\nicefrac{1}{2}\cdot l_M\subst{n}{n+1}\subst{b}{b-2}\right)\\
	&\eeq [ n \le 0 ] \cdot b  +  \nicefrac{1}{2}\cdot [0 < n < M] \cdot \left(b+2n\right) \cdot \left(\left(1 - \nicefrac{n-1}{M}\right)  +  \left(1-\nicefrac{n+1}{M}\right)\right)\\
	&\eeq [n \le 0] \cdot b  +  [0 < n < M] \cdot (b+2n) \cdot \left(1-\nicefrac{n}{M}\right)\\
	&\eeq [n < 0] \cdot b  +  [0 \le n \le M]\cdot(b+2n) \cdot \left(1-\nicefrac{n}{M}\right)\\
	&\eeq l_M~.
\end{align*}%
\endgroup%
Thus, $	l_M \preceq \charfunaked{g}^M(l_M)$ holds.

We now show that $l_M$ is not conditionally difference bounded. 
Let $\cc_{\textnormal{bubody}}$ denote the loop body of $\cc_\textnormal{bu}$ in \cref{ex:additional-ui}.
Recall $l_M$ is conditionally difference bounded if there exists $c\in\RR_{>0}$ such that $\wp{\cc_\textnormal{bubody}}{|l_M-l_M(\pstate)|}\le c$ for any $\pstate$ satisfying the loop guard of $\cc^M_{\textnormal{bu}}$. 
We show that
\begingroup
\allowdisplaybreaks
\begin{align*}
	\Delta l_M \eeq [0 < n < M]\cdot\wp{\cc_{\textnormal{bubody}}}{|l_M-l_M(\pstate)|}
\end{align*}%
is \emph{unbounded}. Since
\begin{align*}
	\Delta l_M & \eeq [0 < n < M] \cdot \frac{1}{2}\left( \left| l_M - l_M(\pstate) \right|\subst{n}{n-1}\subst{b}{b+2} + \left| l_M - l_M(\pstate) \right|\subst{n}{n+1}\subst{b}{b-2} \right)\\
	& \eeq [0 < n < M] \cdot \frac{1}{2} \left( |l_M\subst{n}{n-1}\subst{b}{b+2}-l_M|(\pstate) + |l_M\subst{n}{n+1}\subst{b}{b-2}-l_M|(\pstate) \right)\\
	& \eeq [0 < n < M] \cdot \frac{1}{2} \left(\left|(b+2n)\frac{1}{M}\right|(\pstate) + \left|(b+2n)\frac{1}{M}\right|(\pstate)\right)	\\
	& \eeq [0 < n < M] \cdot \frac{|b+2n|}{M}(\pstate)~.
\end{align*}%
\endgroup%
$\Delta l_M$ is unbounded as $b$ can be arbitrarily large over the guard $(0<n<M)$. $l_M$ is thus not conditionally difference bounded.  

\subsection{Checking Subinvariance for~\textnormal{\cref{ex:continuous}}}\label{app:check_continuous}

We present detailed calculations for checking subinvariance of $l_M$, i.e., $l_M \preceq \charfunaked{g}^M(l_M)$. Recall
\begin{align*}
	l_M \eeq [n<0] \cdot b + [0 \leq n \leq M] \cdot \left(b+\nicefrac{1}{2} \cdot n\right) \cdot \left(1 - \nicefrac{n}{M}\right)~.
\end{align*}%
By definition of $\charfunaked{g}^M$, we have
\begingroup
\allowdisplaybreaks
\begin{align*}
	\charfunaked{g}^M\left(l_M\right)
	=\,& [n\le 0]\cdot b  +  [0 < n < M]\cdot\\
	&\int_0^1\left(p \cdot \!\int_0^1\! l_M\subst{n}{n-1}\subst{b}{b+r_1}\dif\,r_1+(1-p) \cdot \!\int_0^1\! l_M\subst{n}{n+1}\subst{b}{b-r_2}\dif\,r_2 \right)\!\dif\,p\\
	=\,& [n\le 0]\cdot b + [0 < n < M]\cdot\left(\nicefrac{1}{2}\cdot l_M\subst{n}{n-1}\subst{b}{b+\nicefrac{1}{2}}+\nicefrac{1}{2}\cdot l_M\subst{n}{n+1}\subst{b}{b-\nicefrac{1}{2}} \right)\\
	=\,& [n\le 0]\cdot b + [0 < n < M] \cdot \left(b+\nicefrac{1}{2}\cdot n\right)\cdot\left(1-\nicefrac{n}{M}\right)\\
	=\,& [n<0] \cdot b + [0 \leq n \leq M] \cdot \left(b+\nicefrac{1}{2} \cdot n\right) \cdot \left(1 - \nicefrac{n}{M}\right) \eeq l_M~.
\end{align*}%
\endgroup%

\subsection{Checking Subinvariance for~\textnormal{\cref{ex:nested_loop}}}\label{app:check_nl}

We present detailed calculations for checking subinvariance of $l_M$. To this end, we first prove that
\begingroup
\allowdisplaybreaks
\begin{align}\label{eq:inner_loop_lowerbound}
		\wp{{\cc^M_{\textnormal{inner}}}}{h} &\ssucceq 
		\begin{aligned}[t]
			&[\left(k-n < -a\right)\vee \left(k-n> b\right)]\cdot h \,+ \\
			&[-a\leq k-n\leq b]\cdot \left( \frac{b-(k-n)}{b+a} \cdot h\subst{k}{n-a} +  \frac{(k-n)+a}{b+a} \cdot h\subst{k}{n+b} \right)
		\end{aligned}\notag\\
	&\ddefeq l_{\textnormal{inner}}
\end{align}%
\endgroup%
holds for any bounded postexpectation $h\in \Expectations$, where $\cc^M_{\textnormal{inner}}$ is the inner loop of $\cc^M_{\textnormal{nl}}$, i.e. $$\cc^M_{\textnormal{inner}} = \WHILEDO{-a< k-n < b}{\ASSIGN{k}{k-1}~[\nicefrac{1}{2}]~\ASSIGN{k}{k+1}}~.$$
Since $\cc^M_{\textnormal{inner}}$ terminates after at most $a$ steps of consecutively decreasing $k$ with probability $(\nicefrac{1}{2})^{a}$, we know it terminates almost-surely by \cref{lem:exponential_decrease}. Moreover, $l_{\textnormal{inner}}$ is bounded, thus, by \cref{thm:lower_bound_hark,thm:sufficient_conditions_uniform}, $l_{\textnormal{inner}}\preceq \wp{{\cc^M_{\textnormal{inner}}}}{h}$ if $l_{\textnormal{inner}}$ is a subinvariant of $\cc^M_{\textnormal{inner}}$\! with respect to postexpectation $h$. In fact, 
\begingroup
\allowdisplaybreaks
\begin{align*}
	\charfunaked{h}^{\textnormal{inner}}\left(l_\textnormal{inner}\right)&\eeq
	\begin{aligned}[t]
		&[\left(k-n\leq -a\right)\vee \left(k-n\geq b\right)]\cdot h +\\ 
		& [-a<k-n<b]\cdot \left( \nicefrac{1}{2}\cdot l_{\textnormal{inner}}\subst{k}{k-1} + \nicefrac{1}{2}\cdot l_{\textnormal{inner}}\subst{k}{k+1} \right)
	\end{aligned}\\
	&\eeq
	\begin{aligned}[t]
		&[\left(k-n\leq -a\right)\vee \left(k-n\geq b\right)]\cdot h\; +\\ 
		& [-a < k-n<b]\cdot \left(\frac{b-(k-n)}{b+a} \cdot h\subst{k}{n-a} +  \frac{(k-n)+a}{b+a} \cdot h\subst{k}{n+b}  \right)
	\end{aligned}\\
	&\eeq 
	\begin{aligned}[t]
		&[\left(k-n< -a\right)\vee \left(k-n> b\right)]\cdot h\; +\\ 
		&[-a \leq k-n \leq b]\cdot \left(\frac{b-(k-n)}{b+a} \cdot h\subst{k}{n-a} +  \frac{(k-n)+a}{b+a} \cdot h\subst{k}{n+b}  \right)\\
	\end{aligned}\\
	&\eeq l_\textnormal{inner} ~.
\end{align*}%
\endgroup%
It follows that $l_\textnormal{inner}$ is a lower bound on $\wp{{\cc^M_{\textnormal{inner}}}}{h}$. We now continue the computation in \cref{ex:nested_loop} to check the subinvariance of $l_M$:
\begingroup
\allowdisplaybreaks
\begin{align*}
	\charfunaked{g}^M\left(l_M\right) &\eeq [n\leq 0\vee n\geq M]\cdot [n\leq 0] + [0<n<M]\cdot \wp{\cc^M_{\textnormal{body}}}{l_M}\\
	&\eeq 
	\begin{aligned}[t]
		&[n\leq 0] + [0<n<M]\cdot\\
		&\wp{\COMPOSE{\ASSIGN{k}{n}}{\cc^M_{\textnormal{inner}}}}{[k<n] \cdot l_M\subst{n}{n-1} + [k\geq n]\cdot l_M\subst{n}{n+1} }\\
	\end{aligned}\\	
	&\eeq 
	\begin{aligned}[t]
		&[n\leq 0] + [0<n<M]\cdot\\
		&\wp{\ASSIGN{k}{n}}{\wp{{\cc^M_{\textnormal{inner}}}}{[k<n] \cdot l_M\subst{n}{n-1} + [k\geq n]\cdot l_M\subst{n}{n+1} }}\\
	\end{aligned}\\
	&\eeq 
	\begin{aligned}[t]
		&[n\leq 0] + [0<n<M]\cdot\\
		&\Big(\underbrace{\wp{{\cc^M_{\textnormal{inner}}}}{[k<n] \cdot l_M\subst{n}{n-1} + [k\geq n]\cdot l_M\subst{n}{n+1} }}_{\textnormal{apply \cref{eq:inner_loop_lowerbound}}}\Big)\subst{k}{n}\\
	\end{aligned}\\
	&\eeq [n\leq 0] + [0<n<M]\cdot \left( \left(\nicefrac{b}{a}\right)^n -\left(\nicefrac{b}{a}\right)^M\right)\\
	&\ssucceq [n<0] + [0\leq n\leq M]\cdot \left( \left(\nicefrac{b}{a}\right)^n -\left(\nicefrac{b}{a}\right)^M\right)\\
	&\eeq l_M~.
\end{align*}%
\endgroup%
It follows that $l_M$ is a subinvariant.

\end{document}